\documentclass[12pt]{iopart}
\usepackage{iopams}  
\newcommand{\bm}[1]{\mbox{\boldmath$#1$}}
\newcommand{\be}{\begin{equation}}
\newcommand{\ee}{\end{equation}}
\newcommand{\bea}{\begin{eqnarray}}
\newcommand{\eea}{\end{eqnarray}}
\newcommand{\non}{\nonumber}
\newcommand{\ra}{\rangle}

\newtheorem{df}{Definition}
\newtheorem{th1}[df]{Theorem}
\newtheorem{lem}[df]{Lemma}
\newtheorem{conj}[df]{Conjecture}
\newtheorem{prop}[df]{Proposition}
\newtheorem{cor}[df]{Corollary}
\newtheorem{lem2}{Lemma}[section]
\begin{document}
\article[XXZ Bethe states as highest weight]{}
{Regular XXZ Bethe states at roots of unity 
as highest weight vectors of the $sl_2$ loop algebra}
\author{Tetsuo Deguchi}                     
\date{}
\address{Department of Physics, Ochanomizu University, \\
2-1-1 Ohtsuka, Bunkyo-ku,Tokyo 112-8610, Japan
} 
\ead{deguchi@phys.ocha.ac.jp} 
\begin{abstract} 
We show that every regular Bethe ansatz eigenvector 
of the XXZ spin chain at roots of unity is a highest weight vector  
of the $sl_2$ loop algebra, for some restricted sectors  
with respect to eigenvalues 
of the total spin operator $S^Z$, and evaluate explicitly 
the highest weight in terms of the Bethe roots.   
We also discuss whether a given regular Bethe state in the sectors  
generates an irreducible representation or not.  
In fact, we present such a regular Bethe state in the inhomogeneous case 
that generates a reducible Weyl module. 
Here, we call a solution of the Bethe ansatz 
equations which is given by a set of distinct and finite rapidities  
 {\it regular Bethe roots}. 
We call a nonzero Bethe ansatz eigenvector with regular Bethe roots  
 a {\it regular Bethe state}. 
\end{abstract}
%

\maketitle

 \setcounter{equation}{0} 
 \renewcommand{\theequation}{1.\arabic{equation}}

\section{Introduction}

The XXZ spin chain is one of the most important exactly solvable 
quantum systems.  
 The Hamiltonian under the periodic boundary conditions is given by 
\be 
{\cal H}_{\rm XXZ} =  
{\frac 1 2} \sum_{j=1}^{L} \left(\sigma_j^X \sigma_{j+1}^X +
 \sigma_j^Y \sigma_{j+1}^Y + \Delta \sigma_j^Z \sigma_{j+1}^Z  \right) \, . 
\label{hxxz}
\ee
Here we define parameter $q$ from the XXZ  coupling $\Delta$ 
by  $\Delta= (q+q^{-1})/2$. 
When $q$ is a root of unity,  
the XXZ Hamiltonian commutes with the generators of 
the $sl_2$ loop algebra \cite{DFM}.   
Let $q_0$ be a root of unity satisfying 
$q_0^{2N}=1$ for an integer $N$. 
We introduce operators $S^{\pm(N)}$ as follows
\begin{eqnarray}
S^{\pm(N)}  
&=&  
\sum_{1 \le j_1 < \cdots < j_N \le L}
q_0^{{N \over 2 } \sigma^Z} \otimes \cdots \otimes q_0^{{N \over 2} \sigma^Z}
\otimes \sigma_{j_1}^{\pm} \otimes
q_0^{{(N-2) \over 2} \sigma^Z} \otimes  \cdots \otimes q_0^{{(N-2) \over 2}
\sigma^Z} \nonumber \\
 & & \otimes \sigma_{j_2}^{\pm} \otimes 
 q_0^{{(N-4) \over 2} \sigma^Z} \otimes
\cdots
\otimes \sigma^{\pm}_{j_N} \otimes q_0^{-{N \over 2} \sigma^Z} 
\otimes \cdots \otimes q_0^{-{N \over 2} \sigma^Z}  \, . 
\label{sn}
\end{eqnarray}
They are derived from the $N$th power 
of the generators $S^{\pm}$ of the quantum group $U_q(sl_2)$. 
We define  $T^{(\pm)}$ by the complex conjugates of $S^{\pm(N)}$,  
i.e. $T^{\pm (N)} = \left( S^{\pm (N)} \right)^{*}$. 
The operators,  $S^{\pm(N)}$ and $T^{\pm (N)}$, 
generate the $sl_2$ loop algebra, $U(L(sl_2))$. 
Let us consider sectors 
with respect to eigenvalues of the total spin operator $S^Z$.  
We call the sector $S^Z \equiv 0$  (mod $N$) {\it sector A}.  
 Here the value of $S^Z$ is given by an integral multiple of $N$. 
It was shown that the following commutation relations hold 
in sector A \cite{DFM}:
 \begin{eqnarray}
{[}S^{\pm(N)},{\cal H}_{\rm XXZ} 
{]}={[}T^{\pm(N)}, {\cal H}_{\rm XXZ} {]}=0 \, . 
\label{sthcomm}
\end{eqnarray} 

Let us assume that rapidities, ${t}_1, {t}_2, \ldots, {t}_R$, satisfy   
the Bethe ansatz equations at a given value of $q$ as follows:  
\be 
\left( {\frac {\sinh({t}_j + \eta)} 
 {\sinh({t}_j - \eta)}} \right)^L 
= \prod_{k=1; k \ne j}^{M} 
{\frac {\sinh({t}_j - {t}_k + 2 \eta)} 
{\sinh({t}_j - {t}_k - 2 \eta)}} \, , 
\quad \mbox{for} \, \, j=1, 2, \ldots, R.  
\label{BAE}
\ee
Here we have defined parameter $\eta$  by $q=\exp(2\eta)$.    
We call such rapidities $t_j$  {\it Bethe roots} at $q$. 
If Bethe roots are finite and distinct, we call them {\it regular}. 
Let $B(t)$ denote the $B$ operator of the algebraic Bethe ansatz with 
rapidity $t$, and $|0 \ra$ the vacuum state 
(see e.g. Ref. \cite{Korepin}).     
It is known that the Bethe state, $B(t_1) \cdots B(t_R)| 0 \ra$, 
gives an eigenvector of the XXZ Hamiltonian,
if $t_j$ are Bethe roots, i.e. they satisfy eqs. (\ref{BAE}) 
 \cite{8VABA}. 
We also call it the XXZ Bethe state.  
We call a nonzero Bethe state with regular Bethe roots {\it regular}. 
For a root of unity, $q_0$, we define $\eta_0$ by  $q_0=\exp(2 \eta_0)$. 
We now formulate the highest weight conjecture \cite{DFM,FM1,FM2,Odyssey}
as follows: every regular Bethe state at $q_0$   
should be a highest weight vector of the $sl_2$  loop algebra.

\par Let us now define  
highest weight vectors of the $sl_2$ loop algebra.  
The generators of $U(L(sl_2))$, ${  x}_k^{\pm}$ and 
${ h}_k$ ($k \in {\bf Z}$), satisfy the defining relations:  
\be 
[{h}_j, {x}_{k}^{\pm} ] = \pm 2 {x}_{j+k}^{\pm} \, , \quad 
[{x}_j^{+}, {x}_k^{-} ] = {h}_{j+k} \, , 
\quad {\rm for} \, j, k 
\in {\bf Z} \, . 
\label{CDR}
\ee
Here $[{  h}_j, {  h}_{k} ]=0$ and 
$[{  x}_j^{\pm}, {  x}_k^{\pm}] =0$ 
for $j, k \in {\bf Z}$.  
In a representation of $U(L(sl_2))$, a vector $\Omega$ is called 
{\it a highest weight vector} if $\Omega$ 
is annihilated by generators ${ x}_{k}^{+}$ 
for all integers $k$ and such that 
$\Omega$ is a simultaneous eigenvector of every generator 
 ${ h}_k$ ($k\in {\bf Z}$)  
\cite{Chari,Chari-P0,Chari-P1,Chari-P2,Drinfeld}: 
\bea 
{  x}_k^{+} \Omega &= & 0 \, , \quad {\rm for} \, \, k \in {\bf Z} \, , 
\label{eq:annihilation} \\ 
{  h}_{k} \Omega & = & {  d}_k \Omega \, , 
\quad {\rm for} \, \, k \in {\bf Z} \, . 
\label{eq:Cartan}
\eea
Here, the set of eigenvalues ${ d}_k$ 
is called the {\it highest weight}. 
We call a representation 
{\it highest weight} 
if it is generated by a highest weight vector.  
We denote it by $U \Omega$, where $\Omega$ is 
the highest weight vector and $U$ denotes $U(L(sl_2))$. 
We assume in the paper that $U \Omega$ is finite-dimensional.  
We can show that weight ${ d}_0$ 
is given by a non-negative integer, which we denote by $r$,  
and also that $\Omega$ is a simultaneous eigenvector 
of operators $({x}_{0}^{+})^k 
({  x}_{1}^{-})^k/(k!)^2$, i.e. $({x}_{0}^{+})^k 
({  x}_{1}^{-})^k/(k!)^2 \, \Omega = \lambda_k \Omega$ 
for $k= 1, 2, \ldots, r$.  
In terms of the sequence of eigenvalues $\lambda_k$:      
${\lambda}=(\lambda_1, \lambda_2, \ldots, \lambda_r)$,  
we define highest weight polynomial ${\cal P}^{\lambda}(u)$ 
\cite{Criterion}  by 
\be 
{\cal P}^{\lambda}(u) =\sum_{k=0}^{r} \lambda_k \, (-u)^k \,  .     
\label{polynomialP}
\ee
 If $U \Omega$ is irreducible, 
the highest weight polynomial ${\cal P}^{\lambda}(u)$ corresponds  
to the Drinfeld polynomial. 
It was shown that every irreducible representation  
is highest weight and characterized by the Drinfeld polynomial 
\cite{Chari-P1}. However, $U \Omega$ may be reducible.  
We shall show that it is indeed the case in some physical application. 
Here we note that a reducible representation has no 
Drinfeld polynomial but the highest weight polynomial.

\par 
 Recently, for the XXZ spin chain at roots of unity, 
 Fabricius and McCoy have made important observations 
 on the highest weight conjecture \cite{FM1,FM2,Odyssey}. 
Through the algebraic Bethe ansatz 
it was suggested  \cite{Odyssey}  
that any given XXZ Bethe state in sector A  
should be highest weight.   
 Let  ${\tilde t}_1,  {\tilde t}_2, \ldots, {\tilde t}_R$ 
be a set of regular Bethe roots at $q_0$.  
We introduce function $Y(v)$ as  
\be 
Y(v) =  \sum_{\ell=0}^{N-1} 
{\frac {(\sinh(v- (2\ell +1) \eta_0) )^L}  
{\prod_{j=1}^{R} \sinh(v- {\tilde t}_j - 2 \ell  \eta_0) 
\sinh(v- {\tilde t}_j - 2 (\ell+1) \eta_0)} } \, .
\label{Y(v)}
\label{eq:Y(v)}
\ee
It follows from the Bethe ansatz equations 
(\ref{BAE}) at $q_0$ that  
$Y(v)$ is a Laurent polynomial 
of variable $u=\exp(-2Nv)$ \cite{Odyssey}.  
We call it the Fabricius-McCoy polynomial 
of the XXZ Bethe state, 
and denote it by $P^{\rm FM}(u)$. 
Furthermore, it was conjectured  \cite{Odyssey} that  it 
should be equivalent to a `Drinfeld polynomial' $P(u)$  
through the following relation: 
\be 
 P^{FM}(u) =  A \, u^{-r/2} \, P(u) \, .   
\label{PY}
 \ee 
Here $A$ gives the normalization.  
However, it has not been shown 
whether a given XXZ Bethe state is highest weight  
or whether it generates an irreducible representation.

In the paper we  prove the highest weight conjecture 
for regular Bethe states in sectors A and B. 
Here sector B denotes such a sector $S^Z \equiv N/2$ (mod $N$) 
for odd $N$. It is the first result of the paper. 
Furthermore, we discuss how far conjecture (\ref{PY}) is valid.  
In fact, we shall show that the Fabricius-McCoy polynomial corresponds 
to the highest weight polynomial, and also that there is 
such a regular Bethe state 
in sector A that generates a reducible representation.   
This gives the second result.

The first result is summarized as follows.    
Let $| R \ra$ be a regular Bethe state at $q_0$ 
with $R$ down-spins in sectors A or B. 
We have 
$|R \ra = B({\tilde t}_1) B({\tilde t}_2) \cdots B({\tilde t}_R) | 0 \ra$ 
with  regular Bethe roots ${\tilde t}_j$  of $| R \ra$. 
By the algebraic Bethe ansatz, we shall derive the following:  
\bea 
S^{+(N)} \, | R \ra  =  T^{+(N)} \, | R \ra 
 & = & 0 \, ,   
\label{annST} \\
\left(S^{+(N)} \right)^k \left(T^{-(N)} \right)^{k}/(k!)^2 
\, \, | R \ra & = & 
{\cal Z}_k^{+} \, | R \ra \, , \quad 
\quad \mbox{for} \quad k \in {\bf Z}_{> 0} \, .  
\label{diagST}
\eea
Constants ${\cal Z}_k^{+}$ will be expressed in terms of ${\tilde t}_j$, 
explicitly. 
Here we assume that a given set of regular Bethe roots at $q_0$ 
makes an isolated solution of the Bethe ansatz equations (\ref{BAE}). 
We shall show 
that relations (\ref{annST}) and (\ref{diagST}) are sufficient to have  
conditions equivalent to (\ref{eq:annihilation}) and (\ref{eq:Cartan}).  
It will thus follow that $|R \ra$ is highest weight.

We remark that relations (\ref{annST}) generalize 
the $SU(2)$ symmetry of the XXX spin chain.  
As was  shown by Takhtajan and Faddeev \cite{TF},  
regular XXX Bethe states are 
highest weight vectors of the spin $SU(2)$ symmetry. 
We shall discuss it in \S 5.3.

Interestingly, novel spectral degeneracy similar  
to that of  the XXZ spin chain at $q_0$  
appears in the transfer matrix of the eight-vertex model 
at roots of unity 
\cite{Missing,CSOS,FM-8vertex,FM-8vertex2,Elliptic-Current}, 
which are related to some restricted 
IRF models \cite{Missing,CSOS}. 
Some of the degenerate eigenvectors were first  discussed 
by Baxter \cite{Baxter73,Baxter}. 
The elliptic degeneracy is discussed systematically 
by using $Q$ matrices \cite{FM-8vertex,FM-8vertex2}. 
There are some relevant researches 
on the $sl_2$ loop algebra symmetry 
of the XXZ spin chain at $q_0$ \cite{Andrei,Tarasov,Korff-Q,Korff-Q05}. 
The higher rank loop algebra symmetry  
 has been discussed for various trigonometric  
vertex models \cite{Korff-McCoy}. 
The $sl_2$ loop algebra and its subalgebra symmetries have been derived  
from the XXZ chain under twisted boundary conditions at $q_0$ 
\cite{twisted,Korff-twisted}.

The paper consists of the following.  
In \S 2, we make a summary of the main results. 
 In \S 2.1, we specify roots of unity conditions in 
definition \ref{df:roots}, and 
we formulate a main statement on the highest weight conjecture 
in theorem \ref{th:main}. 
We shall discuss its proof in \S 5.  
In \S 2.2, for a given regular Bethe state at $q_0$ in sector A or B,   
we express the highest weight polynomial 
${\cal P}^{\lambda}(u)$ in terms of the regular Bethe roots.  
We present in \S 2.3 
an irreducibility criterion of a highest weight representation.   
We discuss that regular Bethe states should generate Weyl modules,  
which may be reducible.  
In \S 3, we introduce the algebraic Bethe ansatz  
and define the inhomogeneous transfer matrix of the six-vertex model. 
In \S 4 we review    
the $sl_2$ loop algebra symmetry of the XXZ spin chain 
through the algebraic Bethe ansatz.   
In \S 5, we give an outline of the proof of 
theorem \ref{th:main}. 
We show it for the inhomogeneous transfer matrix of 
 the six-vertex model at roots of unity. In fact, the derivation 
is also valid  for the homogeneous case, i.e. for the XXZ spin chain.
Theorem \ref{th:main} is derived from propositions 
\ref{prop:ann-property} and \ref{prop:diag-property} 
through lemma \ref{lem:scheme} by assuming conjecture \ref{conj:main}.  
Here, propositions \ref{prop:ann-property} and \ref{prop:diag-property} 
are derived from lemmas \ref{lem:ann} and \ref{lem:kN}, respectively.  
In \S 6, we prove lemmas \ref{lem:ann} and \ref{lem:kN} explicitly. 
In \S 7, we show that for a regular Bethe state at $q_0$ in sector A or B,   
the Fabricius-McCoy polynomial $P^{\rm FM}(u)$ (\ref{eq:Y(v)}) 
is equivalent to the highest weight polynomial. 
 We show some examples of highest weight polynomials, 
 and give such a regular Bethe state that generates a reducible Weyl module.  
In \S 8, we give a concluding remark.

%
%

 \setcounter{equation}{0} 
 \renewcommand{\theequation}{2.\arabic{equation}}

\section{Summary of the main results}

\subsection{Main theorem on the highest weight conjecture}


We call $q$ {\it generic} if it is not a root of unity. 
We define a root of unity as follows.      
\begin{df}[Roots of unity conditions]
We say that $q_0$ is a root of unity with $q_0^{2N}=1$, if one of the 
following three conditions holds:  
(1) $q_0$ is a primitive $N$th root of unity with $N$  odd ($q_0^N=1$);  
(2) $q_0$ is a primitive $2N$th root of unity with $N$ odd ($q_0^N=-1$); 
(3) $q_0$ is a primitive $2N$th root of unity with $N$ even ($q_0^N=-1$). 
We call conditions (1) and (3)  type I, condition (2)  type II.   
\label{df:roots}
\end{df} 

In the case of sector A 
where $S^Z \equiv 0$ (mod $N$), we assume that $q_0$ is a root of unity 
with $q_0^{2N}=1$, in the paper.   In the case of sector B where   
 $S^Z \equiv N/2$ (mod $N$) with $N$ odd, we assume that   
$q_0$ is a primitive $N$th root of unity with $N$ odd ($q_0^N=1$).  

Here we note that 
the same conditions of roots of unity have been 
discussed in Refs. \cite{Baxter,FM-8vertex,FM-8vertex2}.  
Let us express $q_0$ as $q_0 = \exp( \sqrt{-1} {\pi m}/{N}) $. 
In terms of $m$ and $N$, 
roots of unity conditions (1), (2) and (3) are expressed as follows: 
(1) $N$ is odd and $m$ even; (2) $N$ is odd and $m$ odd; (3) $N$ is even  
($m$ odd by definition).

We now formulate a theorem on the highest weight conjecture in the case of    
 the inhomogeneous transfer matrix of the six-vertex model.  
We shall  define it in \S 3. 
Let $t_1, t_2, \ldots, t_R$ be a set of regular Bethe roots 
at a given value of $q$ in the inhomogeneous case.   
It is known that the Bethe state, $B(t_1) \cdots B(t_R) | 0 \ra$,  
gives an eigenvector of the inhomogeneous transfer matrix at $q$, 
if $t_j$ satisfy the Bethe ansatz equations (\ref{BAE-inhomo}). 
Here, eqs. (\ref{BAE-inhomo}) generalize 
 eqs. (\ref{BAE}) in the inhomogeneous case.

\begin{conj} 
Let  $q_0$ be a root of unity with $q_0^{2N}=1$.  
Every set of regular Bethe roots at $q=q_0$  
gives an isolated solution of the Bethe ansatz equations 
(\ref{BAE-inhomo}).  
\label{conj:main} 
\end{conj}

Assuming conjecture \ref{conj:main} 
we shall show in \S 5 the following.  
\begin{th1}  
Every regular Bethe state at $q=q_0$ 
in sectors A and B is  highest weight.  
\label{th:main} 
\end{th1}

\subsection{Highest weight polynomials of  regular Bethe states}

 Let us denote the inhomogeneous parameters by 
$\xi_{\ell} $ for $\ell= 1, 2, \ldots, L$,  
where  $L$ is the lattice size. 
We define functions ${\phi}_{\xi}^{\pm}(x)$  by   
\be 
{\phi}_{\xi}^{\pm}(x) = \prod_{\ell=1}^{L} (1- x e^{\pm 2 \xi_{\ell}}) \, .   
\label{eq:phi-inhomo}
\ee
Recall that $t_1, t_2, \ldots, t_R$ 
are regular Bethe roots at a given value of $q$ in the inhomogeneous case.  
We define ${F}^{\pm}(x)$ by  
\be
{F}^{\pm}(x) =\prod_{j=1}^{R} (1 - x \exp(\pm 2 t_j)) \, . 
\label{eq:F(x)}
\ee

Let $| R \ra $ be a regular Bethe state  at $q_0$ in sectors A or B 
in the inhomogeneous case and 
${\tilde t}_1, {\tilde t}_2, \ldots, {\tilde t}_R $ 
 the regular Bethe roots.  
 At $q=q_0$, we express ${F}^{\pm}(x)$ as 
${\tilde F}^{\pm}(x)=\prod_{j=1}^{R} (1 - x \exp(\pm 2 {\tilde t}_j))$.  
We consider the following series with respect to $x$: 
\be 
{\frac {\phi_{\xi}^{\pm}(x)} {{\tilde F}^{\pm}(x q_0) 
{\tilde F}^{\pm}(x q_0^{-1})} } 
=  \sum_{m=0}^{\infty} {\tilde{\chi}}^{\pm}_{\xi,m} \, x^m \,  \quad 
{\rm for} \quad |x| < {\rm min}\{ |e^{\pm 2 {\tilde t}_j} | \} .  
\label{eq:df-chi}
\ee
Here, we define coefficients $\tilde{\chi}_{\xi,m}^{\pm}$ by 
power series (\ref{eq:df-chi}).   
Explicitly, we have     
\bea   
{\tilde{\chi}}_{\xi,m}^{\pm} & = & \sum_{\rho=0}^{{\rm min}(L, m)}  
(-1)^{\rho}  
\sum_{1 \le j_1 < \cdots < j_{\rho} \le L}
\exp \left( \pm \sum_{k=1}^{\rho} 2 \xi_{j_k} \right) \non \\ 
& & \quad \times \, 
 \sum_{n_1 + \cdots + n_R =m - \rho} 
e^{\pm \sum_{j=1}^{R} 2 n_j \tilde{t}_j } \, 
\prod_{j=1}^{R} [n_j + 1]_{q_0} \, , 
\label{diag-inhomo}
\eea
where symbol $\sum_{n_1 + \cdots + n_R =m-\rho}$ denotes the sum  over all 
nonnegative integers $n_1, n_2, \ldots, n_R$ satisfying  
$n_1 + \cdots + n_R =m-\rho$.  When $R=0$, we set $\rho=m$.

\begin{prop} Let $q_0$ be a root of unity with $q_0^{2N}=1$.   
For a regular Bethe state $| R \ra$ at $q_0$ 
in sector A or B in the inhomogeneous case 
such as given in theorem \ref{th:main},   
 weight $d_0$ is given by $r=(L-2R)/N$, and  
 eigenvalues ${\lambda}_k$ of (\ref{polynomialP}) are given by  
\be 
\lambda_k= \left\{ 
\begin{array}{cc} 
 (-1)^{kN} {\tilde{\chi}}_{\xi,kN}^{+}  \, ,  
&  {\rm if} \, \,  q_0 \, \, {\rm is} \, \, {\rm of} \, \, {\rm type} 
\, {\rm I} \, , \\
  (-1)^{k(N-1)} {\tilde{\chi}}_{\xi,kN}^{+}  \, , 
& {\rm if} \, \,  q_0 \, \, {\rm is} \, \, {\rm of} \, \, {\rm type} 
\, \, \rm{II} \, . 
\end{array} 
\right.
\label{eq:cor}
\ee
\label{cor:lambda}
\end{prop}
 
We thus obtain 
the highest weight polynomial ${\cal P}^{\lambda}(u)$ 
of $|R \ra$.   
 For type I we have 
\be 
{\cal P}^{\lambda}(u) 
 =  \left\{ 
\begin{array}{cccc}  
\sum_{k=0}^{r} {\tilde{\chi}}_{\xi, kN}^{+} \, 
u^k & {\rm for} & {\rm odd} \, 
N & (q_0^{N}=1) \,   \\
\sum_{k=0}^{r} {\tilde{\chi}}_{\xi, kN}^{+} \, 
(-u)^k & {\rm for} & {\rm even}\, 
N & (q_0^{N}=-1) \, ,    
\end{array}
\right. 
 \label{PR1-inhomo}
\ee
and for type II 
\be 
{\cal P}^{\lambda}(u)  
=  \sum_{k=0}^{r} {\tilde{\chi}}_{\xi, kN}^{+} \, (-u)^{k}  
\qquad 
(N: \mbox{odd}; \, q_0^{N}=-1) \, .   
\label{PR2-inhomo}
\ee

\subsection{An irreducibility criterion 
of highest weight representations}

In order to obtain the degenerate multiplicity of 
a regular Bethe state at $q_0$, 
it is fundamental to derive the dimension of a given 
highest weight representations \cite{Poincare,Criterion}.  
In fact, 
every finite-dimensional representation of $U(L(sl_2))$ 
should be given by a collection of finite-dimensional 
highest weight representations.

Recall that we assume in the paper that $U\Omega$ is finite-dimensional.  
Let ${\cal P}^{\lambda}(u)$ be the highest weight polynomial.   
We introduce parameters $a_k$ by 
\be 
{\cal P}^{\lambda}(u) = \prod_{k=1}^{s} (1 - a_k u)^{m_k} \, .  
\label{eq:factorization}
\ee
Here $a_1, a_2, \ldots, a_s$ are distinct, and their  
 multiplicities are given by $m_1, m_2, \ldots, m_s$, respectively,        
where $r=m_1 + \cdots + m_s$. 
We now introduce {\it highest weight parameters}  
${\hat a}_i$ for $i=1, 2, \ldots, r$, as follows:  
\be 
{\hat a}_i = a_k \quad {\rm if } \, \, m_1+ m_2 + \cdots + m_{k-1} < i \le  
m_1+ \cdots + m_{k-1} + m_{k} \, . 
\label{eq:hat-a}
\ee
It can be shown that ${\hat a}_j$  are nonzero \cite{Criterion}.   
We thus have three equivalent expressions 
for the highest weight ${ d}_k$ of $\Omega$:  
sequence $\lambda$ of eigenvalues $\lambda_k$,  
polynomial ${\cal P}^{\lambda}(u)$, 
and parameters ${\hat a}_j$ 
(i.e. parameters $a_j$ 
with multiplicities $m_j$).

It was shown by Chari and Pressley \cite{Chari-P3} that 
corresponding to each irreducible finite-dimensional representation 
with  highest weight  ${\cal P}^{\lambda}(u)$   
there exists a unique finite-dimensional highest weight 
module $W$ 
such that any  finite-dimensional highest weight module $V$ 
with highest weight  ${\cal P}^{\lambda}(u)$ 
is a quotient of $W$. The modules $W$ are called Weyl modules
 \cite{Chari-P3}. 
Furthermore, it was shown that a Weyl module is irreducible 
if and only if the polynomial ${\cal P}^{\lambda}(u)$ has distinct roots 
\cite{Chari-P3}. 
Thus, if the highest weight parameters ${\hat a}_j$ 
of $V$ are distinct, $V$ is irreducible,   
and the polynomial ${\cal P}^{\lambda}(u)$ becomes  
the Drinfeld polynomial of $V$.  
We have ${\rm dim} \, V = 2^s$.

However, highest weight parameters ${\hat a}_j$ are not always distinct. 
It is shown that the highest weight 
representation generated by $\Omega$, i.e. $U \Omega$, 
is irreducible 
if and only if the following  holds \cite{Criterion}:  
\be 
\sum_{j=0}^{s} (-1)^{s-j} \mu_{s-j} {  x}_{j}^{-} \Omega = 0  \, ,  
\label{eq:cond}
\ee
where $\mu_{k}$ $(k=1, 2, \ldots, s)$ are given by 
\be 
\mu_{k} = \sum_{1 \le i_1 < \cdots < i_k \le s} a_{i_1} \cdots a_{i_k}  
\, . 
\ee
We note that every irreducible representation 
has no invariant subspace under the action of $U(L(sl_2))$ 
except for trivial cases. 
If $U \Omega$ is irreducible, the dimension is given by \cite{Chari} 
\be 
{\rm dim} U \Omega = \prod_{j=1}^{s} (m_j+1) . 
\ee

Through criterion (\ref{eq:cond}) 
we shall show such a regular Bethe state 
in the inhomogeneous case that generates a 
reducible representation (see  \S 7.2.3). 

We shall show  that 
for a given regular Bethe state in sectors A or B, 
the Fabricius-McCoy polynomial (\ref{eq:Y(v)}) corresponds to 
 the highest weight polynomial evaluated in (\ref{PR1-inhomo}) 
and (\ref{PR2-inhomo}) (see corollary \ref{cor:PFM}). 
It thus follows that conjecture (\ref{PY}) is valid in sectors A and B 
if the highest weight parameters ${\hat a}_j$ are distinct. 
However, if $U \Omega$ is reducible, conjecture (\ref{PY}) is not valid, 
since it has no Drinfeld polynomial.

We have a conjecture that regular Bethe states at $q_0$ 
in sectors A and B should generate Weyl modules. 
A majority of regular Bethe states 
should have distinct highest weight parameters, 
 while most of those with degenerate ones should generate 
such reducible representations that are equivalent to Weyl modules.

%
%

 \setcounter{equation}{0} 
 \renewcommand{\theequation}{3.\arabic{equation}}

\section{The transfer matrix of the six-vertex model}

\subsection{$R$ matrix and $L$ operator of the algebraic Bethe ansatz}
\par 
In order to fix the notation, let us introduce the algebraic 
Bethe ansatz \cite{8VABA,TF,Korepin}.  
We define the $R$ matrix of the XXZ spin chain by 
\be
R(z-w)  
= \left(
\begin{array}{cccc} 
f(w-z) &   0 & 0 & 0 \\
0 &   g(w-z) & 1 & 0 \\
0 &  1 & g(w-z) & 0 \\
0 &   0 & 0 & f(w-z)
\end{array} 
\right) \, ,  
\ee
where $f(z-w)$ and $g(z-w)$ are given by 
\be 
f(z-w)= {\frac {\sinh(z-w-2 \eta)} {\sinh(z-w)}} \, , \quad
g(z-w)= {\frac {\sinh(-2 \eta)} {\sinh(z-w)}} \, . 
\label{fg}
\ee
We recall that  parameter $2 \eta$ is related to $q$  by $q=\exp(2\eta)$. 
We now introduce $L$ operators for the XXZ spin chain.  
Let $V_n$ be two-dimensional vector spaces for $n=0, 1, \ldots, L$. 
We define an operator-valued matrix $L_n(z)$ by 
\be 
 L_n(z) =  \left(
\begin{array}{cc}  
\sinh \left( z \, I_n + \eta \sigma_n^z \right) 
& \sinh 2 \eta \, \sigma_n^{-} \\
\sinh 2 \eta \, \sigma_n^{+}  
& \sinh \left( z \, I_n - \eta \sigma_n^z \right) 
\end{array} 
\right)  \, . 
\label{eq:Loperator}
\ee
Here   
$L_n(z)$ is a matrix acting on the auxiliary vector space $V_0$, 
where $I_n$ and $\sigma_n^a$ ($a=z, \pm$) are operators 
acting on the $n$ th vector space $V_n$.    
The symbol $I$ denotes the two-by-two identity matrix, and  
 $\sigma^{\pm}$ denote $\sigma^{+}= E_{12}$ and $\sigma^{-} = E_{21}$,    
where they satisfy relations $E_{ij} E_{k \ell} = \delta_{j, k} E_{i \ell}$ 
for $i,j,k, \ell=1,2$.  
Here $\delta_{j,k}$ denotes the Kronecker delta. 
The symbols, $\sigma^x, \sigma^y, \sigma^z$ denote the Pauli matrices.

Let us introduce the monodromy matrix with 
inhomogeneous parameters $\xi_n$   
\be
T(z; \{ \xi_n \} )=  
L_{L}(z-\xi_{L}) \cdots L_2(z-\xi_2) L_1(z-\xi_1) \, . 
\label{monodromy-inhomo}
\ee
We call  $T(z; \{ \xi_n \} )$ the {\it inhomogeneous monodromy matrix}.  
In terms of the $R$ matrix and the monodromy matrices, 
the Yang-Baxter equation is expressed as 
\be
R(z-w) \left( T(z; \{ \xi_n \}) \otimes T(w;  \{ \xi_n \}) \right) 
 = \left( T(w; \{ \xi_n \}) \otimes T(z; \{ \xi_n \}) \right) R(z-w) .  
\label{RLL} 
\ee
We express the matrix elements 
of the inhomogeneous monodromy matrix $T(z; \{ \xi_n \} )$ as   
\be 
T(z; \{ \xi_n \} ) = 
\left( \begin{array}{cc}
A(z; \{\xi_n \}) & B(z; \{ \xi_n \} ) \\
C(z; \{\xi_n \}) & D(z; \{ \xi_n \} ) 
\end{array} 
\right) \, . 
\ee
 From the Yang-Baxter equation for  $T(z; \{ \xi_n \} )$ 
we have the commutation relations  
such as $B(w_1; \{ \xi_n \}) B(w_2; \{ \xi_n \} ) 
= B(w_2; \{ \xi_n \})   B(w_1; \{ \xi_n \})$ 
and  
\bea 
A(w_1; \{ \xi_n \}) B(w_2; \{ \xi_n \}) 
& = & f(w_1-w_2) B(w_2; \{ \xi_n \})  A(w_1; \{ \xi_n \})  \non \\ 
& & - g(w_1-w_2) B(w_1; \{ \xi_n \})  A(w_2; \{ \xi_n \}) . \label{CR}
\eea 
Here parameters $w_j$ are arbitrary.  
Hereafter, we suppress the inhomogeneous parameters, $\xi_n$ for 
operators $A, B, C$ and $D$. 
We denote $B(w_j; \{ \xi_n \})$  simply by $B(w_j)$. 

Through the commutation relations such as (\ref{CR}) 
we have for $n \in {\bm Z}_{\ge 0}$ the following: 
\bea 
& & A(w_0) \, B(w_1) \cdots B(w_n) =   
\left( \prod_{j=1}^{n}  f(w_0-w_j) \right) \, 
B(w_1) \cdots B(w_n) \, A(w_0) 
\non \\
& & - \displaystyle{ \sum_{j=1}^{n} g(w_0-w_j) \prod_{k \ne j}^{n} f(w_j-w_k)} 
B(w_1) \cdots B(w_{j-1}) B(w_{0}) \times \non \\
& & \qquad \times B(w_{j+1}) \cdots B(w_{n}) 
A(w_{j}) \, .  
\label{ABBB}    
\eea

\subsection{The inhomogeneous transfer matrix of the six-vertex model 
and the Bethe states}

We define the 
{\it inhomogeneous transfer matrix} of the six-vertex model,   
$\tau_{6V}(z; \{ \xi_n \})$, by 
\be 
\tau_{6V}(z; \{ \xi_n \} )= {\rm tr}  \, T(z; \{ \xi_n \}) 
= A(z) + D(z) \, . 
\label{tm-inhomo} 
\ee
When all the inhomogeneous parameters, $\xi_n$, are 
set to be zero, 
we call $\tau_{6V}(z; \{ \xi_n=0 \} )$ 
the homogeneous transfer matrix of the six vertex model and denote it 
simply as  $\tau_{6V}(z)$.  It is invariant under lattice translation. 
The XXZ Hamiltonian is given by  
the logarithmic derivative of the transfer matrix,  $\tau_{6V}(z)$, 
at $z=\eta$:  
\bea 
& &  {\sinh 2 \eta}  \times  \, 
{\frac d {dz} } \log \tau_{6V}(z)|_{z=\eta}  
=  H_{XXZ} + {\frac  L 2} \cosh 2 \eta  \, . 
\label{logXXZ}
\eea 
Here we note that the XXZ coupling $\Delta$ is given by 
$\cosh 2\eta$.

We denote by $|0 \ra$ the vector with all spins up. 
We have    
\be 
A(z) \, | 0 \ra = a_{\xi}^{6V}(z) \, | 0 \ra \, , \quad 
D(z) \, | 0 \ra = d_{\xi}^{6V}(z) \, | 0 \ra \, .  
\ee
Here  ${a_{\xi}^{6V}(z)}$ and ${d_{\xi}^{6V}(z)}$ are given by  
\be 
a_{\xi}^{6V}(z)= \prod_{n=1}^{L} \sinh(z-\xi_n + \eta) \, , \quad 
d_{\xi}^{6V}(z) = \prod_{n=1}^{L} \sinh(z-\xi_n - \eta) \, . 
\ee
It is shown \cite{8VABA} that the vector 
$B(t_1) B(t_2) \cdots B(t_R) | 0 \ra$ is  
an eigenvector of the inhomogeneous transfer matrix 
$\tau_{6V}(z; \{ \xi_n \})$ 
if  rapidities $t_1, t_2, \ldots, t_R$ satisfy the Bethe ansatz equations  
\be 
{\frac {a_{\xi}^{6V}(t_{j})} {d_{\xi}^{6V}(t_j)} } 
= \prod_{k=1, k \ne j}^{R} 
 {\frac {f(t_k-t_j)} {f(t_j-t_k)}}  \, , 
\quad {\rm for} \, j= 1, 2, \cdots, R \, .  
\label{BAE-inhomo}
\ee  
In the same way as the homogeneous case for (\ref{BAE}), 
we call the eigenvector 
$B(t_1) B(t_2) \cdots B(t_R) | 0 \ra$ 
{\it the Bethe ansatz eigenvector} or 
{\it the Bethe state}, briefly. 
Here we call 
$t_1, t_2, \ldots, t_R$,  the {\it Bethe roots}, 
and call them regular if they are finite and distinct.   
Furthermore, we call the Bethe state {\it regular}, 
if it is nonzero and the Bethe roots are regular.

%
%

 \setcounter{equation}{0} 
 \renewcommand{\theequation}{4.\arabic{equation}}

\section{The $sl_2$ loop algebra symmetry at roots of unity}

\subsection{Generators of the quantum groups}

The quantum affine algebra $U_q(\hat{sl_2})$ 
is an associative algebra over ${\bf C}$ generated by  
$e_i^{\pm}, K_i^{\pm}$ for $i=0,1$ with the following relations: 
\bea 
K_i K_i^{-1} & = & K_i K_i^{-1} = 1 \, , \quad 
K_i e_i^{\pm} K_i^{-1}  =  q^{\pm 2} e_i^{\pm} \, ,  \quad 
K_i e_j^{\pm} K_i^{-1}  =  q^{\mp 2} e_j^{\pm}  
\quad (i \ne j) \, , \non \\
{[} e_i^{+}, e_j^{-} {]} & = & \delta_{i,j} \, 
{\frac   {K_i - K_i^{-1}}  {q- q^{-1}} } \, ,     \non \\
(e_i^{\pm})^{3} e_j^{\pm} & - & [3]_q \, (e_i^{\pm})^{2} e_j^{\pm} e_i^{\pm}   
+ [3]_q \, e_i^{\pm} e_j^{\pm} (e_i^{\pm})^2 -   
 e_j^{\pm} (e_i^{\pm})^3 = 0 \quad (i \ne j) \, . 
\label{defrl}
\eea
Here $q$ is generic, and $[n]_q$ denotes the $q$-integer of an integer $n$: 
$[n]_q=(q^n -q^{-n})/(q-q^{-1})$. 
The algebra $U_q(\hat{sl_2})$ is also a Hopf algebra over ${\bf C}$ 
with comultiplication
\bea 
\Delta (e_i^{+}) & = & e_i^{+} \otimes K_i + 1 \otimes e_i^{+}  \, , 
 \quad 
\Delta (e_i^{-})  =  e_i^{-} \otimes 1 + K_i^{-1} \otimes e_i^{-} \, ,  \non \\
\Delta(K_i) & = & K_i \otimes K_i  \, , 
\eea 
and antipode:  
$S(K_i)=K_i^{-1} \, , S(e_i)= - e_i^{+} K_i^{-1} \, , 
S(e_i^{-}) = - K_i e_i^{-} $.  

The quantum algebra $U_q(sl_2)$ is an associative Hopf algebra over ${\bf C}$ 
generated by elements $e^{\pm}$ and $K$ with the same defining relations of the Hopf algebra as those for $e_1^{\pm}$ and $K_1$  of $U_q(\hat{sl_2})$.

We now introduce evaluation representations for $U_q(\hat{sl_2})$ 
\cite{Jimbo}. 
For a given nonzero complex number $a$ there is a homomorphism 
of algebras  $\varphi_a$: $U_q(\hat{sl_2}) \rightarrow U_q({sl_2})$  
such that  
$\varphi_a(e_0^{\pm}) = q^{\mp 1} a^{\pm 1}e^{\mp}$,  
$\varphi_a(e_1^{\pm}) = e^{\pm}$, $\varphi_a(K_0)=K^{-1}$,  
and $\varphi_a(K_1) =K$. 
Let us denote by $(\pi, V)$ a representation of an algebra ${\cal A}$ 
such that $\pi(x)$ give linear maps on vector space $V$ 
for $x \in {\cal A}$. 
For a given finite-dimensional representation $(\pi_V, V)$ of $U_q(sl_2)$ 
 we have a finite-dimensional representation $(\pi_{V(a)}, V(a))$ 
of $U_q(\hat{sl_2})$  through homomorphism $\varphi_a$, i.e. 
$\pi_{V(a)}(x)= \pi_V(\varphi_a(x))$ for $x \in U_q(\hat{sl_2})$. 
We call $(\pi_{V(a)}, V(a))$ or $V(a)$ 
the {\it evaluation representation} of $V$ and  nonzero parameter $a$ 
the {\it evaluation parameter} of $V(a)$.

We consider such finite-dimensional representations 
of $U_q(\hat{sl_2})$ where  
 $K_i$ are equivalent to $q^{H_i}$ with diagonal matrices $H_i$ 
 for $i=0,1$.  
In the representations we denote by $K_i^{1/2}$ 
the square root of $K_i$. 
We introduce the following operators for $i=0,1$:
\be 
{\hat e}_i^{+} = q^{n_i} K_i^{-1/2} e_i^{+} \, , \quad 
{\hat e}_i^{-} = q^{-n_i} e_i^{-}  K_i^{1/2}\, .  
\label{eq:hat-e}
\ee 
Here $n_i$ ($i=0,1$) are arbitrary. 
The operators ${\hat e}_i^{\pm}$ 
satisfy the same defining relations 
(\ref{defrl}) with $e_i^{\pm}$ and  the following comultiplication 
\cite{Jimbo-review}: 
\be 
\Delta({\hat e}_i^{\pm}) = {\hat e} \otimes K_i^{1/2} + K_i^{-1/2} \otimes 
{\hat e}_i^{\pm} \, . 
\ee
Taking the comultiplication $L-1$ times we have 
\be 
\Delta^{(L-1)}({\hat e}_i^{\pm}) = \sum_{j=1}^{L} 
(K_i^{-1/2} )^{\otimes (j-1)} \otimes {\hat e}_i^{\pm} \otimes 
(K_i^{1/2} )^{\otimes (L-j)} \, . 
\ee

\par 
Let us denote by $(\pi_1, V_1)$ 
such a two-dimensional irreducible representation  
 of $U_q(sl_2)$ where generators $e^{\pm}$ and $K$ are represented by 
the Pauli matrices $\sigma^{\pm}$ and $q^{\sigma^Z}$, respectively, i.e. 
we have $\pi_1(e^{\pm}) = \sigma^{\pm}$ and 
$\pi_1(K)=q^{\sigma^Z}$.  
 For the evaluation representation  $V_1(a)$  we take  
the $L$th tensor product, $V_1(a)^{\otimes L}$. 
Setting $a=q$ we denote 
 by $S^{\pm}$ and $T^{\pm}$ the matrix representations of 
generators ${\hat e}_0^{\mp}$ and  ${\hat e}_1^{\pm}$ 
acting on $V_1(q)^{\otimes L}$, respectively. Here we note     
$S^{\pm}= \pi_{V_1(q)} \otimes \cdots \otimes \pi_{V_1(q)}
(\Delta^{(L-1)}({\hat e}_0^{\mp})))$.   
Explicitly, we have 
\bea 
S^{\pm} &= & \sum_{j=1}^{L} 
q^{\sigma^Z/2} \otimes \cdots \otimes q^{\sigma^Z/2} \otimes 
\sigma_j^{\pm} \otimes q^{-\sigma^Z/2} \otimes \cdots \otimes q^{-\sigma^Z/2} 
\, , \non \\  
T^{\pm} & = & \sum_{j=1}^{L} 
q^{-\sigma^Z/2} \otimes \cdots \otimes q^{-\sigma^Z/2} \otimes 
\sigma_j^{\pm} \otimes q^{\sigma^Z/2} \otimes \cdots \otimes q^{\sigma^Z/2}
\, . 
\eea
Here we have set $n_0=n_1=1/2$. The symbol $\sigma_j^{\pm}$ 
denote the Pauli matrices $\sigma^{\pm}$ acting 
on the $j$th component of the tensor product. 
We denote by $q^{S^Z}$ 
the matrix representation of $K_1^{1/2}$ acting 
on the tensor product $V_1(q)^{\otimes L}$. We have  
\be 
q^{S^Z} =  (q^{\sigma^Z/2}) ^{\otimes L}= q^{\sigma^Z/2} \otimes \cdots 
\otimes q^{\sigma^Z/2} \, . 
\ee
We recall that  $S^Z$ denotes the $Z$-component of the total spin 
operator,   $S^Z= \sum_{j=1}^{L} \sigma_j^{Z}/2$.

\par 
We define the $q$-factorial of $n$ 
by $[n]_q! = [n]_q [n-1]_q \cdots [1]_q$. 
Here, $q$ is generic and not a root of unity. 
We introduce the following notation for the $n$th power of an operator $X$ 
divided by the $q$-factorial of $n$:      
\be 
\left( X \right)^{(n)}_{q} = \left( X \right)^n/[n]_q! \, . 
  \ee   
It is easy to show the following \cite{DFM}:
\bea 
\left(T^{\pm} \right)^{(n)}_{q}
& = & 
\sum_{1 \le j_1 < \cdots < j_n \le L}
q^{-{n \over 2 } \sigma^Z} \otimes \cdots \otimes q^{-{n \over 2} \sigma^Z}
\otimes \sigma_{j_1}^{\pm} \otimes
q^{-{(n-2) \over 2} \sigma^Z} \otimes  \cdots \otimes q^{-{(n-2) \over 2}
\sigma^Z}
\nonumber \\
 & & \otimes \sigma_{j_2}^{\pm} \otimes q^{-{(n-4) \over 2} \sigma^Z} \otimes
\cdots
\otimes \sigma^{\pm}_{j_N} \otimes q^{{n \over 2} \sigma^Z} \otimes \cdots
\otimes q^{{n \over 2} \sigma^Z}  \, . 
\label{tn}
\eea
We derive operators $S^{\pm(N)}$ and $T^{\pm(N)}$ 
defined by (\ref{sn}) of \S 1 through the following limit:  
\be 
S^{\pm(N)} = \lim_{q \rightarrow q_0} 
 \left( S^{\pm} \right)^{(N)}_{q} \, , \quad 
T^{\pm(N)} =  \lim_{q \rightarrow q_0} 
\left( T^{\pm} \right)^{(N)}_{q} \, .  
\ee
Here we recall that $q_0$  denotes a root of unity satisfying $q_0^{2N}=1$.

%
\subsection{Quantum group generators through infinite rapidities}
%
%

Let us define function $n_{\xi}(z)$  by 
\be 
  n_{\xi}(z)= \prod_{j=1}^{L} \sinh(z-\xi_j) \, ,   
\ee
and function ${\hat g}(z)$  by 
\be  
 {\hat g}(z) = 
\left\{
\begin{array}{cc}  
2 \exp(-z) \sinh 2\eta  & ({\rm Re} \, z > 0 ) \, , \\
-2 \exp(z) \sinh 2\eta  & ({\rm Re} \, z < 0 ) \, . 
\end{array} 
\right. 
\ee 
We normalize operators $A(z), \ldots, D(z)$ as follows:  
\bea 
{\hat A}(z) & = & A(z)/n_{\xi}(z) \, , \quad 
{\hat B}(z) = B(z)/({\hat g}(z) n_{\xi}(z)) \, , \non \\  
{\hat D}(z) & = & D(z)/n_{\xi}(z) \, , \quad 
{\hat C}(z) = C(z)/({\hat g}(z) n_{\xi}(z))  \, . 
\label{normalization-inhomo}
\eea
Taking the limit of infinite rapidities 
for the inhomogeneous case,  we have 
\bea 
{\hat A}(\pm \infty) & = &  q^{\pm S^Z} \, , 
\quad {\hat B}(\infty) = V^{-} T^{-} V^{+} \, ,  \quad  
{\hat B}(-\infty) = V^{+} S^{-} V^{-}\, , 
 \non \\    
{\hat D}(\pm \infty) & = & q^{\mp S^Z} \,  , \quad 
 {\hat C}(\infty) 
= V^{+} S^{+} V^{-} \, , \quad  
{\hat C}(-\infty) =  V^{-} T^{+} V^{+}\, ,  
\label{infnite2}
\eea
where $V^{\pm}$ are given by the following diagonal matrices \cite{twisted}: 
\be 
\left( V^{\pm} \right)^{j_1, \cdots, j_L}_{k_1, \cdots, k_L} = 
\exp\left(\pm \sum_{i=1}^{L} \xi_i j_i \right) 
\delta_{j_1, k_1} \cdots \delta_{j_L, k_L} \, ,  
\label{eq:V-diag}
\ee
where $j_{\ell}, k_{\ell}= 1,2$ for $\ell =1, 2, \ldots L$. 
Let us define operators $S^{\pm}_{\xi}$ and $T^{\pm}_{\xi}$ by 
\be 
 S^{\pm}_{\xi} = V^{+} S^{\pm} V^{-} \, , 
\quad  T^{\pm}_{\xi} = V^{-} T^{\pm} V^{+} \, .  
\ee

We can show that ${S}^{\pm}_{\xi}$,   ${T}^{\pm}_{\xi}$ and 
 $q^{S^{Z}}$ satisfy the defining relations of $U_{q}(L(sl_2))$ 
 through the evaluation homomorphism \cite{Jimbo}.   
Recall generators ${\hat e}_i^{\pm}$ for $i=0, 1$ defined 
by (\ref{eq:hat-e}) with $n_0=n_1=1/2$.  
In the tensor product $\otimes_{j=1}^{L} V_1(q_j) = 
V_1(q e^{2 \xi_1}) \otimes V_1(q e^{2 \xi_2}) \otimes 
\cdots \otimes V_1(q e^{2 \xi_L})$,  
generators ${\hat e}_i^{\pm}$ for $i=0, 1$ are related to 
$S^{\pm}_{\xi}$ and $T^{\pm}_{\xi}$ as follows 
$$
S^{\pm}_{\xi} = V^{-} \otimes_{j=1}^{L} \pi_{V_1(q_j)} 
\Delta^{(L-1)}({\hat e}_0^{\mp}) V^{+} \, , 
\quad  T^{\pm}_{\xi} = V^{-} \otimes_{j=1}^{L} \pi_{V_1(q_j)} 
\Delta^{(L-1)}({\hat e}_1^{\pm}) V^{+} \, .   
$$
Here $q_j=q \exp 2 \xi_j$ for $j=1, 2, \ldots, L$. 
Thus, ${S}^{\pm}_{\xi}$ and ${T}^{\pm}_{\xi}$ 
satisfy the same defining relations of $U_{q}(L(sl_2))$ 
as generators ${\hat e}_0^{\mp}$ and ${\hat e}_1^{\pm}$, respectively.

\subsection{Complete $N$-strings}

Let $N$ be a positive integer.  
\begin{df}[Complete $N$-string] 
We call a set of rapidities $z_j$ {\it a complete $N$-string}, 
if they have the following relation:  
\be 
z_j= \Lambda + \eta(N+1 - 2j) \quad (j=1, 2, \ldots, N) \, . 
\label{N-string}
\ee
We call the parameter 
$\Lambda$ the {\it center of the $N$-string}. 
\end{df} 

Setting $w_j=z_j$ with $z_j$ being the complete $N$-string, 
we have the following:   
\bea 
& & \prod_{k=1,k \ne j}^{N} f(z_{j}-z_{k}) = 
\left\{ 
\begin{array}{cc}  
 0    &    (j \ne N) \\ 
{[N]_q}  &    (j=N) 
\end{array} \right. \, , 
\quad 
\prod_{k=1, k \ne j}^{N} f(z_{k}-z_{j}) = 
\left\{ 
\begin{array}{cc}  
 {[N]_q}  &    (j = 1 ) \\ 
 0 &     (j \ne 1) 
\end{array} \right. \, . \non \\ 
& & \quad \label{factN}
\eea
Applying (\ref{factN}) into (\ref{ABBB}) with $n=N$,  
and  sending the center $\Lambda$ of the 
complete $N$-string (\ref{N-string}) to infinity,  
we have 
\be 
A(w_{0}) (T^{-}_{\xi})^N  = q^N (T^{-}_{\xi})^N \, A(w_{0}) - [N]_q \, 
B(w_{0}) (T^{-}_{\xi})^{N-1} q^{S^Z} 
e^{w_0} \, .  
\ee
Similarly, we have 
\be 
D(w_{0}) (T^{-}_{\xi})^N  =  q^{-N} (T^{-}_{\xi})^N \, D(w_{0})  
- [N]_q \, B(w_{0}) 
(T^{-}_{\xi})^{N-1} q^{-S^Z} (-e^{w_0}) 
\, .  
\ee
We thus have 
\bea 
\left( A(w_{0}) + D(w_{0}) \right) \, \left( T^{-}_{\xi} \right)^{(N)}_q 
& = & \left( T^{-}_{\xi} \right)^{(N)}_q \, \left( q^{N} A(w_{0}) 
+ q^{-N} D(w_{0}) \right) 
\non \\ 
& &  - e^{w_0} B(w_{0}) (T^{-}_{\xi})_q^{(N-1)} 
(q^{S^Z} - q^{-S^Z}) \, .
\label{loop0}
\eea
Taking the limit: $\Lambda \rightarrow - \infty$ 
we have the commutation relation for $S^{-}_{\xi}$. 
Similarly, we derive 
commutation relations for $(S^{+}_{\xi})^{N}$ and $(T^{+}_{\xi})^{N}$.  
In summary we have the following: 
\bea 
\left( A(w_{0}) + D(w_{0}) \right) \, 
\left( S^{\pm}_{\xi} \right)^{(N)}_q 
& = & \left( S^{\pm}_{\xi} \right)^{(N)}_q \, 
\left( q^{-N} A(w_{0}) + q^{N} D(w_{0}) \right) 
\non \\ 
& + & e^{\pm w_0} X(w_{0}) (S^{\pm}_{\xi})_q^{(N-1)} (q^{S^Z} - q^{-S^Z}) 
\, , \label{loop-inhomo1} \\
\left( A(w_{0}) + D(w_{0}) \right) \, \left( T^{\pm}_{\xi} \right)^{(N)}_q 
& = & \left( T^{\pm}_{\xi} \right)^{(N)}_q \, 
\left( q^{N} A(w_{0}) + q^{-N} D(w_{0}) \right) 
\non \\
& - & e^{\mp w_0} X(w_{0}) (T^{\pm}_{\xi})_q^{(N-1)} 
(q^{S^Z} - q^{-S^Z}) \, ,  
\label{loop-inhomo2}
\eea 
where $X(w_{0})= C(w_{0})$ for $S^{+}_{\xi}$ and $T^{+}_{\xi}$,  
while $X(w_{0})= B(w_{0})$ for $S^{-}_{\xi}$ and $T^{-}_{\xi}$.

\subsection{$S^{\pm (N)}_{\xi}$ and $T^{\pm (N)}_{\xi}$ 
as generators of the $sl_2$ loop algebra}

Let us recall that in sector A  
where $S^{Z} \equiv 0 $ (mod $N$) 
$q_0$ is a root of unity with $q_0^{2N}=1$, 
(cf.  definition \ref{df:roots}),  
while in sector B where $S^{Z} \equiv N/2$ (mod $N$) with $N$ odd 
 $q_0$ is a primitive $N$ th root of unity.

It follows from (\ref{loop-inhomo1}) and (\ref{loop-inhomo2}) 
that $S_{\xi}^{\pm (N)}$ and $T^{\pm (N)}_{\xi}$ (anti-)commute with 
the transfer matrix of the six-vertex model  
$\tau_{6V}(z; \{ \xi_n \})$ in the cases of sector A and B:  
\bea
S^{\pm (N)}_{\xi} \, \tau_{6V}(z; \{ \xi_n \}) & = & 
q_0^N \tau_{6V}(z; \{ \xi_n \}) \, S^{\pm (N)}_{\xi}  \, , \non \\ 
T^{\pm (N)}_{\xi} \, \tau_{6V}(z; \{ \xi_n \}) & = & 
q_0^N \tau_{6V}(z; \{ \xi_n \}) \, T^{\pm (N)}_{\xi} \, . 
\label{loop-inhomo}
\eea 
 It is readily derived 
 from the (anti-)commutation relations (\ref{loop-inhomo})  
that the operators $S^{\pm (N)}$  and $T^{\pm (N)}$ commute 
with the XXZ Hamiltonian in the cases of sector A and B. 
Here we recall that the XXZ Hamiltonian $H_{XXZ}$ is given by 
the logarithmic derivative of the homogeneous transfer matrix 
$\tau_{6V}(z)$.

We now show that $S^{\pm (N)}_{\xi}$ and $T^{\pm (N)}_{\xi}$ 
generate the $sl_2$ loop algebra \cite{DFM}.  
When  $q_0$ is of type I, we set    
\bea
E_0^{+} & = & T^{-(N)}_{\xi}, \quad E_0^{-}=T^{+(N)}_{\xi}, 
\quad E_1^{+}=S^{+(N)}_{\xi}, \quad E_1^{-}=S^{-(N)}_{\xi}, \quad 
\non \\ 
& & - H_0 =  H_1 = {\frac 2 N}  S^Z \, .    
\label{id1-inhomo}
\eea
When $q_0$ is of type II, we set 
\bea
E_0^{+} & = & \sqrt{-1} \, T^{-(N)}_{\xi}, \,  E_0^{-} =  \sqrt{-1} \,  
T^{+(N)}_{\xi}, \,
 E_1^{+}=  \sqrt{-1} \,  S^{+(N)}_{\xi}, \, \non \\
& & E_1^{-} =  \sqrt{-1} \,  S^{-(N)}_{\xi}, \,  
  - H_0 = H_1=  {\frac 2 N}  S^Z  .   
\label{imaginary-inhomo}
\eea
Here $\sqrt{-1}$ denotes the square root of $-1$.  
(See also (A.13) of Ref. \cite{Missing}.)  
Then, operators $E_j^{\pm}, H_j$ for $j=0,1$,  
satisfy the defining relations of the $sl_2$ loop algebra 
$U(L(sl_2))$ \cite{DFM}: 
\bea 
& &  H_0 + H_1  = 0  , \, 
{\rm [} H_i, E_j^{\pm} {\rm ]}  = \pm a_{ij} E_j^{\pm}  ,  \, 
  \quad (i,j = 0, 1) 
 \label{Cartan} \\ 
& &  {\rm [} E_i^{+}, E_j^{-} {\rm ]}  =  \delta_{ij} H_{j} \, , 
  \qquad (i,j = 0, 1) \label{EF} \\
& & {\rm [} E_i^{\pm},  {\rm [} E_i^{\pm}, {\rm [} E_i^{\pm}, 
 E_j^{\pm} {\rm ]}   {\rm ]}   {\rm ]} = 0 , 
\, (i,j=0,1, \, i \ne j) \, . 
\label{Serre} 
\eea 
Here, the Cartan matrix $(a_{ij})$ of $A_1^{(1)}$ is defined  by 
\be 
\left( 
\begin{array}{cc}  
a_{00} & a_{01} \\
a_{10} & a_{11} 
\end{array} 
\right) 
 = \left( 
\begin{array}{cc} 
2 & -2 \\
-2 & 2 
\end{array}
\right) \, . 
\ee 

We obtain relations (\ref{Cartan}), (\ref{EF}), and (\ref{Serre})  
from the fact that $S^{\pm}_{\xi}$ and $T^{\pm}_{\xi}$ are generators 
of the quantum group $U_q({\hat{sl}}_2)$. 
The Serre relations (\ref{Serre}) hold if $q_0$ is a 
primitive $2N$th root of unity,  
or a primitive $N$th root of unity with $N$ odd \cite{DFM}. 
We derive it through the higher order quantum Serre relations 
due to Lusztig \cite{Lusztig}. 
The Cartan relations (\ref{Cartan}) hold for generic $q$.  
Relation (\ref{EF}) holds for the identification (\ref{id1-inhomo})
when $q_0$ is a root of unity of type I, 
and for the identification (\ref{imaginary-inhomo}) 
when $q_0$ is a root of unity of type II. 
In the case of sector A ($S^Z \equiv 0$ (mod $N$) 
and $q_0$ is a root of unity with $q_0^{2N}=1$ )  
we have the commutation relation \cite{DFM}  
\be 
[ S^{+(N)}_{\xi}, S^{-(N)}_{\xi}] 
= (-1)^{N-1} q^{N} \, {\frac 2 N} S^Z  \, .  
\ee
Here the sign factor $(-1)^{N-1} q^{N}$ is given by  1 or $-1$ 
 when $q$ is a root of unity of type I or II, respectively. 
In the case of  sector B ($S^Z \equiv N/2$ (mod $N$) with $N$ odd   
and $q_0$ a primitive $N$th root of unity),  
we have the commutation relation:  
\be 
[ S^{+(N)}_{\xi}, S^{-(N)}_{\xi}] =  {\frac 2 N} S^Z \, .   
\ee

Using the auto-morphism of the loop algebra:  
$\theta(E_0^{\pm})=E_1^{\pm}$ and   
$\theta(H_0)=  H_1$, 
we derive another identification.  
For instance, for the type I case we may set    
\bea
E_0^{+} & = & S^{+(N)}_{\xi}, \, E_0^{-}=S^{-(N)}_{\xi}, \, 
E_1^{+}=T^{-(N)}_{\xi}, \, 
E_1^{-}=T^{+(N)}_{\xi}, \non \\
& & \quad  H_0=-H_1= {\frac 2 N}  S^Z  .    
\label{id2-inhomo}
\eea
The identification (\ref{id2-inhomo}) with 
$\xi_n=0$ for all $n$ is given in Ref. \cite{DFM}.

Generators $x_{k}^{\pm}$ and $h_k$ for $k \in {\bm Z}$ 
 satisfying the defining relations (\ref{CDR}) are the classical limits of 
the Drinfeld generators \cite{Drinfeld,Chari-P1}.  
However, we also call them Drinfeld generators for simplicity. 
There is an isomorphism between the Drinfeld generators   
 and the Chevalley generators as follows \cite{Drinfeld,Chari-P1}: 
\be
 E_1^{\pm} \mapsto {  x}_0^{\pm}  \, , \quad 
  E_0^{+} \mapsto {  x}_1^{-} \, , \quad  
E_0^{-} \mapsto {  x}_{-1}^{+} \, , \quad 
 -H_0 =  H_1 \mapsto {  h}_0 \, . 
\label{isom}
\ee
 For roots of unity of type I, 
i.e.  $q_0$ is a primitive $N$th root of unity with $N$ odd ($q_0^N=1$) 
or $q$ is a $2N$th primitive root of unity with $N$ even ($q_0^N=-1$),  
through isomorphism (\ref{isom}) and identification   
(\ref{id1-inhomo}) we have the following correspondence:    
\bea 
{  x}_0^{+} & = & S^{+(N)}_{\xi} \, , \quad  
{  x}_0^{-} = S^{-(N)}_{\xi} \, , \quad  
{  x}_{-1}^{+} = T^{+(N)}_{\xi}  \, , \quad  
{  x}_{1}^{-} = T^{-(N)}_{\xi} \, , \quad  
\non \\
& & {  h}_{0} = {\frac 2 N} S^Z \, . 
\label{corr}
\eea
Here relations (\ref{corr}) 
are valid both in the sector $S^Z \equiv 0$ (mod $N$) 
and in the sector $S^Z \equiv N/2$ (mod $N$).  
 For roots of unity of type II, i.e. 
when $q_0$ is a $2N$th primitive root of unity with $N$ odd ($q_0^N=-1$),   
through the isomorphism (\ref{isom}) and the identification   
(\ref{imaginary-inhomo}) we have the following:    
\bea 
{  x}_0^{+} & = & \sqrt{-1} \, S^{+(N)}_{\xi} \, , \quad  
{  x}_0^{-} = \sqrt{-1} \, S^{-(N)}_{\xi} \, , \quad  
{  x}_{-1}^{+} = \sqrt{-1} \, T^{+(N)}_{\xi}  \, , \quad  
\non \\
& & {  x}_{1}^{-} = \sqrt{-1} \,  T^{-(N)}_{\xi} \, , \quad  
{  h}_{0} = {\frac 2 N} S^Z \, . 
\label{corr2}
\eea

Let the symbol $U_{q}^{\rm res}(g)$ denote 
the algebra generated by the $q$-divided powers 
of the Chevalley generators of a Lie algebra $g$ such as 
$(e_j^{\pm})^{(N)}_q$  \cite{Chari-P2}. 
The correspondence of the algebra $U_{q_0}^{\rm res}(g)$ 
at a root of unity, $q_0$, to the Lie algebra $U(g)$ was     
obtained essentially 
through the machinery introduced by Lusztig \cite{Modular,Lusztig} 
both for finite-dimensional simple Lie algebras   
and infinite-dimensional affine Lie algebras. In fact, 
by using the higher order quantum Serre relations \cite{Lusztig}, 
it has been shown 
that the affine Lie algebra $U(\hat{sl}_2)$ 
is generated by  $(e_j^{\pm})^{(N)}_{q_0}$ at  roots of unity.  
However, in the case of the affine Lie algebras $\hat{g}$,  
the highest weight conditions for the Drinfeld generators 
are different from those for the Chevalley generators.   
Through the highest weight vectors of the Drinfeld generators,  
finite-dimensional representations  
were discussed by Chari and Pressley for 
$U_{q_0}^{\rm res}(\hat{g})$ \cite{Chari-P2}.

%
%

 \setcounter{equation}{0} 
 \renewcommand{\theequation}{5.\arabic{equation}}

\section{The outline of the proof of the highest weight conjecture}

\subsection{Sufficient conditions of a highest weight vector}
 
%
\begin{lem}
Suppose that ${x}_{0}^{\pm}$, ${x}_{-1}^{+}$,  
 ${  x}_{1}^{-}$  and  ${h}_{0}$ 
satisfy the defining relations of $U(L(sl_2))$, 
and ${ x}_k^{\pm}$ and ${h}_k$ ($k \in {\bm Z}$) 
are generated from them.  
If a vector $|\Phi \ra$ satisfies the following:      
\bea 
& & {  x}_0^{+} | \Phi \ra  =   
{  x}_{-1}^{+} | \Phi \ra =0 \, , \label{ann-10} \\
& & {  h}_0 | \Phi \ra = r | \Phi \ra  \, , 
\label{diag-0} \\
& & ({  x}_0^{+})^{(n)} ({  x}_{1}^{-})^{(n)} 
 | \Phi \ra  =   
\lambda_n \, | \Phi \ra \quad   
\mbox{for} \, \, n = 1, 2, \ldots, r \, ,  
 \label{diag-101}
\eea 
where $r$ is a nonnegative integer and 
$\lambda_n$ are complex numbers.   Here $(X)^{(n)}$ 
denotes $X^n/n!$.  
Then $|\Phi \ra$ is highest weight, i.e. we have   
 \bea 
 {  x}_k^{+} |\Phi \ra & = & 0 \, \quad   ( k \in {\bm Z}) \, , 
   \label{ann}  \\ 
{  h}_k | \Phi \ra & = & {  d}_k | \Phi \ra \, \quad 
 ( k \in {\bm Z} ) \, , \label{diag}
\eea
where ${d}_k$ are complex numbers. 
\label{lem:scheme} 
\end{lem} 
Lemma \ref{lem:scheme} will be shown in Appendix A.  
Here we note that conditions 
(\ref{ann-10}), (\ref{diag-0}) and (\ref{diag-101}) are also 
necessary for $| \Phi \ra $ to be highest weight.

We shall derive theorem \ref{th:main} on 
the highest weight conjecture through lemma \ref{lem:scheme}. 
Conditions (\ref{ann-10}) and (\ref{diag-101}) correspond to 
(\ref{annST}) and (\ref{diagST}), respectively. 
We call conditions (\ref{ann-10}) and (\ref{diag-101}) 
{\it annihilation property} and {\it diagonal property}, 
respectively. We note that 
conditions (\ref{ann}) and (\ref{diag}) 
correspond to (\ref{eq:annihilation}) 
and (\ref{eq:Cartan}), respectively.

%
%
\subsection{Isolated solutions of the Bethe ansatz equations}
 
As far as the Bethe ansatz equations are concerned,   
every isolated solution at a root of unity,  $q_0$, is continuously  
extended  to a solution at generic $q$ near $q_0$. Note that 
the Bethe ansatz equations (\ref{BAE}) (and (\ref{BAE-inhomo})) 
 are expressed in terms of rational functions of $q$ of finite degree. 

We now assume conjecture \ref{conj:main}.   
Let ${\tilde t}_1, {\tilde t}_2, \ldots, {\tilde t}_R$ be 
a set of regular Bethe roots at $q_0$ forming 
an isolated solution of Bethe ansatz equations.   
It follows from conjecture \ref{conj:main} that 
there exist such regular Bethe roots 
at $q$, $t_1, t_2, \ldots, t_R$,  
that approach the regular Bethe roots at $q_0$, 
${\tilde t}_1, {\tilde t}_2, \ldots, {\tilde t}_R$, respectively, 
when $q$ goes to $q_0$. 
The regular Bethe state $|R \ra$ at $q_0$  
associated with ${\tilde t}_j$ is thus given by 
\be 
 |R \ra = B({\tilde t}_1, \eta_0) B({\tilde t}_2, \eta_0) \cdots 
B({\tilde t}_R, \eta_0) | 0 \ra \, . 
\ee
Here we recall $q=\exp(2 \eta)$ 
 and the $\eta$-dependence has been explicitly expressed as 
$B(w, \eta)$.  
Let us denote by $|R \ra_{q}$ the regular Bethe state at $q$ 
\be 
|R \ra_{q} = B(t_1, \eta) B(t_2, \eta) 
\cdots B(t_R, \eta) | 0 \ra \, .  
\ee 
Then, we have the following:   
\be 
 | R \ra = \lim_{q \rightarrow q_0} | R \ra_{q}  \, . 
\ee

We remark that 
conjecture \ref{conj:main} is supported by an extensive  study of numerical 
solutions of the Bethe ansatz equations near 
roots of unity \cite{FM1}. For $R=1$, 
we can show it explicitly with 
analytic expressions of Bethe roots in terms of $q$.

%
%
\subsection{Annihilation property}
\par 
Let us discuss the derivation of relations 
(\ref{annST}) (i.e. (\ref{ann-10})). 
We construct operators $S_{\xi}^{+(N)}$ and 
$T_{\xi}^{+(N)}$ as follows:
\be 
 S_{\xi}^{+(N)} 
 =  \lim_{q \rightarrow q_0}  
{\frac 1 {[N]_{q} !}} ({\hat{C}}({\infty}, \eta))^N \, , \, 
 T_{\xi}^{+(N)} 
 =  \lim_{q \rightarrow q_0}  
{\frac 1 {[N]_{q} !}} ({\hat{C}}(-{\infty}, \eta))^N. 
\ee
We  evaluate the action of the operator $S_{\xi}^{+(N)}$ on the 
vector $|R \ra$ by the limiting procedure: 
\be 
S_{\xi}^{+(N)} \, | R \ra  
=  \lim_{q \rightarrow q_0} 
\left\{ {\frac 1 {[N]_{q} !}}  
 \left(  {\hat C}(\infty, \eta) \right)^N \,  
  B(t_1, \eta) \cdots B(t_R, \eta) \, | 0 \ra \right\} \, . 
\ee

Let us recall that $L$ is the lattice size 
and $R$ the number of regular Bethe roots. 
We denote by  $\Sigma_R=\{1, 2, \ldots, R \} $ the set of 
indices of the regular Bethe roots, $t_1, t_2, \ldots, t_R$.  
For a given set $S$ we denote by $|S|$ the number of elements. 
We express by $\sum_{A \subset B}^{|A|=n}$ the sum over all 
such subsets $A$ of $B$ that have $n$ elements.  
The following lemma will be shown in \S 6.

\begin{lem} Let  $t_1, t_2, \ldots, t_R$ be regular Bethe roots 
at generic $q$. For  a given positive integer  $N_c$, we have  
\bea 
& &  {\frac 1 {[N_c]_q!}} 
\left({\hat C}(\pm \infty) \right)^{N_c} \, B(t_1) B(t_2) 
\cdots B(t_R) \, | 0 \ra \non \\ 
& = & \sum_{S_{N_c} \subseteq \Sigma_R}^{|S_{N_c}|={N_c}}  
 \prod_{\ell \in \Sigma_R \setminus S_{N_c}}  B(t_{\ell}) \,  
|0 \ra \, \exp(\pm \sum_{j \in S_{N_c}} t_j ) \, 
\prod_{j \in S_{N_c}} \left(  a_{\xi}^{6V}(t_{j})  
\prod_{k \in \Sigma_R \setminus S_{N_c}} f(t_j-t_k) \right)   \non \\
& & 
\times 
\, (-1)^{N_c} \, (q^{\pm 1}-q^{\mp 1})^{N_c}  \, \times  
\prod_{\ell=0}^{{N_c}-1} [{\frac L 2} - R + {N_c} - \ell]_q \, . 
\label{SB}
\eea
\label{lem:ann}
\end{lem}

Assuming conjecture \ref{conj:main} we have the following: 
\begin{prop}   
(i) When $L$ is even and  
 $q_0$  a root of unity with $q_0^{2N}=1$,     
 every regular Bethe state 
$|R \ra$ at $q_0$ is annihilated by operators $S^{+(N)}$ and $T^{+(N)}$ 
 in any sector of $S^Z \in {\bf Z}$:  
\be 
S_{\xi}^{+(N)} \, |R \ra  = 
T_{\xi}^{+(N)} \, |R \ra = 0 \, .  \label{annS}
\ee
(ii) When $L$ is odd, $N$ is odd and  
 $q_0$  a primitive $N$th root of unity,     
every regular Bethe state $|R \ra$ at $q_0$ 
is annihilated by operators $S_{\xi}^{+(N)}$ and $T_{\xi}^{+(N)}$ 
in any sector of $S^Z$, where $S^Z$ takes half-integers.  
\label{prop:ann-property}
\end{prop}   
\par \noindent {\bf Proof.}
Let us put $N_c=N$ in (\ref{SB}). 
In the case of even $L$, 
the product: $\prod_{\ell=0}^{N-1} [L/2 - R + N - \ell]_{q_0}$ 
vanishes if $q_0^{2N}=1$. 
Thus, by taking the limit  $q \rightarrow q_0$, 
 it follows from (\ref{SB})  that operators $S^{+(N)}$ and $T^{+(N)}$ 
annihilate the regular Bethe state $| R \ra $.     
In the case of odd $L$, 
the product: $\prod_{\ell=0}^{N-1} [L/2 - R + N - \ell]_{q_0}$ 
vanishes if $q_0^{N}=1$. 
It follows that 
operators $S^{+(N)}$ and $T^{+(N)}$ 
annihilate the regular Bethe state $| R \ra $. 
\hfill \opensquare

We note that when $q=\pm 1$ and $N=1$, expression 
(\ref{SB}) leads to another 
 proof for the spin $SU(2)$ invariance 
of the XXX Bethe states shown in Ref. \cite{TF}. 
In fact, the right hand side of  (\ref{SB}) vanishes 
when $q=\pm 1$ and $N=1$, since factor $q-q^{-1}$ vanishes.

%
%
\subsection{Diagonal property}

 Let us consider the limiting procedure of sending rapidities $z_j$ 
and $z_k$ to infinity 
in the products of operators such as ${\hat B}(z_j)$ and ${\hat C}(z_k)$.    
 Since they are matrices of finite sizes, 
the infinite limiting procedure does not depend on the order of 
sending arguments $z_j$ to infinity. For instance, we have 
\be 
\lim_{z_1 \rightarrow \infty} 
\left( \lim_{z_2 \rightarrow \infty} {\hat C}(z_1) {\hat B}(z_2)\right) 
= \left( \lim_{z_1 \rightarrow \infty} {\hat C}(z_1) \right)
\left( \lim_{z_2 \rightarrow \infty} {\hat B}(z_2) \right) \, . 
\label{limiting}
\ee
Each of the matrix elements of operators 
${\hat B}(z_j)$ and ${\hat C}(z_j)$ is written as a sum of products 
of $2 \times 2$ matrices such as $\sinh(z_j \pm \eta \sigma_n^z)/\sinh z_j$ 
and $\sinh 2\eta \sigma^{\pm}_n/\sinh z_j$. Here we recall 
normalization (\ref{normalization-inhomo}).

In addition to regular Bethe roots at generic $q$,  
$t_1, t_2, \ldots, t_R$,    
we introduce  $kN$ rapidities, $z_1, z_2, \ldots, z_{kN}$, 
forming a complete $kN$-string: 
$z_j = \Lambda +(kN+1-2j)\eta$ for $j=1, 2, \ldots, kN$.   
Here we recall definition (\ref{N-string}) of the complete $N$-string. 
We calculate the action of $(S_{\xi}^{+(N)})^{k}(T_{\xi}^{-(N)})^{k}$ 
on the Bethe state at $q_0$, $| R \ra = 
B({\tilde t}_1) \cdots B({\tilde t}_R) | 0 \ra$,  
as follows:  
\bea 
& &  \left( S_{\xi}^{+(N)} \right)^k \, \left(T_{\xi}^{-(N)} \right)^k  
\, | R \ra 
 = \lim_{q \rightarrow q_0} \Big( \lim_{\Lambda \rightarrow \infty}   
{\frac 1 {([N]_{q} !)^k}} \, ({\hat C}(\infty))^{kN} \, \non \\
& &  \quad \times \, {\frac 1 {([N]_{q} !)^k}} \, 
{\hat B}(z_1, \eta) \cdots {\hat B}(z_{kN}, \eta_n) \, 
 B(t_1, \eta) \cdots B(t_R, \eta) \, | 0 \ra  
\Big) \, .  
\label{eq:STR}
\eea
Associated with the limit: $\Lambda \rightarrow \pm \infty$,  
we define $\epsilon_{\nu}^{\pm}$  by  
\be 
\epsilon_{\nu}^{\pm} = \exp(\mp 2 z_{\nu}) 
= \epsilon_0^{\pm} q^{\pm 2 \nu } \, , 
\quad {\rm for} \quad \nu = 0, 1, \ldots, kN. 
\label{epsilon}
\ee 
Here $\epsilon_0^{\pm}=\exp(\mp \Lambda \mp (kN +1)\eta)$. 
We expand ${\hat B}(z_{\nu})$ at infinities: 
\be 
{\hat B}(z_{\nu}) = \sum_{n=0}^{\infty} {\hat b}_{\xi,n}^{\pm} 
 (\epsilon_{\nu}^{\pm})^{n} \, 
\quad (\Lambda \rightarrow \pm \infty).   
\label{eq:Bexp}
\ee
Infinite series (\ref{eq:Bexp}) is convergent 
if  $\epsilon_{\nu}^{\pm}$ is small enough.  
The matrix ${\hat B}(z_{\nu})$ has 
a finite number of matrix elements which are given by 
sums of products of $2 \times 2$ matrices such as 
$\sinh(z_j - \xi_n \pm \eta \sigma_n^z)/\sinh (z_j - \xi_n)$ and 
$\sinh 2 \eta \sigma_n^{\pm} /\sinh (z_j - \xi_n)$.    

Let $t_1, t_2, \ldots, t_R$ be 
a set of regular Bethe roots at generic $q$.  
We define $s_j^{\pm}(x)$ by 
\be 
s_j^{\pm}(x)= 1-x \exp(\pm 2t_j) \, \quad {\rm for} \quad j=1, 2, \ldots, R .  
\ee
For a subset $J$ of $\Sigma_R$, i.e. $J \subset \Sigma_R$,  
we define $F_J^{\pm}(x)$ by 
\be  
F_J^{\pm}(x) = \prod_{\ell \in \Sigma_R \setminus J} s_j^{\pm}(x) \, . 
\ee
When $J$ is empty,  
$F_J^{\pm}(x)$ reduces to $F^{\pm}(x)$ defined in (\ref{eq:F(x)}).  
We define $X_{\xi, J}^{\pm}(x)$  by  
\bea
X_{\xi, J}^{\pm}(x) & = & 
{\frac {\phi^{\pm}_{\xi}(x q^{\pm \rho})} 
{F_J^{\pm}(x q^{\mp ({\rho}+1)}) F_J^{\pm}(x q^{\pm ({\rho}+1)} )}}  
\prod_{\ell=1}^{\rho} 
{\frac 
 {\displaystyle \phi^{\pm}_{\xi}(x q^{\pm (2\ell-{\rho}-1)} )} 
 {\displaystyle \phi^{\pm}_{\xi}(x q^{\pm (2\ell-{\rho})} )} } \times  
\non \\
& & \quad \times \prod_{j \in J} (s_j^{\pm}(x q^{\mp ({\rho}-1)}) 
s_j^{\pm}(x q^{\pm ({\rho}-1)} ))^{-1} . 
\eea
Here $\rho$ is given by $|J|$. 
We define $\chi^{\pm; J}_{\xi, n}$ by 
the following series expansion: 
\be 
X_{\xi,J}^{\pm}(x) = \sum_{n=0}^{\infty} 
\chi^{\pm; J}_{\xi, n} \, x^n \, \quad (|x| \ll 1)  . 
\label{eq:chi_n-inhomo}
\ee

We shall show in \S 6 the following. 
\begin{lem}
Let $t_1, t_2, \ldots, t_R$ be a set of regular Bethe roots 
at generic $q$ in the inhomogeneous case.   
Sending  center $\Lambda$ of the $kN$-complete string 
to $\pm \infty$, we have  
\bea
& & \left( {\frac 1 {[N]_q!}} ({\hat C}(\pm \infty))^N \right)^k \, 
B(t_{1}) B(t_{2}) \cdots B(t_{R}) \, 
{\frac 1 {([N]_q!)^k}} \, {\hat B}(z_{1}) {\hat B}(z_{2}) \cdots 
{\hat B}(z_{kN}) | 0 \ra \quad   
\non \\ 
& & =  \left( {\frac {[kN]_q!} {([N]_q!)^k}} \right)^2 \, 
\sum_{{\rho}=0}^{kN} \sum_{J \subset \Sigma_R}^{|J|={\rho}}
 \sum_{n_0=0}^{(kN-{\rho})\Theta({\rho-1})} 
\left( 
\sum_{n_1+ \cdots + n_{\rho}=n_0} q^{\pm \sum_{j=1}^{\rho} 2j(n_j +1) } 
 \prod_{k=1}^{\rho} {\hat b}_{\xi,n_{k}}^{\pm}  
\right)  
\non \\
& & \times 
\prod_{j \in \Sigma_R \setminus J} B(t_{j}) | 0 \ra \, 
 \exp \left(\pm \sum_{j \in J} t_j \right) 
\, \, q^{\mp (n_0+{\rho})({\rho}+1)}  
 \left( \prod_{j \in J} a_{\xi}^{6V}(t_j) 
\prod_{\ell \in \Sigma_R \setminus J} f(t_j-t_{\ell}) \right) \non \\
& & \times (-1)^{kN} (q^{\pm 1}-q^{\mp 1})^{\rho} 
 \sum_{\ell=0; \ell \le \rho}^{kN-{\rho}-n_0} (-1)^{\ell} \, 
\chi^{\pm; J}_{\xi, kN-{\rho}-n_0-\ell} \, 
 \sum_{L \subset J}^{|L|=\ell} \, \exp(\pm \sum_{j \in L} 2t_j) 
\non \\
& & \, \times \, {\frac 1 {[{\rho}]_q!}} \prod_{i=0}^{\ell-1} [L/2-R-kN+i]_q 
\prod_{j=0}^{{\rho}-\ell-1} [L/2- R -kN -j]_q \quad 
  + O(\epsilon_{0}^{\pm} ) \, . 
\label{final-kN}
\eea
\label{lem:kN}
Here $\Theta(x)=1$ for $x \ge 0$ and $\Theta(x)=0$ for $x < 0$.  
The symbol $\sum_{n_1+ \ldots + n_{\rho}=n_0}$ denotes the sum 
over such nonnegative integers $n_1, n_2, \ldots, n_{\rho}$ such 
that their sum is given by $n_0$.   
\end{lem} 

In expansion (\ref{final-kN}), 
we call such terms with $\rho > 0$ {\it off-diagonal terms}, 
and  the term with ${\rho}=0$ the {\it diagonal term}. 

Assuming conjecture \ref{conj:main} we have the following:  
\begin{prop} 
Let $q_0$ be a root 
of unity with $q_0^{2N}=1$. 
In the cases of sector A and sector B we have   
\bea 
(S_{\xi}^{+(N)})^{(k)} (T_{\xi}^{-(N)})^{(k)}  \, | R \ra & = & (-1)^{kN} 
{\tilde{\chi}}_{\xi, kN}^{+} \, 
| R \ra  \, , 
\non \\
(T_{\xi}^{+(N)})^{(k)} (S_{\xi}^{-(N)})^{(k)} \, | R \ra & = & (-1)^{kN} 
{\tilde{\chi}}_{\xi, kN}^{-}  \, 
| R \ra  \, , \quad {\mbox{\rm for}} \, \, k \in {\bf Z}_{\ge 0} \, . 
\label{eq:diag-prop}
\eea
\label{prop:diag-property}
\end{prop} 
%
\par \noindent {\bf Proof.}  Let us recall $S^Z=L/2-R$. 
We have two cases 
whether $\rho \equiv 0$ (mod $N$) or not. 
When $\rho \ne 0$ (mod $N$), we have the vanishing product:    
\be 
\lim_{q \rightarrow q_0} \left( {\frac 1 {[{\rho}]_{q} !}} 
\prod_{i=0}^{\ell-1} [S^Z-kN+i]_{q} 
\prod_{j=0}^{{\rho}-\ell-1} [S^Z-kN -j]_{q}  \right) =0 \, . 
\ee 
When $\rho \equiv 0$ (mod $N$) and $\rho>0$, then ${\rho}=tN$ 
 for an integer $t$ with $0 < t \le k$. 
 We have  
\be 
\sum_{n_1 + \cdots + n_{tN}=n_0} q_0^{\pm \sum_{j=1}^{tN} 2j n_j} 
\prod_{k=1}^{tN} {\hat b}^{\pm}_{\xi, n_{k}} \, | 0 \ra = 0 \, , 
\quad {\rm for} \, \, n_0 \ge 0 \, . \label{v-p-n}
\ee
We now show (\ref{v-p-n}). 
When $q=q_0$, a root of unity with $q_0^{2N}=1$,  
we have the vanishing product of $B$ operators with their arguments  
given by a complete $tN$-string, as follows \cite{Tarasov-cyclic}:   
\be 
{\hat B}(z_1){\hat B}(z_2) 
\cdots {\hat B}(z_{tN}) | 0 \ra = 0 \, . 
\label{vanishing}
\ee 
Expanding ${\hat B}(z_1){\hat B}(z_2) \cdots {\hat B}(z_{tN}) | 0 \ra$ 
 with respect to $\epsilon_0^{\pm}$, we have   
$$  
\sum_{n_0=0}^{\infty} (\epsilon_0^{\pm})^{n_0} \sum_{n_1+ \cdots + n_{tN}=n_0} 
q_0^{\pm \sum_{j=1}^{tN} 2 j n_j} \prod_{k=1}^{tN} 
{\hat b}_{\xi,n_k}^{\pm} \, | 0 \ra \, = 0.  
$$
Therefore, all the off-diagonal terms vanish in (\ref{final-kN}). 
Sending $q$ to $q_0$, a root of unity with $q_0^{2N}=1$, 
we have  relations (\ref{eq:diag-prop}) of proposition 
\ref{prop:diag-property}.   
\hfill \opensquare

Here we note that we can show (\ref{vanishing}) also 
by the algebraic Bethe ansatz   
calculating all the matrix elements of 
the product of $B$ operators acting on the vacuum \cite{Korepin}.  

%

We obtain coefficients ${\tilde \chi}_{\xi, kN}^{\pm}$ of 
proposition \ref{prop:diag-property} (and proposition \ref{cor:lambda}), 
evaluating ${\chi}_{\xi, kN}^{\pm; J}$,  
which is defined by (\ref{eq:chi_n-inhomo}),  
at $q_0$ where we set  
$J= \emptyset$, $\rho=0$ and $t_j={\tilde t}_j$ for all $j$. 
Expression (\ref{diag-inhomo}) of ${\tilde \chi}_{\xi, kN}^{\pm}$ is 
derived through the following series 
expansion with respect to small $x$: 
\be 
{\frac 1 {(1 - x q \exp( 2t_j)) (1 - x q^{-1} \exp( 2 t_j) )}} 
 = \sum_{k=0}^{\infty} [k+1]_q \, x^k e^{ 2 k t_j} \, . 
\ee

We derive expression (\ref{eq:cor}) 
of eigenvalues $\lambda_k$ through relations 
(\ref{corr}) and (\ref{corr2}) for type I and II, respectively. 
They connect the two sets of generators, 
$S^{+ (N)}_{\xi}$ and  $T^{- (N)}_{\xi}$,    
and $x^{+}_{0}$ and  $x_1^{-}$, respectively.

%
%
\subsection{Derivation of theorem \ref{th:main}}

It follows from lemma \ref{lem:scheme} that we obtain theorem \ref{th:main} 
(and proposition \ref{cor:lambda}) 
from propositions \ref{prop:ann-property} and \ref{prop:diag-property}.  
Here we recall that by assuming conjecture \ref{conj:main},  
propositions \ref{prop:ann-property} and \ref{prop:diag-property} 
are derived from lemmas \ref{lem:ann} and \ref{lem:kN},  
which are shown in \S 6.3 and \S 6.4, respectively.

\subsection{On higher spin generalizations}

By introducing  higher dimensional representations of $L$ operators,   
the inhomogeneous transfer matrix of the six-vertex model,  
$\tau_{6V}(z;  \{ \xi_n \} )$,  
is generalized into that acting on 
the tensor product of higher dimensional vector spaces. 

In the cases of sector A and sector B 
we show that the generalized inhomogeneous transfer matrix has     
the $sl_2$ loop algebra symmetry at $q=q_0$. 
Here we employ the standard fusion method, and hence the  
higher spin generalization almost corresponds 
to a special case of the inhomogeneous one.  
The derivation of the $sl_2$ loop algebra symmetry for the    
generalized inhomogeneous transfer matrix is  parallel to 
that of \S 3.  The symmetry operators are derived from the $N$th powers 
of $B$ and $C$ operators by taking the infinite rapidity limit  
and then sending $q$ to $q_0$.  
We then show similarly as in \S 6 
that all regular Bethe states at $q_0$ 
are highest weight vectors 
in the cases of sector A and sector B.   
%

%
%
\setcounter{section}{5}
 \setcounter{equation}{0} 
 \renewcommand{\theequation}{6.\arabic{equation}}

\section{Explicit derivation of regular Bethe vectors being highest weight}

%
%

\subsection{Important Relations for Bethe roots}
\begin{lem}
Let $t_1$, $t_2$, $\ldots$, $t_R$ be a set of 
 Bethe roots at  a given value of $q$, and 
$S$ be a subset of $\Sigma_R= \{ 1,2, \cdots, R \}$. 
 For any pair of  sets $J_A$ and $J_B$ such that $J_A \cup J_B = S$ and 
 $J_A \cap J_B = \emptyset$,  we have the following:  
\bea 
& & \prod_{j \in J_A} 
\left( a_{\xi}^{6V}(t_j) \prod_{k \in \Sigma_R \setminus S} f(t_j-t_k) \right) 
\, \prod_{j \in J_A} \prod_{k \in J_B} f(t_j-t_k) 
\non \\ 
& = & \prod_{j \in J_A} 
\left( {d}_{\xi}^{6V}(t_{j}) \prod_{k \in \Sigma_R \setminus S} 
f(t_k-t_j) \right)  
\, \prod_{j \in J_A} \prod_{k \in J_B} f(t_k-t_j)    \, .  
\label{ABBA}
\eea  
Furthermore, we have  
\be 
\prod_{j \in S} a_{\xi}^{6V}(t_j) \prod_{k \in \Sigma_R \setminus S} 
 f(t_j- t_k)  
= \prod_{j \in S} d_{\xi}^{6V}(t_j) 
\prod_{k \in \Sigma_R \setminus S}  f(t_k-t_j)  \, . 
\label{affd}
\ee
\label{lem:BAE}
\end{lem} 
\par \noindent {\bf Proof.}
The first relation (\ref{ABBA}) is derived from the Bethe ansatz equations  
(\ref{BAE-inhomo}). The second relation (\ref{affd}) follows from 
 (\ref{ABBA}). 
\hfill \opensquare 

%
%

\subsection{Fundamental formula of the algebraic BA with infinite rapidities}
%

Through the commutation relations such as (\ref{CR}), 
which are derived from the Yang-Baxter equation,   
it was shown in Ref. \cite{Korepin}:  
\bea 
& & C(w_0) B(w_1) \cdots B(w_n) = B(w_1) B(w_2) \cdots B(w_n) C(w_0) \non \\
& & \quad + \, \sum_{j=1}^{n} B(w_1) \cdots B(w_{j-1}) B(w_{j+1})  
\cdots B(w_n) \, g(w_0-w_j) \times \non \\ 
& & \times \left\{ A(w_0) D(w_j) \prod_{k \ne j} f(w_0-w_k) f(w_k-w_j) 
- A(w_{j}) D(w_0) \prod_{k \ne j} f(w_k-w_0) f(w_j-w_k) \right \} 
\non 
\eea
\bea
& &  - \, \sum_{1 \le j < k \le n} 
B(w_0) B(w_1) \cdots B(w_{j-1}) B(w_{j+1})  
\cdots B(w_{k-1}) B(w_{k+1}) \cdots B(w_n) \times \non \\ 
& &  \times g(w_0-w_j) g(w_0-w_k) \{ A(w_{j}) D(w_k) f(w_j- w_k) 
\prod_{\ell \ne j, k} f(w_j-w_{\ell}) f(w_{\ell}-w_k)   
\non \\ 
& & + A(w_k) D(w_{j}) f(w_k-w_j) 
\prod_{\ell \ne j, k} f(w_k-w_{\ell}) f(w_{\ell}-w_j) \} \, . 
\label{CBBB}
\eea
Here  parameters $w_j$ for $j=0, 1, \ldots, n$ are arbitrary.

We denote by  $\Sigma_M$ 
  the set of $M$ letters:  $\Sigma_M= \{ 1, 2, \cdots, M \}$. 
 For a  set ${S}$ we express by $|{S}|$ the number of elements.  
The symbol ${\cal S}(n)$ denotes 
the symmetric group on $n$ letters such as $\{1, 2, \ldots, n \}$. 
 For a finite set $\Sigma$ we define ${\rm Sym}(\Sigma)$,  
 the symmetric group acting on the set $\Sigma$, as follows:   
Let $m$ be the number of elements of $\Sigma$. 
Then, each element of ${\rm Sym}(\Sigma)$ gives a one-to-one map: 
$\Sigma_m \rightarrow \Sigma$.   
 For instance,  we have ${\cal S}(n)={\rm Sym}(\Sigma_n)$. 

Recall that $w_j$ for $j=1, 2, \ldots, n$ are arbitrary parameters. 
For $S_n= \{j_1, j_2, \ldots, j_n \} \subset \Sigma_M$   
we introduce the following symbols: 
\bea 
\alpha_{\xi}^{\pm; \Sigma_M \setminus S_n}(z) & = & a_{\xi}^{6V}(z) \, 
q^{\mp L/2} 
\prod_{\ell \in \Sigma_M \setminus S_n} q^{\pm 1} f(z-w_{\ell}) \, , 
\non \\
{\bar \alpha}_{\xi}^{\pm; \Sigma_M \setminus S_n}(z) & = & 
d_{\xi}^{6V}(z) \, q^{\pm L/2} 
\prod_{\ell \in \Sigma_M \setminus S_n} q^{\mp 1} f(w_{\ell}-z) \, . 
\eea

\begin{df}
Let $S_n$ be a subset of $\Sigma_M$ with $n$ elements:   
$S_n= \{j_1, j_2, \ldots, j_n \} \subset \Sigma_M$. 
 For a given $P \in {\cal S}(n)$,  
we denote by  $S_{\ell}^{P}$ 
 the set $\{j_{P1}, \ldots, j_{P \ell} \}$ for $\ell=1, 2, \ldots, n$.   
Then we define $\Delta(\xi)_{S_n; \Sigma_M}^{\pm}$ by    
\bea  
\Delta(\xi)_{S_n; \Sigma_M}^{\pm} 
%
%
& = & \sum_{P \in {\cal S}(n)} \prod_{\ell=1}^{n} 
 \Bigg( \alpha_{\xi}^{\pm; \Sigma_M \setminus S_n}(w_{j_{P \ell}}) 
\prod_{k \in S_n \setminus S_{\ell}^{P}} q^{\pm 1} f(w_{j_{P \ell}}-w_k) 
\non \\ 
& &  - {\bar \alpha}_{\xi}^{\pm; \Sigma_M \setminus S_n}(w_{j_{P \ell}}) 
\prod_{k \in S_n \setminus S_{\ell}^{P}} q^{\mp 1} f(w_{k}-w_{j_{P \ell}}) 
\Bigg) \, . 
\label{CF0}
\eea 
\label{df:Delta} 
\end{df}
Here we note that $\Sigma_M \setminus S_{\ell}^{P} = 
(\Sigma_M \setminus S_n) \cup (S_n \setminus S_{\ell}^{P})$, 
 for $P \in {\cal S}(n)$ and $\ell =1, 2, \ldots, n$.     
We shall sometimes express  $\Delta(\xi)_{S_n; \Sigma_M}^{\pm}$ as 
$\Delta(\xi)_{S_n}^{\pm; \Sigma_M \setminus S_n}$.

\begin{lem}    
 Let  $w_j$ be arbitrary parameters for $j \in \Sigma_M$. 
We have the following: 
\be
\left( {\hat C}({\pm \infty}) \right)^{n} 
\, \prod_{\ell \in \Sigma_M} B(w_{\ell}) | 0 \ra 
= \sum_{S_n \subset \Sigma_M}^{|S_n|=n}   
\prod_{\ell \in \Sigma_M \setminus S_n}  B(w_{\ell})   
|0 \ra \, \exp(\pm \sum_{j \in S_n} w_j) \, 
\Delta(\xi)_{S_n; \Sigma_M}^{\pm} \, . 
\label{CCC-new}
\ee 
Here $\sum_{S_n \subset \Sigma_M}^{|S_n|=n}$ denotes  
the sum over all such subsets $S_n$ of $\Sigma_M$ 
that have $n$ elements.   
\end{lem} 
\par \noindent {\bf Proof.}
%
 Formula (\ref{CCC-new}) is derived through induction on $n$. 
 First, sending $w_0$ to infinity in equation (\ref{CBBB}), 
 we have the following:  
\bea 
& &{\hat C}({\pm \infty}) \, B(w_1) \cdots B(w_{M}) | 0 \ra  =  
\sum_{j=1}^{M} B(w_1) \cdots B(w_{j-1}) B(w_{j+1}) \cdots B(w_{M})   
| 0 \ra  \, e^{\pm w_j} \non \\
& \times & 
\left(a_{\xi}^{6V}(w_j) \, q^{ \mp L/2} \prod_{k \ne j} q^{\pm 1} f(w_j-w_k)
- d_{\xi}^{6V}(w_j) \, q^{\pm L/2}
\prod_{k \ne j} q^{\mp 1} f(w_k-w_j)   
  \right) \, . 
\label{formulaCB} 
\eea
This gives the case of $n=1$. 
Let us assume the case of $n$. 
Multiplying both hand sides of equation (\ref{CCC-new}) in the case of $n$ 
by ${\hat C}(\pm \infty)$, 
applying equation (\ref{formulaCB}) to the product of $B$ operators 
in the right hand side,  
we have equation (\ref{CCC-new}) in the case of $n+1$. 
\hfill \opensquare 

Let  $m$ and $n$ be  nonnegative integers satisfying $m \ge n$.   
We define the $q$-binomial coefficient by 
\be 
\left[
\begin{array}{c} 
m \\
n
\end{array} 
 \right]_q 
 = {\frac {[m]_q} { [m-n]_q [n]_q!} } \, . 
\label{q-binomial}
\ee

\begin{lem} 
Let $S_n$ be a subset of $\Sigma_M=\{ 1, 2, \cdots, M \}$ 
with $n$ integers. We express it as $S_n =\{j_1, j_2, \ldots, j_n\}$.   
Let $w_j$ for $j=1, 2, \ldots, n$ be arbitrary parameters. 
We have 
\bea 
\Delta(\xi)_{S_n; \Sigma_M}^{\pm} & = & 
\sum_{P \in {\cal S}(n)}  
\sum_{k=0}^{n} (-1)^k \left[ 
\begin{array}{c} 
n \\
k 
\end{array}
 \right]_q q^{\pm n(n-1)/2} q^{\mp (n-1)k} 
\prod_{1 \le \ell \le n-k} 
\alpha_{\xi}^{\pm; \Sigma_M \setminus S_n}(w_{j_{P \ell}}) \non \\ 
& \times &  
\prod_{n-k < \ell \le n} 
{\bar \alpha}_{\xi}^{\pm; \Sigma_M \setminus S_n}(w_{j_{P \ell}})   
\prod_{1 \le \ell < m \le n} f(w_{j_{P \ell}}-w_{j_{P m}})  \, .  
\label{CF1}
\eea
\label{lem:MainCF}
\end{lem}
The proof of lemma \ref{lem:MainCF} will be given in Appendix C.

%
%

\subsection{Proof of lemma \ref{lem:ann}}

\par \noindent {\bf Proof.} 
In formula (\ref{CF1}) we put $n=N_c$ and 
set parameters $w_j$ as 
\be
w_j= t_j \qquad  {\rm for} \quad j=1, 2, \ldots, R.   
\ee
Let us specify permutation $P \in {\cal S}(N_c)$ as follows: 
 First, we define disjoint sets $I$ and $K$ 
by $I=\{ P1, P2, \ldots, P(N_c-k) \}$ and 
 $K=\{ P(N_c-k+1), \ldots, P(N_c-1), P N_c \}$. Here we have 
$I \cup K = \Sigma_{N_c} = \{1, 2, \ldots, N_c \}$.   
Secondly, we define $P_I \in {\rm Sym}(I)$  
by $P_I \, j = P j$ for $j=1, 2, \ldots, {N_c}-k$, 
and $P_K \in {\rm Sym}(K)$ by $P_K \, j = P(j + (N_{c}-k))$ 
for $j=1, 2, \ldots, k$. 
Then, permutation  $P$ is given by   
$P j = P_I j$ for $1 \le j \le N_c-k$ 
and $P j = P_K (j - (N_c-k)) $ for $N_c-k +1  \le j \le N_c$.      
We therefore express the sum over all permutations as follows:  
\be 
\sum_{P \in {\cal S}({N_c})} = 
\sum_{I \cup K = \Sigma_{N_c}}^{|I|={N_c}-k, |K|=k}  
\sum_{P_I \in {\rm Sym}(I)} \sum_{P_K \in {\rm Sym}(K)}  \, . 
\label{eq:I+K}
\ee
Here, $\sum_{I \cup K = \Sigma_{N_c}}^{|I|= {N_c} -k, |K|=k}$ 
denotes the sum over all decompositions of $\Sigma_{N_c}$ 
into disjoint sets $I$ and 
$K$ where the numbers of elements are fixed such that   
$|I|={N_c}-k$ and $|K|=k$. 
Let $J_I$ and $J_K$ be $J_I=\{j_{\ell} | \ell \in I\}$  
and $J_K=\{j_{\ell} | \ell \in K \}$, respectively.  
We have  
$$
\prod_{\ell=1}^{{N_c}-k} 
\alpha_{\xi}^{\pm; \Sigma_R \setminus S_{N_c}}(w_{j_{P_I \, \ell}})   
= \prod_{j \in J_I} \alpha_{\xi}^{\pm; \Sigma_R \setminus S_{N_c}}(t_j) 
\quad {\rm for} \, \,   P_I \in {\rm Sym}(I)
$$ 
Similarly, we have 
$$
\prod_{\ell={N_c}-k+1}^{N_c} 
{\bar \alpha}_{\xi}^{\pm; \Sigma_R \setminus S_{N_c}}
(w_{j_{P_K (\ell-(N_c-k))}})  
= \prod_{j \in J_K} 
{\bar \alpha}_{\xi}^{\pm; \Sigma_R \setminus S_{N_c}}(t_j) 
\quad {\rm for} \, \,  P_K \in {\rm Sym}(K)
$$ 
Thus, $\Delta(\xi)_{S_{N_c}; \Sigma_R}^{\pm} \, q^{\mp {N_c}({N_c}-1)/2}$  
is given by    
\bea 
& & 
\sum_{k=0}^{N_c} (-1)^k 
\left[ 
\begin{array}{c} 
{N_c} \\
k 
\end{array} 
 \right]_q 
q^{\mp ({N_c}-1)k}  
 \sum_{I \cup K = \Sigma_{N_c}}^{|I|={N_c}-k,|K|=k} \prod_{j \in J_I} 
 \alpha_{\xi}^{\pm; \Sigma_R \setminus S_{N_c}}(t_j)  
\non \\
& \times & 
\prod_{\ell \in J_K} 
 {\bar \alpha}_{\xi}^{\pm; \Sigma_R \setminus S_{N_c}}(t_{\ell}) 
\prod_{j \in J_I}  \prod_{\ell \in J_K} f(t_j-t_{\ell}) \, \non \\
& \times & 
 \sum_{P_I \in {\rm Sym}(I) } 
  \prod_{1 \le \ell < m \le N-k}  f(t_{j_{P_I \ell}}-t_{j_{P_I m}}) 
  \sum_{P_K \in {\rm Sym}(K)} 
  \prod_{1 \le \ell < m \le k} f(t_{j_{P_K \ell}}-t_{j_{P_K m}})  \, . 
\non 
\eea
Applying formula (\ref{CF2}) to the sums over ${\rm Sym}(I)$ and 
${\rm Sym}(K)$,  
we have factors $[{N_c}-k]_q!$ and $[k]_q!$, respectively. Thus, we have 
\bea 
& & \Delta(\xi)_{S_{N_c}; \Sigma_R}^{\pm} =  
q^{\pm {N_c}({N_c}-1)/2} \, [{N_c}]_q! \,  
\sum_{k=0}^{N_c}  
(-1)^k q^{\mp ({N_c}-1)k}  \times  \non \\ 
& & \times \sum_{I \cup K = \Sigma_{N_c}}^{|I|={N_c}-k,|K|=k} 
\prod_{j \in J_I}  
\alpha_{\xi}^{\pm; \Sigma_R \setminus S_{N_c}}(t_j) 
 \prod_{\ell \in J_K}
  {\bar \alpha}_{\xi}^{\pm; \Sigma_R \setminus S_{N_c}}(t_{\ell}) 
\left( \prod_{j \in J_I}  \prod_{\ell \in J_K} f(t_j-t_{\ell}) \right) \, .
\non 
\eea
We apply lemma \ref{lem:BAE} 
to the product of $\alpha_{\xi}$'s or ${\bar{ \alpha}}_{\xi}$'s, 
and we make use of formula (\ref{CF3}) where $S_{N_c}$ and $J_K$ 
give $\Sigma_m$ and $S_n$ of (\ref{CF3}), respectively.  
Through the $q$-binomial formula (\ref{CFqb}),  
we obtain (\ref{SB}) for  generic $q$.   
\hfill \opensquare 

%
%
%
%

\subsection{Derivation of  lemma \ref{lem:kN}}  

%
\subsubsection{A complete $kN$-string as additional rapidities}
%

Let us recall that  the action of 
$(S_{\xi}^{+(N)})^{k}(T_{\xi}^{-(N)})^{k}$ 
on the Bethe state $| R \ra$ is formulated through (\ref{eq:STR}), 
where  $t_1, t_2, \ldots, t_R$ are regular 
Bethe roots at generic $q$, and  
$z_1, z_2, \ldots, z_{kN}$ are additional $kN$ rapidities
forming a complete $kN$-string: 
$z_j = \Lambda +(kN+1-2j)\eta$ for $j=1, 2, \ldots, kN$.

We now consider a complete $N_c$-string, where  
$N_c$ corresponds to the number 
of rapidities in the complete string, i.e. $N_c=kN$.     
Hereafter, we set parameters $w_j$ as follows: 
\be 
w_j= t_j \quad {\rm for}  \quad 1 \le j  \le R;  \qquad 
w_{j+R} = z_j \quad {\rm for} \quad  1 \le j \le N_c \, . 
\ee
We write index $R+j$ as $\underline{j}$, i.e.  
 $ w_{\underline{j}} = w_{R+j}$ for $1 \le j \le N_c$. 
Here $\Sigma_R=\{1, 2, \ldots, R \} $ gives the set of 
indices of Bethe roots, $t_1, t_2, \ldots, t_R$.    
We denote by  $Z_{N_c}$ the set of indices of 
 $N_c$ rapidities $z_j$:
\be 
 Z_{N_c} = \{R+1, R+2, \cdots, R+  N_c \}= \{ \underline{1}, \underline{2}, 
\cdots, \underline{N_c} \} \, .  
\ee
The set of all indices, 
$\Sigma_{R+N_c}=\{ 1, 2, \ldots, R, R+1, \ldots, R+N_c \}$,  
is given by the union of $\Sigma_R$ and $Z_{N_c}$, i.e. 
$\Sigma_{R+N_c} = \Sigma_R \cup Z_{N_c}$. 

 Let  $S_{N_c}$ be a subset of $\Sigma_{R+N_c}$ 
with $N_c$ elements. We define two disjoint sets $J$ and $W$ by 
$J = S_{N_c} \cap \Sigma_R$ and $W=S_{N_c} \cap Z_{N_c}$, respectively.  
Let $\rho$ be the number of elements of $J$, i.e. ${\rho}=|J|$.  
We express elements of $J$ as $j_{\ell}$ for $\ell=1, 2, \ldots, \rho$, 
and put them in increasing order: $j_1 < j_2 < \cdots < j_{\rho}$. 
Set $W$ is given by $S_{N_c} \setminus J$, and it is the set of suffices 
 for  rapidities $z_j$  in $S_{N_c}$. 
In order to specify subset $S_{N_c}=J \cup W$, 
we specify set $Z_{N_c} \setminus W$.  
We express elements of $Z_{N_c} \setminus W$  
by $\underline{\nu_1}, \underline{\nu_2}, \ldots, 
\underline{\nu_{\rho}}$, and put them in increasing order: 
$\nu_1 < \nu_2 < \ldots < \nu_{\rho}$.  
Here we note that 
 $\Sigma_{R+N_c} \setminus S_{N_c} =(\Sigma_R \setminus J) 
\cup (Z_{N_c} \setminus W)$.

We normalize $\Delta(\xi)_{S_{N_c}; \Sigma_{R+N_c}}^{\pm}$ as      
$ {\hat{\Delta}(\xi)}_{S_{N_c}; \Sigma_{R+N_c}}^{\pm} 
= \Delta(\xi)_{S_{N_c}; \Sigma_{R+N_c}}^{\pm}/
{\prod_{\underline{\nu} \in W}  n_{\xi}(z_{\nu})}$.   
We express ${\hat \Delta}(\xi)^{\pm}_{S_{N_c}; \Sigma_{R+N_c}}$ as  
${\hat \Delta}(\xi)^{\pm; \nu_1, \cdots, \nu_{\rho}}_{j_1, \cdots, j_{\rho}}$, 
for simplicity.   We have 
\be
 {\hat \Delta(\xi)}_{S_{N_c}}^{{\pm}; \Sigma_{R+{N_c}} \setminus S_{N_c}} 
= {\hat \Delta}(\xi)_{S_{N_c}}^{\pm; (\Sigma_R \setminus J) 
\cup (Z_{N_c} \setminus W)}  = 
{\hat \Delta}({\xi})_{j_1, j_2, \ldots, j_{\rho}}
^{{\pm}; \nu_1, \nu_2, \ldots, \nu_{\rho}} \, . 
\ee
%
Hereafter we write $\epsilon_0^{+}$ simply as $\epsilon_0$
and ${\hat \Delta}(\xi)_{j_1, j_2, \ldots, j_{\rho}}
^{+; \nu_1, \nu_2, \ldots, \nu_{\rho}}$ as 
${\hat \Delta}(\xi)_{j_1, j_2, \ldots, 
j_{\rho}}^{\nu_1, \nu_2, \ldots, \nu_{\rho}}$.

Taking advantage of the complete $N_c$-string $z_j$'s, 
we can show the following:   
\begin{lem} If $q$ is generic and $\rho >0$, 
${\hat{\Delta}(\xi)}_{j_1, \cdots, j_{\rho}}
^{\nu_1, \cdots, \nu_{\rho}}$ vanishes  
unless  $\nu_1, \nu_2, \ldots, \nu_{\rho}$ 
are given by $\nu+1, \nu+2, \ldots, \nu+{\rho}$, respectively, 
for an integer $\nu$ with $0 \le \nu \le {N_c}-{\rho}$.  
\label{lem:main}
\end{lem} 
Lemma \ref{lem:main} will be shown in Appendix D. 

Substituting $n$ and $M$ of formula  (\ref{CCC-new})  by 
$N_c$ and $R+N_c$, respectively,    
we  have   for generic $q$ the following: 
\bea   
& & \left({\hat C}(\infty) \right)^{N_c} B(t_1) 
\cdots B(t_R) {\hat B}(z_1) 
\cdots {\hat B}(z_{N_c}) | 0 \ra 
 = \sum_{{\rho}=0}^{{N_c}} \sum_{J \subset \Sigma_R}^{|J|={\rho}}  
\sum_{W \subset Z_{N_c}}^{|W|={N_c}-{\rho}} \times \non \\
&  & \times 
 \prod_{{\underline{\nu}} \in Z_{N_c} \setminus W} 
  {\hat B}(z_{\nu})  
\prod_{\ell \in \Sigma_R \setminus J} 
B(t_{\ell}) | 0 \ra \, 
\exp( \sum_{j \in J} t_j) 
{\frac  {\exp( \sum_{{\underline {\nu}} \in W} 2 z_{\nu})}  
{(q-q^{-1})^{N_c-{\rho}}} } \,   
{\hat{\Delta}}(\xi)_{S_{N_c}}^{+; \Sigma_{R+N_c} \setminus S_{N_c}} \non \\
& = & \sum_{{\rho}=0}^{N_c} \quad  (\epsilon_0)^{{\rho}-{N_c}} 
\sum_{1 \le j_1 < \cdots < j_{\rho} \le R} \quad   
\sum_{1 \le \nu_1 < \cdots < \nu_{\rho} \le N_c} \, 
{\hat B}(z_{\nu_1}) \cdots 
{\hat B}(z_{\nu_{\rho}}) 
\prod_{\ell \in \Sigma_R \setminus J}  
B(t_{\ell}) | 0 \ra   \non \\
& & \times \, 
q^{2 \nu_1 + \cdots + 2 \nu_{\rho}}  
{\hat{\Delta}(\xi)}_{j_1, \cdots, j_{\rho}}^{\nu_1, \cdots, \nu_{\rho}} 
\, \times \,  
e^{\sum_{j \in J} t_j}
{\frac {q^{- N_c (N_c +1)}} {(q-q^{-1})^{N_c -{\rho}}} } \, . 
\label{expand}
\eea
Here we assume that when ${\rho}=0$, the sums 
$\sum_{1 \le \nu_1 < \cdots < \nu_{\rho} \le N_c}$ 
and $\sum_{J \subset \Sigma_R}$ are given by 1, respectively.
Recall that  the term of $\rho=0$ in (\ref{expand}) is called 
the diagonal term, and the terms of $\rho > 0$ are called  
off-diagonal terms.

%
%
\begin{prop}
Let  $t_1, t_2, \ldots, t_R$ be regular Bethe roots  
at generic $q$, and  
 $z_1, z_2,\ldots, z_{N_c}$ form     
 a complete $N_c$-string with center $\Lambda$. We have  
\bea
& & ({\hat C}(\infty))^{N_c} \, 
B(t_{1}) \cdots B(t_{R}) {\hat B}(z_{1}) \cdots 
{\hat B}(z_{N_c}) | 0 \ra 
\non \\
& & = \prod_{\ell=1}^{R} B(t_{\ell}) | 0 \ra \, \epsilon_{0}^{-{N_c}} \, 
{\hat{\Delta}}(\xi)_{Z_{N_c}}^{+; \Sigma_{R+N_c} \setminus Z_{N_c}} 
\, {\frac  {q^{-{N_c} ({N_c}+1)}} {(q-q^{-1})^{N_c} }} \,  \non \\
&  & + \sum_{{\rho}=1}^{N_c} \sum_{J \subset \Sigma_R}^{|J|={\rho}} 
\sum_{n_0=0}^{\infty} 
\epsilon_{0}^{n_0+{\rho}-N_c} 
\sum_{n_1 + \cdots + n_{\rho} = n_0} 
{\hat b}_{\xi, n_1}^{+} \cdots {\hat b}_{\xi, n_{\rho}}^{+} 
\prod_{j \in \Sigma_R \setminus J}  
B(t_j) |0 \ra \,  \non \\
& & \quad \times \, \Sigma({\hat {\Delta}}_{\xi})_J \, \times \, 
e^{\sum_{j \in J} t_j} \, {\frac  {q^{-{N_c}({N_c}+1)}}
 {(q-q^{-1})^{{N_c}-{\rho}} }} \, 
 q^{\sum_{\ell=1}^{\rho} 2 \ell(n_{\ell} +1) }  
\label{a-sum} 
\eea
where $\Sigma({\hat {\Delta}}_{\xi})_J$ is given by 
\be 
\Sigma({\hat {\Delta}}_{\xi})_J = 
\sum_{\nu =0}^{{N_c}-{\rho}} q^{2\nu(n_0+{\rho})} 
{\hat{\Delta}(\xi)}_{j_1, \cdots, j_{\rho}}^{\nu+1, \cdots, \nu+{\rho}} \, .  
\label{Dsum}
\ee
Here, $\sum_{J \subset \Sigma_R}^{|J|={\rho}}$ denotes the sum 
over all such subsets of  $\Sigma_R$ that have $\rho$ elements, where  
$J=\{j_1, \ldots, j_{\rho}\}$,     
and $\sum_{n_1 + \cdots + n_{\rho} = n_0}$ the sum over 
all nonnegative integers $n_1, n_2, \ldots, n_{\rho}$ 
satisfying the condition: $n_1 + n_2 + \cdots + n_{\rho} = n_0$. 
\label{prop:Dsum}
\end{prop} 
\par \noindent {\bf Proof.}
For off-diagonal terms  in (\ref{expand}) we expand the 
 product of $B$ operators, 
${\hat B}(z_{\nu_1}) \cdots {\hat B}(z_{\nu_{\rho}})$, 
in the power series of $\epsilon_0$ through (\ref{eq:Bexp}).  
We have the infinite sum over  
all nonnegative integers $n_1, n_2, \ldots, n_{\rho}$ 
satisfying the condition: $n_1 + n_2 + \cdots + n_{\rho} = n_0$. 
It follows from lemma \ref{lem:main} that we have    
\bea 
\sum_{1 \le \nu_1 < \cdots < \nu_{\rho} \le {N_c}} 
q^{2 \nu_1 (n_1+1) + \cdots + 2 \nu_{\rho} (n_{\rho}+1)} 
{\hat{\Delta}(\xi)}_{j_1, \cdots, j_{\rho}}^{\nu_1, \cdots, \nu_{\rho}} 
= q^{\sum_{\ell=1}^{\rho} 2 \ell(n_{\ell} +1)} \,
\Sigma(\Delta_{\xi})_J .  \non 
\eea
Making use of it, we derive the expression (\ref{a-sum}). 
\hfill \opensquare

%
%

Let us consider the term of ${\rho}=0$ in expansion (\ref{expand}).  
It leads to the eigenvalue of 
$(S_{\xi}^{+(N)})^{k} (T_{\xi}^{-(N)})^{k}$  by  
putting $N_c=kN$ and sending $q$ to a root of unity $q_0$. 
\begin{lem} 
The coefficient of the diagonal term  of (\ref{expand}) 
is given by  
\be 
\epsilon_{0}^{- N_c} \,  
{\frac {q^{-N_c(N_c+1)}} {(q-q^{-1})^{N_c}} } 
{\hat{\Delta}}(\xi)_{Z_{N_c}}^{+; \Sigma_{R+ N_c} \setminus Z_{N_c}} 
= (-1)^{N_c} \, {\chi}_{\xi, N_c}^{+} \, 
\left( [N_c]_q! \right)^2 + O( \epsilon_0) \, . 
\ee 
\label{lem:diagonal_term} 
\end{lem} 
\par \noindent {\bf Proof.}
Making use of formula (\ref{CF1}) at generic $q$ 
we have 
\bea 
& & {\hat{\Delta}}(\xi)_{Z_{N_c}}^{+; \Sigma_{R+{N_c}} \setminus Z_{N_c}}
  =  [{N_c}]_q!  \, 
\displaystyle{\frac 
{\prod_{\ell=1}^{{N_c}-1} \phi_{\xi}^{+}(\epsilon_0 q^{2 \ell+1})}  
{\prod_{\ell=1}^{N_c} \phi_{\xi}^{+}(\epsilon_0 q^{2 \ell})}  
}
\, F^{+}(\epsilon_0) F^{+}(\epsilon_0 q^{2{N_c}+2}) 
\non \\
& & \times \sum_{j=0}^{N_c} (-1)^j \, 
\left[
\begin{array}{c} 
{N_c} \\
j
\end{array} 
 \right]_q
q^{{N_c}({N_c}-1)/2 - ({N_c}-1)j} \, 
\displaystyle{\frac 
{ \phi_{\xi}^{+}(\epsilon_0 q^{2j+1})}  
{F^{+}(\epsilon_0 q^{2j}) F^{+}(\epsilon_0 q^{2j+2})}  } \, . 
\label{Delta0}
\eea 
We expand the last line of (\ref{Delta0}) 
in terms of $\epsilon_0$.  It is given by 
the series of (\ref{eq:chi_n-inhomo}) 
with $J= \emptyset$ and $\rho=0$.  
By noting $\prod_{\ell=0}^{{N_c}-1} (1- q^{2m -2\ell})=0$ 
for $0 \le m < N_c$, 
${\hat{\Delta}}(\xi)_{Z_{N_c}}^{+; \Sigma_{R+{N_c}} \setminus Z_{N_c}}$ 
is given by   
\be 
\epsilon_0^{N_c} \, {\chi}_{\xi, N_c}^{+} \, (-1)^{N_c} q^{{N_c}({N_c}+1)} 
([{N_c}]_q!)^2 (q-q^{-1})^{N_c} \,  + 
O(\epsilon_0^{{N_c}+1} ) \, . 
\ee 
\hfill \opensquare 

%
%
%
%

We now consider the off-diagonal terms of (\ref{expand}). 
Let $J_I$ and $J_K$ be disjoint sets such that 
$J_I \cup J_K =J$ where $|J|=\rho$. 
We define $G_{J_I,J_K}^{\pm}(x)$ by 
\be
 G_{J_I, J_K}^{\pm}(x) 
 = \prod_{j \in J_I} s_j^{\pm}(x q^{\mp(\rho-1)}) 
\prod_{j \in J_K} s_j^{\pm}(x q^{\pm(\rho-1)}) \, . 
\ee

Let us define subsequences. 
We consider two sequences of numbers,  
$a_1, a_2, \ldots, a_m$ and $b_1, b_2, \ldots, b_n$,  and assume that 
$\{b_1, b_2, \ldots, b_n\} \subset \{ a_1, a_2, \ldots, a_m \}$ and 
$n \le m$.  
We define sequence $i(j)$ 
by $a_{i(j)} = b_j$ for $j=1, 2, \ldots, n$. 
We say that $b_1, b_2, \ldots, b_n$ is a subsequence 
of $a_1, a_2, \ldots, a_m$, 
if $1 \le i(1) < i(2) < \cdots < i(n) \le m$.

%
%
\begin{prop} 
Let  $J$ be a subset of $\Sigma_R$. We denote it as follows: 
 $J=\{j_1, j_2, \ldots, j_{\rho} \}$    
where $j_1 < j_2 < \ldots < j_{\rho}$. 
Then,  
$\Sigma({\hat{\Delta}}_{\xi})_{J}$ 
is evaluated at generic $q$  as follows:  
\bea 
& & \sum_{\nu=0}^{N_c-\rho} q^{2\nu(n_0+ \rho)} 
{\hat{\Delta}(\xi)}_{j_1, \cdots, j_{\rho}}^{\nu+1, \cdots, \nu+{\rho}}  
=  [{N_c}]_q! 
\left[ 
\begin{array}{c}
{N_c} \\ 
 \rho  
\end{array} 
\right]_q
 q^{{N_c}({N_c}-1)/2 } 
q^{-(S^Z-N_c) \rho}  
\non \\ 
& \times &
\prod_{j \in J} \left( a_{\xi}^{\rm 6V}(t_j) 
\prod_{\ell \in \Sigma_R \setminus J} f(t_j - t_{\ell})  \right) 
 F_J^{+} (\epsilon_{{N_c}+1}) F_J^{+} (\epsilon_0)  
  \displaystyle{\frac 
{\prod_{\ell=1}^{{N_c}-1} \phi_{\xi}^{+}(\epsilon_{\ell} q)} 
{\prod_{\ell=1}^{N_c} \phi_{\xi}^{+}(\epsilon_{\ell})} } 
\non \\
&  \times & 
\sum_{\sigma=0}^{\rho} (-1)^{\sigma} q^{2 (S^Z-N_c) \sigma} 
q^{-({\rho}-1) \sigma}  
 \sum_{J_I \cup J_K=J}^{|J_I|={\rho}-{\sigma}, |J_K| = {\sigma}} \, 
\left( \prod_{j \in J_I} \prod_{\ell \in J_K} f(t_{\ell} - t_j) \right) 
\non \\
& & \quad \times 
\prod_{j \in J_I} s_j^{+}(\epsilon_0) 
\prod_{j \in J_K} s_j^{+}(\epsilon_{{N_c}+1}) 
\non \\
& \times &  \sum_{\nu=0}^{N_c- \rho} 
(-1)^{\nu} q^{-({N_c}-{\rho}-1-2 n_0){\nu}} 
\left[
\begin{array}{c} 
{N_c} -{\rho} \\
\nu 
\end{array} 
\right]_q
\, X_{\xi,J}^{+}(\epsilon_{\nu} q^{{\rho}+1}) \, 
G_{J_I,J_K}^{+}(\epsilon_{\nu} q^{{\rho}+1}) 
\non \\
\label{DSZ}
\eea
Here,  
$\sum_{J_I \cup J_K=J}^{|J_I|={\rho}-{\sigma}, |J_K| = {\sigma}}$ 
denotes the sum over 
all pairs of disjoint sets $J_I$ and $J_K$ such that 
the sum of  $J_I$ and $J_K$ gives $J$, and  
 $J_I$ and $J_K$ have ${\rho}-{\sigma}$ and ${\sigma}$ elements, 
 respectively.  
\end{prop} 
%
%
\par \noindent {\bf Proof.} 
When $(\nu_1, \nu_2, \cdots, \nu_{\rho}) = 
(\nu+1, \nu+2, \cdots, \nu+{\rho})$, we have      
$Z_{N_c} \setminus W=\{ \underline{\nu+1}, \underline{\nu+2}, 
\ldots, \underline{\nu+{\rho}} \}$. The set $S_{N_c}  = J \cup W $ 
 is given by the following: 
\be 
S_{N_c} = \{ j_1, j_2, \ldots, j_{\rho} \} \cup 
\{\underline{1}, \underline{2}, \ldots, \underline{\nu}, 
\, \underline{\nu+{\rho}+1}, 
\ldots, \underline{{N_c}-1}, \underline{N_c} \} \, .  
\label{eq:S_N}
\ee
Let us put elements of $S_{N_c}$ in increasing order as  
$i_1 < i_2 < \cdots < i_{N_c}$. The first ${\rho}$ elements  
$i_1, i_2, \ldots, i_{\rho}$ are 
given by $j_1, j_2, \ldots, j_{\rho}$, respectively, and  
 $i_{{\rho}+1}, i_{{\rho}+2}, \ldots, i_{{\rho}+\nu}$ by 
$\underline{1}, \underline{2}, \ldots, \underline{\nu}$, respectively, and   
$i_{{\rho}+\nu+1}, i_{{\rho}+\nu+2}, \ldots ,i_{N_c}$ by 
$\underline{\rho+\nu+1},  \ldots, \underline{N_c-1}, \underline{N_c}$, 
respectively.   
By making use of formula (\ref{CF1}) 
with $S_{N_c}$ given by (\ref{eq:S_N}),   
${\hat {\Delta}}(\xi)_{j_1, \ldots, j_{\rho}}^{\nu+1, \ldots, \nu+{\rho}}$ 
is expressed as follows: 
\bea 
& &  \sum_{\kappa=\nu}^{\nu+{\rho}} (-1)^{\kappa} 
q^{{N_c}({N_c}-1)/2 - ({N_c}-1)\kappa} 
\left[ 
\begin{array}{c}
{N_c} \\ 
\kappa 
\end{array} 
\right]_q  \quad 
\left(\prod_{\beta=1}^{\nu} n_{\xi}(z_{\beta}) 
\prod_{\gamma=\nu+\rho+1}^{N_c} n_{\xi}(z_{\gamma}) 
\right)^{-1}
\times \non \\
&  \times & 
\sum_{P \in {\cal S}(N_c)}  
\prod_{\ell=1}^{{N_c}-\kappa} 
\alpha_{\xi}^{+; \Sigma_{R+N_c} \setminus S_{N_c}}(w_{i_{P \ell}}) 
\prod_{\ell={N_c}-\kappa+1}^{N_c} 
{\bar \alpha}_{\xi}^{+; \Sigma_{R+N_c} \setminus S_{N_c}}(w_{i_{P \ell}}) 
\non \\
& & \times 
\prod_{1 \le \ell < m \le {N_c}} f( w_{i_{P \ell}} - w_{i_{P m}}) . 
\label{ksum}
\eea
In (\ref{ksum}),  the range of $\kappa$ has been reduced 
from $\kappa= 0, 1,, \ldots, {N_c}$ into 
$\kappa= \nu, \nu+1, \ldots, \nu+{\rho}$.     
We first note that $f(z_{j}-z_{j+1})=0$ for $j=1, \ldots, \nu-1$ 
and for $j=\nu+{\rho}, \ldots, {N_c}-1$.  
Hence the product $\prod_{\ell < m} f_{i_{P \ell}, i_{P m}}$ vanishes  
unless sequence $i_{P1}, i_{P2}, \ldots, i_{PN}$ contains two 
decreasing subsequences $\underline{N_c}, \underline{{N_c}-1}, 
\ldots, \underline{\nu+{\rho}+1}$ and $\underline{\nu}, \underline{\nu-1}, 
\ldots, \underline{1}$, i.e. 
unless $P^{-1} \ell > P^{-1} m$ for 
${\rho}+\nu+1 \le \ell < m \le {N_c}$ and 
$P^{-1} \ell > P^{-1} m $ for 
${\rho}+1 \le \ell < m \le {\rho}+\nu $. 
Here we define $I$ and $K$ by $I=\{P1, P2, \ldots, P({N_c}-\kappa) \}$ 
and $K=\{ P({N_c}-\kappa+1), \ldots, P({N_c}-1), P{N_c} \}$, respectively.  
Secondly, we show that the summand of equation (\ref{ksum}) vanishes 
unless $\kappa \ge \nu$. We note that    
$\alpha_{\xi}^{+; \Sigma_{R+N_c} \setminus S_{N_c}}(z_{\nu})=0$ 
for ${S_{N_c}}$ of (\ref{eq:S_N}), 
and $\underline{\nu}=i_{{\rho}+\nu}$.  
The summand of (\ref{ksum}) therefore vanishes if  $\rho+\nu \in I$, 
and we consider only such $P$ where subsequence  
${\rho}+\nu, {\rho}+\nu-1, \ldots, {\rho}+1$ is contained in sequence 
$P({N_c}-\kappa+1), \ldots, P({N_c}-1), P{N_c}$.  
We therefore have $\kappa \ge \nu$.    
Thirdly, we show that the summand of (\ref{ksum}) vanishes  
unless $\kappa \le \nu+{\rho}$, in the same way 
as the case of $\kappa \ge \nu$. 
Here we recall that 
$\bar{\alpha}_{\xi}^{+; \Sigma_{R+N_c} \setminus S_{N_c}}
(z_{\nu+\rho+1})=0$ for ${S_{N_c}}$ of (\ref{eq:S_N}), and  
$i_{\rho+\nu+1}={\underline{\rho+\nu+1}}$.  
Thus, we have shown the reduction of the range of $\kappa$. 
Furthermore, we have shown that  
 the summand of (\ref{ksum}) vanishes unless 
\be  
\{\rho+\nu+1, \ldots, {N_c}-1, {N_c} \} \subset I \, , \quad   
\{{\rho}+1, {\rho}+2, \ldots, {\rho}+\nu \} \subset K .  
\label{eq:IK}
\ee
We now consider the sum over $P \in {\cal S}(N_c)$ in (\ref{ksum}). 
Let us introduce disjoint subsets $S_I$ and $S_K$ of $S_{N_c}$ as follows:   
$S_I = \{i_{P1}, i_{P2}, \ldots, i_{P({N_c}-\kappa)} \}$ and 
$S_K = \{i_{P({N_c}-\kappa+1)}, \ldots, i_{P({N_c}-1)}, i_{P{N_c}} \}$. 
Then, we have 
\be 
S_I = J_I \cup  
\{ \underline{{N_c}}, \underline{{N_c}-1}, \ldots, 
\underline{\nu+{\rho}+1} \} 
 \, , \quad 
S_K = J_K \cup \{\underline{\nu}, \ldots, \underline{2}, 
\underline{1} \}  \, . 
\label{eq:SIK}
\ee
Here, $J_I$ and $J_K$ are  disjoint subsets of $J$ 
such that $J=J_I \cup J_K$. 
Let us denote by ${\sigma}$ 
the number of elements of $J_K$, i.e. 
$|J_K|=\sigma$ and $|J_I| = {\rho}-{\sigma}$.  
We express $\sum_{P \in {\cal S}(N_c)}$ of (\ref{ksum}) by 
the sum over disjoint sets $I$ and $K$ with  
 $I \cup K = \Sigma_{N_c}$, such as shown in (\ref{eq:I+K}). 
 It follows from (\ref{eq:SIK}) that the sum 
$\sum_{I \cup K}^{|I|=N_c-\kappa, |K|=\kappa}$ reduces to the sum  
$\sum_{J_I \cup J_K = J}^{|J_I|=\rho-\sigma, |J_K|=\sigma}$. 
Here we note that $\kappa=\sigma+ \nu$.   
Similarly as in the proof of lemma \ref{lem:ann}, calculating the sums over 
$P_I \in {\rm Sym}(I)$ and over $P_K \in {\rm Sym}(K)$, we have 
factors $[N_c-\kappa]! $ and $[\kappa]!$, respectively. 
Applying formula (\ref{ABBA}) of lemma \ref{lem:BAE}, we show the following:  
\bea 
& & \prod_{j \in J_I} 
\alpha_{\xi}^{+; \Sigma_R \setminus J}(t_j)  
\prod_{k \in J_K} 
{\bar \alpha}_{\xi}^{+; \Sigma_R \setminus J}(t_k)  
\prod_{j \in J_I} \prod_{k \in J_K} f(t_j - t_k)  
\non \\
& = & q^{(L/2-R+\rho)(2 \sigma - \rho)} 
\left( \prod_{j \in J} a_{\xi}^{\rm 6V}(t_j) 
\prod_{\ell \in \Sigma_R \setminus J} 
f(t_j - t_{\ell}) \right) \prod_{j \in J_I} \prod_{k \in J_K} f(t_k - t_j)   
\non 
\eea
Expressing sum (\ref{ksum}) over $\kappa$ 
as that of ${\sigma}$ ($\kappa=\sigma + \nu$), 
we obtain expression (\ref{DSZ}). 
\hfill \opensquare 

Let us  expand  $G_{J_I,J_K}^{\pm}(\epsilon)$ 
with respect to small parameter $\epsilon$:  
\be 
G_{J_I,J_K}^{\pm}(\epsilon) = \sum_{\ell=0}^{\rho} 
 (-1)^{\ell} G_{\ell}^{\pm; J_I,J_K} \, \epsilon^{\ell}
\ee
The coefficients $G_{\ell}^{\pm; J_I,J_K}$ for $\ell \le \rho$ 
are explicitly given by 
\be 
G_{\ell}^{\pm; J_I,J_K} =
 \sum_{\ell_I=0}^{\ell} q^{\pm({\rho}-1)(\ell_K - \ell_I) } 
\sum_{L_I \subset J_I}^{|L_I|= \ell_I} \exp(\pm \sum_{j \in L_I} 2t_j)   
\sum_{L_K \subset J_K}^{|L_K|= \ell_K} \exp(\pm \sum_{j \in L_K} 2t_j) \, . 
\label{defGC}
\ee
Here $\ell_K=\ell-\ell_I$, and $G_{\ell}^{\pm; J_I, J_K}=0$ 
for $\ell > \rho$.   

%
%
\begin{lem} 
Sum (\ref{Dsum}),  
$\Sigma({\hat \Delta}_{\xi})_J = 
\sum_{\nu=0}^{N-{\rho}} q^{2\nu(n_0+{\rho})} 
{\hat{\Delta}}_{j_1, \ldots, j_{\rho}}^{\nu+1, \ldots, \nu+{\rho}}$,   
is expanded in terms of $\epsilon_0$ as follows:  
\bea 
& &  (\epsilon_{0})^{{N_c}-{\rho}-n_0} \, 
(-1)^{N_c} q^{{N_c}({N_c}+1) - ({\rho}+1)(n_0+{\rho})} (q-q^{-1})^{N_c} 
\left([{N_c}]_q! \right)^2 \non \\ 
& \times & \left( \prod_{j \in J}
 a_{\xi}^{\rm 6V}(t_j) 
\prod_{\ell \in \Sigma_R \setminus J} f(t_j - t_{\ell}) \right) 
 \sum_{\ell=0, \, \ell \le {N_c}-{\rho}-n_0}^{\rho} 
(-1)^{\ell} \chi^{+; J}_{\xi, {N_c}-{\rho}-n_0-\ell} \non \\  
&\times & {\frac 1 {[{\rho}]_q!}} \prod_{i=0}^{\ell-1} [S^Z-{N_c}+i]_q 
\prod_{j=0}^{{\rho}-\ell-1} [S^Z-{N_c} -j]_q  
 \sum_{L \subset J}^{|L|=\ell} \exp(\sum_{j \in L} 2t_j ) \non \\ 
& & \quad + \, O((\epsilon_0)^{{N_c}-{\rho}-n_0+1}) \, . 
\eea
\label{lem:Dsum}
\end{lem} 
We note that all the possibly divergent 
terms with order of $\epsilon_0^{n_0+{\rho}-{N_c}}$ in (\ref{a-sum}) 
do not diverge since $\Sigma({\hat \Delta}_{\xi})_J$ 
is of order of $(\epsilon_0)^{{N_c}-{\rho}-n_0}$ in the 
limit $\epsilon_0 \rightarrow 0$.   
\par \noindent {\bf Proof.}
We evaluate  sum (\ref{Dsum}) over $\nu$, 
$\Sigma({\hat \Delta}_{\xi})_J$,  
by (B.1) as follows:   
\bea 
& & 
\sum_{\nu=0}^{{N_c}-{\rho}}  (-1)^{\nu} q^{-({N_c}-{\rho}-1-2n_0) \nu} 
\left[
\begin{array}{c} 
{N_c}-{\rho} \\
\nu  
\end{array} 
\right]_q
\, X_{\xi,J}^{+}(\epsilon_{\nu} q^{{\rho}+1}) 
\, G_{J_I,J_K}^{+}(\epsilon_{\nu} q^{{\rho}+1}) 
\non \\
& = & (\epsilon_{0})^{{N_c}-{\rho}-n_0} \, 
(-1)^{{N_c}-{\rho}} (q-q^{-1})^{{N_c}-{\rho}} 
q^{- n_0 ({\rho}+1) + ({N_c}-{\rho})({N_c}+{\rho}+3)/2} 
\non \\
& & \times [{N_c}-{\rho}]_q!  \, 
\sum_{\ell=0}^{\rho}  (-1)^{\ell} 
\chi_{\xi, {N_c}-{\rho}-n_0-\ell}^{+; J} G_{\ell}^{+; J_I, J_K} 
 + O((\epsilon_0)^{{N_c}-{\rho}-n_0+1})  \, . 
\eea
Here, the product 
$\prod_{i=0}^{{N_c}-{\rho}-1} \left( 1- q^{2(j+\ell+n_0-i)} \right)$ 
vanishes for $j+ \ell < {N_c}-{\rho}-n_0$,     
and is given by 
$(-1)^{{N_c}-{\rho}}(q-q^{-1})^{{N_c}-{\rho}} 
[{N_c}-{\rho}]_q !$ $\times q^{({N_c}-\rho)({N_c}-\rho+1)/2}$  
for $j + \ell={N_c}-{\rho}-n_0$.   
Sum (\ref{Dsum}), $\Sigma({\hat \Delta}_{\xi})_J$, is thus given by    
\bea 
&  &  
 \epsilon_0^{{N_c}-{\rho}-n_0} \, \times \, (-1)^{{N_c}-{\rho}}
 q^{N_c({N_c}-1)/2 - {\rho}(S^Z-{N_c)}} 
\left( \prod_{j \in J} a_{\xi}^{\rm 6V}(t_{j}) 
\prod_{\ell \in \Sigma_R \setminus J} 
f(t_j-t_{\ell}) \right) \non \\
\non \\
& \times & q^{-n_0({\rho}+1) + ({N_c}-{\rho})({N_c}+{\rho}+3)/2} 
   {\frac {([{N_c}]_q!)^2} {[{\rho}]_q!} }  
\, (q-q^{-1})^{N_c-\rho}
\, \sum_{\ell=0, \ell \le \rho}^{{N_c}-{\rho}-n_0}  
(-1)^{\ell} \chi_{\xi, {N_c}-{\rho}-n_0 -\ell}^{+; J} 
\non \\ 
& \times &  
\sum_{{\sigma}=0}^{\rho} (-1)^{\sigma} q^{-({\rho}-1){\sigma}} 
q^{2{\sigma}(S^Z-{N_c})} 
\sum_{J_I \cup J_K = J}^{|J_I|={\rho}-{\sigma}, |J_K|={\sigma}}   
\prod_{j \in J_I} \prod_{k \in J_K} f(t_{k}-t_j)  
 \, G_{\ell}^{+; J_I, J_K} 
\non \\ 
& &  + \, O(\epsilon_0^{{N_c}-{\rho}-n_0+1}) \, . 
\label{Dsum2}
\eea
Sum (\ref{Dsum}) over $\nu$ is now reduced into sum (\ref{Dsum2}) 
over ${\sigma}$. Lemma \ref{lem:Dsum} follows from the next lemma 
\ref{lem:CF5}. 
\hfill \opensquare 

\begin{lem}
Let $J$ be a subset of $\Sigma_R$ with $\rho$ elements.  
We have for generic $q$ 
\bea
& & \sum_{{\sigma}=0}^{\rho} 
(-1)^{\sigma} q^{-({\rho}-1){\sigma}} q^{2(S^Z-{N_c}){\sigma}} 
 \sum_{J_I \cup J_K = J}^{|J_I|={\rho}-{\sigma},|J_K|={\sigma}}   
\left( \prod_{j \in J_I} \prod_{k \in J_K} f(t_{k}-t_j) \right) 
 G_{\ell}^{+; J_I, J_K}   \non \\
& = & (-1)^{\rho} q^{-{\rho}({\rho}-1)/2} 
q^{(S^Z-{N_c}){\rho}} (q-q^{-1})^{\rho} \,
\sum_{L \subset J}^{|L|=\ell} 
\exp(\sum_{j \in L} 2t_j)  \non \\
& & \times \prod_{i=0}^{\ell-1} [S^Z-{N_c} + i]_q 
\prod_{j=0}^{{\rho}-\ell-1} [S^Z-{N_c}-j]_q \, . 
\label{CF5}
\eea
\label{lem:CF5}
\end{lem}
The derivation of lemma \ref{lem:CF5} is given in Appendix C.

Substituting the expression of $\Sigma({\hat \Delta}_{\xi})_J$ derived in  
lemma \ref{lem:Dsum} into proposition \ref{prop:Dsum} 
and putting $N_c=kN$, we obtain lemma 
\ref{lem:kN}.

\subsection{Some comments on the highest weight conjectures}

The highest weight conjecture 
has been constructed gradually 
in a series of papers \cite{DFM,FM1,FM2,Odyssey}. 
It is found numerically  \cite{DFM} 
that the degenerate multiplicity of the $sl_2$ loop algebra 
should be given by some power of 2. 
It suggests that the degenerate eigenspace  
corresponds to such an irreducible representation 
forming a tensor product of the spin 1/2 
evaluation representations. 
However,  any connection to the Bethe ansatz was not discussed    
in Ref. \cite{DFM}. 
The spectral degeneracy of the XXZ spin chain at roots of unity 
was carefully compared with numerical solutions of the Bethe ansatz equations 
in Ref. \cite{FM1}, and  degenerate multiplets are explained  
in terms of complete $N$-strings. 
The first version of the highest weight conjecture was 
given in the last paragraph of \S 3 of Ref. \cite{FM2}, 
based on numerical classification 
of degenerate multiplets for $L=12$ and $N=3$. 
Here, the idea of regular Bethe states 
is implicit but should have been known.

In Refs. \cite{FM1,FM2}
 a unique highest weight vector is assigned   
 by such a vector that has the largest value of $S^Z$   
 in a given degenerate multiplet of the $sl_2$ loop algebra.    
We can understand the highest weight conjectures of 
Refs. \cite{FM1,FM2} correctly,  
even without the mathematical definition of highest weight vectors.

%
%
\setcounter{section}{6}
 \setcounter{equation}{0} 
 \renewcommand{\theequation}{7.\arabic{equation}}

\section{Highest weight polynomials and the Drinfeld polynomials}


\subsection{
Equivalence of the Fabricius-McCoy polynomial 
to the highest weight polynomial} 
%

\begin{df}
Let $| R \ra$ be a regular Bethe state at $q_0$ with regular Bethe roots  
${\tilde t}_1, {\tilde t}_2, \ldots, {\tilde t}_R$.    
We define $Y_{\xi}(v)$ by  
\be 
Y_{\xi}(v) =  \sum_{\ell=0}^{N-1} 
{\frac {\prod_{j=1}^{L} (\sinh(v- \xi_j - (2\ell +1) \eta_0) )}  
{\prod_{j=1}^{R} \sinh(v- {\tilde t}_j - 2 \ell  \eta_0) 
\sinh(v- {\tilde t}_j - 2 (\ell+1) \eta_0)} } \, . 
\ee
\end{df}

\begin{prop}
$Y_{\xi}(v)$ is a Laurent polynomial of variable $z=\exp(\mp 2Nv)$ 
with degree $r=(L-2R)/N$ 
in the cases of sector A ($r$ even) and sector B ($r$ odd). 
\end{prop} 
\par \noindent {\bf Proof.}
First, by definition, 
 $Y_{\xi}(v)$ is  a rational function 
 of variable $\exp(\mp 2v)$ with a period $2\eta_0$.  
In sector A where $L$ is even, we have    
$\sinh(v + 2N \eta_0) = q_0^N \sinh v$ with $q_0^N= \pm 1$.   
In sector B where $L$ is odd,  
we have $\sinh(v + 2N \eta_0) =  \sinh v$ since $q_0^N=1$. 
 Thus, it is at least a rational function 
 of variable $\exp(\mp 2Nv)$. 
 Secondly,  $Y_{\xi}(v)$ has no poles,  
since  ${\tilde t}_1, {\tilde t}_2, \ldots, {\tilde t}_R$, 
satisfy the Bethe ansatz equations (\ref{BAE}) at $q_0$. 
 Thirdly, the function $Y_{\xi}(v)$ has the asymptotic behavior: 
$Y_{\xi}(v) \propto \exp\left(\pm(L-2R)v \right)= z^{\mp (L-2R)/2N}$
 for $v \rightarrow \pm \infty$, where  $z=\exp(2Nv)$. Thus, the degree 
 of the Laurent polynomial is given by $(L-2R)/N$.  
\hfill \opensquare 
%

\begin{lem}
When $p$ is an integer with $p \equiv 0$ (mod $N$) 
and $q_0$ a root of unity with $q_0^{2N}=1$,  
or when $p$ is a half-integer with $p \equiv N/2$ (mod $N$) 
and $q_0$  a primitive $N$th root of unity with $N$ odd, 
 we have for any integer $n$  the following: 
\be 
\lim_{q \rightarrow q_0} 
\sum_{\ell=0}^{N-1} q^{\mp(p-n)(2\ell +1)} 
= \left\{ 
\begin{array}{ccccc} 
N q_0^{\mp(p-n)} &  & {\rm for} & n \equiv 0 & ({\rm mod}\, N) \, ,  \\ 
 0 &  & &  {\rm otherwise} &  .  \\ 
\end{array}
\right.
\ee
\label{lem:qsum}
\end{lem}

\begin{prop}
Let $| R \ra$ be a regular Bethe state at $q_0$ 
in sector A or B  in the inhomogeneous case. 
The Laurent polynomial $Y_{\xi}(v)$ of $| R \ra$ 
corresponds to the highest weight polynomial ${\cal P}^{\lambda}(z)$.    
\label{prop:FMP}
\end{prop} 
\par \noindent {\bf Proof.} 
We express $Y_{\xi}(v)(-1)^L e^{\mp (L-2R)v}{2^{(L-2R)}}$ as follows:  
\be 
{\frac {e^{\pm \sum_{j=1}^{R} 2{\tilde t}_j}} 
 {e^{\pm \sum_{k=1}^{L} 2{\xi}_k}}} \, 
 \sum_{\ell=0}^{N-1}  q_0^{\mp(L/2 -R)(2\ell+1) }  
 {\frac {\phi_{\xi}^{\pm} \left( e^{\mp 2v} q_0^{\pm(2\ell+1)} \right) } 
{{\tilde F}^{\pm}(e^{\mp 2v} q_0^{\pm 2\ell}) 
{\tilde F}^{\pm}(e^{\mp 2v} q_0^{\pm 2(\ell+1)})} } \, .  
\ee  
We  expand  
$Y_{\xi}(v)$ in terms of $\exp(\mp 2v)$. Since it is a polynomial 
of $\exp(\mp 2v)$, the infinite sum  reduces 
to a finite sum with an upper bound $r=(L-2R)/N$, where  
the $k$th term  vanishes at $q_0$ unless $k \equiv 0$ (mod $N$) 
 due to lemma \ref{lem:qsum}:       
\be  
Y_{\xi}(v) = {\frac {e^{\pm \sum_{j=1}^{R} 2{\tilde t}_j}} 
{2^{(L-2R)}} } N q_0^{\pm(L/2 -R)}  z^{\mp r/2} 
\sum_{k=0}^{r} {\tilde{\chi}}_{\xi, kN}^{\pm} 
\left( z^{\pm 1} q_0^N \right)^k \, . 
\label{Y-expd}
\ee
Here ${\tilde{\chi}}_{\xi, kN}^{\pm}$ are expressed by      
 equation (\ref{diag-inhomo}) in terms of rapidities,  ${\tilde t}_j$.    
When $q_0$ is a root of unity of type I, we have  
$(z q_0^N)^k =z^k$  for  $N$ odd  ($q_0^N=1$),  
and $(z q_0^N)^k =(-z)^k$  for  $N$ even ($q_0^N=-1$).  
When $q_0$ is a root of unity of type II, we have 
$(q_0^N z)^k = (-z)^k$ for $N$ odd ( $q_0^N=-1$). 
We thus have for type I, 
\be 
\sum_{k=0}^{r} \tilde{\chi}_{\xi, kN}^{+} \left( z q_0^N \right)^k = 
 \left\{ 
\begin{array}{cccc} 
\sum_{k=0}^{r} \tilde{\chi}_{\xi, kN}^{+} z^k &  {\rm for}  
& N: \mbox{odd} & (q_0^N=1) \, ,  \\
\sum_{k=0}^{r} \tilde{\chi}_{\xi, kN}^{+} (-z)^k &  {\rm for} 
& N: \mbox{even} & (q_0^N=-1) \, ,   
\end{array} 
\right. 
\ee
and for type II, 
\be 
\sum_{k=0}^{r} \tilde{\chi}_{\xi, kN}^{+} (q_0^N z)^k 
= \sum_{k=0}^{r} \tilde{\chi}_{\xi, kN}^{+} (-z)^k  
\qquad (N: \mbox{odd} ;\,  q_0^N=-1 ) . 
\ee
Thus, in the cases of sector A and sector B, 
 we have shown   
\be 
 \sum_{k=0}^{r} {\tilde{\chi}}_{\xi, kN}^{+} (q_0^N z)^k 
 = {\cal P}^{\lambda}(z) \, .  
\ee
\hfill \opensquare 

In the homogeneous case where $\xi_n=0$ for all $n$,  
we have the following:  
\begin{cor}
For any given regular XXZ Bethe state $| R \ra$ at $q_0$ 
in sector A or B the Fabricius-McCoy polynomial $P^{\rm{FM}}(u)$ 
corresponds to the highest weight polynomial ${\cal P}^{\lambda}(u)$.  
\label{cor:PFM}
\end{cor}

%
\subsection{Examples of regular Bethe states}
%

\subsubsection{The vacuum state as a regular XXZ Bethe state}

Let us calculate  polynomial ${\cal P}^{\lambda}(u)$ 
 for the vacuum state $| 0 \ra$ 
 where $L=6$, $R=0$, $N=3$ and $q_0^3=1$:  
\be 
{\cal P}^{\lambda}(u) 
= (1-a_1 u) (1-a_2 u ) = 1 - (a_1 + a_2) u + a_1 a_2 u^2 \, . 
\ee
When $N$ is odd and $q_0^N =1$, we have 
$\lambda_k = (-1)^k {\tilde \chi}_{\xi, kN}^{+}$. 
We have  $\lambda_1=6!/(3!)^2 = 20$,   
$\lambda_2 = 6!/(6!0!) = 1$. 
The highest weight parameters are  thus given by  
\be 
{\hat a}_1, {\hat a}_2 =  10 \pm 3 {\sqrt{11}} \, .  
\label{case:plus}
\ee
Since ${\hat a}_1$ and ${\hat a}_2$ are distinct ($m_1=m_2=1$), 
the vacuum state $| 0 \ra$ generates an irreducible representation.   
We thus have $(1+1)^2=4$ as the dimension.

Let us discuss the case where 
 $L=6$, $R=0$, $N=3$ and $q_0^3=-1$.  
When $N$ is odd and $q_0^N =-1$, 
we have $\lambda_{k} =  {{\tilde \chi}}_{\xi, kN}^{+}$. 
We thus have  
\be 
{\hat a}_1, {\hat a}_2 =  - 10 \pm 3 {\sqrt{11}} \, .  \label{case:minus}
\ee
For the XXZ Hamiltonian (\ref{hxxz}) with $L$ even, 
$q$ is mapped to $-q$ by the unitary transformation,  
$\prod_{j=1}^{L/2} \sigma_{2j}^Z$. 
Here, the Hamiltonian ${\cal H}_{XXZ}$ is mapped to $-{\cal H}_{XXZ}$, and 
highest weight parameters  ${\hat a}_j$ are transformed 
to $-{\hat a}_j$ for $j=1, 2$.   

\subsubsection{The regular XXZ Bethe state with one down-spin}

We now discuss  the case 
of $L=8$, $R=1$, $N=3$ and $q_0^3=1$.  
Here $q_0=\exp(\pm 2 \pi \sqrt{-1}/3 )$. 
Let us specify the Bethe root as 
$\exp(2 t_2) = (1- \sqrt{-1} q_0)/(q_0- \sqrt{-1})$. 
Noting $[3\ell+1]_{q_0}=1$,  $[3\ell+2]_{q_0}=-1$ and  $[3\ell]_{q_0}=0$ 
for integers $\ell$, we derive 
${\tilde{\chi}}_3^{+} = - 13(2 - \sqrt{3})$ and 
${\tilde{\chi}}_6^{+} = 7 - 4 \sqrt{3})$. 
We thus have 
\be 
{\cal P}^{\lambda}(u)= 1 - 13(2-\sqrt{3}) u + (7 - 4 \sqrt{3}) u^2 \, , 
\ee
where highest weight parameters ${\hat a}_1$ and ${\hat a}_2$ are given by 
\be 
{\hat a}_1 \, , \, {\hat a}_2 
= {\frac 1 2} (13 \pm \sqrt{165}) (2 -\sqrt{3}) \, .   
\ee

\subsubsection{The case of a reducible Weyl module: 
the inhomogeneous case with degenerate evaluation parameters}

We now discuss such a regular Bethe state that has  
 degenerate highest weight parameters.  
Let us consider the inhomogeneous case of $L=6$, $R=0$, $N=3$ 
and $q_0=\exp(2 \pi i/3)$. We set 
$\xi_1 \ne 0$, while $\xi_2= \xi_3= \xi_4= \xi_5= \xi_6=0$.  
We have  
$\phi^{+}_{\xi}(x) =  (1 - yx) (1-x)^5$, 
where $y=\exp(2 \xi_1)$. 
Expanding $\phi^{+}_{\xi}(x)$, we have 
$\chi_0^{+} = 1$, 
$\chi_3^{+} = 
- 5!/(3!2!) -  
5!/(2!3!) \, y =-(10 + 10 y)$,   and 
$\chi_6^{+} = y$. We have 
 ${\cal P}^{\lambda}(u)= 1 - 10 (1+y) u + y u^2$,  
where parameters ${\hat a}_1$ and ${\hat a}_2$ are given by 
\be 
{\hat a}_1, {\hat a}_2 = 5 \left( 1 +y \pm \sqrt{(1+y)^2 - y/5} \right) \, . 
\ee
We have ${\hat a}_1 ={\hat a}_2$ if and only if 
$y=y_c= (-49 \pm 3 \sqrt{-11})/50$.

When $y \ne y_c$, ${\hat a}_1$ and ${\hat a}_2$ are distinct, 
and hence $U {|0\ra}$ is irreducible. 
We have dim $U |0\ra$ = $(1+1)^2=4$, and ${\cal P}^{\lambda}(u)$
 gives the Drinfeld polynomial. 

When $y=y_c$, however, 
${\hat a}_1$ and ${\hat a}_2$ are degenerate: 
\be 
{\hat a}_1 = {\hat a}_2 = a_1 = {\frac 1 {10}} ( 1 \pm 3 \sqrt{-11}) \, .  
\ee 
Recall that  $U {|0\ra}$ is irreducible if and only if 
$ \left( {  x}_1^{-} - a_1 {  x}_{0}^{-} \right) |0 \ra$ 
vanishes. We apply (\ref{eq:cond}) with $r=2$, 
where generators ${x}_{1}^{-}$ and ${  x}_{0}^{-}$ 
are given by $T_{\xi}^{-(3)}= V^{-} T^{-(3)} V^{+}$ and  
$S_{\xi}^{-(3)}= V^{+} S^{-(3)} V^{-}$, respectively.  
Here we recall (\ref{eq:V-diag}).
The $(1,1,1,2,2,2)$ elements of vectors 
$T_{\xi}^{-(N)} | 0 \ra$ and $S_{\xi}^{-(N)} | 0 \ra$ are  
given by $q_0^{3/2}$ and $q_0^{-3/2}$, respectively, and hence   
 we have $T_{\xi}^{-(N)}|0 \ra  \ne  a_1  S_{\xi}^{-(N)} | 0 \ra$. 
We thus conclude that $U {|0\ra}$ is reducible. 
We now derive dim $U {|0 \ra} = 4$, and hence $U {|0 \ra}$ gives 
a Weyl module. 
In fact, the basis is given by 
$| 0 \ra$, $ {  x}_0^{-} | 0 \ra$, $({  x}_0^{-})^2 | 0 \ra$ and 
$w = ({  x}_1^{-} - a_1 {  x}_0^{-}) | 0 \ra$. 
We also confirm that it is reducible, noting that 
 ${  x}^{+}_k \, w = 0$ for $k \in {\bm Z}$.

%
%
\setcounter{section}{7}
 \setcounter{equation}{0} 
 \renewcommand{\theequation}{8.\arabic{equation}}

\section{Concluding remark}

The highest weight conjecture has been shown in sectors A and B, 
in the paper. 
However, operators commuting with the transfer matrix are 
constructed also in other sectors \cite{DFM}.  
We thus have a conjecture that 
in any sector of $S^Z$, some version of regularized 
 Bethe ansatz eigenvectors should be highest weight   
(see also,  Ref. \cite{Odyssey}). 


{\vskip 1.2cm}
\par \noindent 
{\bf Acknowledgements}

The author would like to thank Profs. M. Jimbo and B.M. McCoy 
for helpful comments. He would also like to thank  
Prof. K. Fabricius for useful comments.  
He is grateful to Prof. A. Nakayashiki for useful 
comments during the RIMS conference ``Solvable Lattice Models 2004'' 
in 20-23 July 2004.  
This work is partially supported by the Grant-in-Aid for 
Young Scientists (A): No. 14702012.

\appendix

%
%

 \setcounter{equation}{0} 
 \renewcommand{\theequation}{A.\arabic{equation}}
\section{Derivation of lemma \ref{lem:scheme} 
for showing highest weight properties}

For a given integer $\ell$, let $U({\cal B}_{\ell})$ 
be such a subalgebra of $U(L(sl_2))$ that is 
generated by ${h}_k, {  x}_{\ell+ k}^{+}$  
and ${  x}_{-\ell+1 +k}^{-}$ for $k \in {\bm Z}_{\ge 0}$. 
We denote by ${\cal B}_{\ell}^{+}$ such a subalgebra of $U({\cal B}_{\ell})$ 
that is generated by ${  x}_{\ell + k}^{+}$ for $k \in {\bm Z}_{\ge 0}$. 
We express by $(X)^{(n)}$ the $n$th power of $X$ divided by 
the factorial of $n$, i.e. $(X)^{(n)}=(X)^{(n)}_{q=1} = X^n / n!$ .

\begin{lem2} 
Let  $\ell$ be an integer. For a given positive integer $n$ 
we have      
\bea
&({\rm A}_n):&  ({  x}_{\ell}^{+})^{(n-1)} ({  x}_{1-\ell}^{-})^{(n)}  
 =  \sum_{k=1}^{n} (-1)^{k-1} {  x}_{k-\ell}^{-} 
({  x}_{\ell}^{+})^{(n-k)} 
({  x}_{1-\ell}^{-})^{(n-k)} \, \mbox{\rm mod} \, 
U({\cal B}_{\ell}) {\cal B}_{\ell}^{+},   \non \\
&({\rm B}_n):&   
({  x}_{\ell}^{+})^{(n)} ({  x}_{1-\ell}^{-})^{(n)}  
 =  {\frac 1 n} \, \sum_{k=1}^{n} (-1)^{k-1} {  h}_{k} 
 ({  x}_{\ell}^{+})^{(n-k)} 
 ({  x}_{1-\ell}^{-})^{(n-k)} \, \mbox{\rm mod} \, 
 U({\cal B}_{\ell}) {\cal B}_{\ell}^{+},  \non \\ 
&({\rm C}_n):&  
  {[} {  h}_j,  ({  x}_{\ell}^{+})^{(m)} 
({  x}_{1-\ell}^{-})^{(m)} {]}  = 0 \, \quad  
\mbox{\rm mod} \, U({\cal B}_{\ell}) {\cal B}_{\ell}^{+} \, \quad 
\mbox{for} \, \, m \le n \, \, {\mbox and} \, \, j \in {\bf Z} \, . \non 
\eea
\label{lem:ABC}
\end{lem2} 
\par \noindent 
{\bf Proof.} We first show the following relations by induction on $n$: 
\bea 
{[} {  h}_1, ({  x}_{\ell}^{+})^{(n)} {]} & = & 
2 {  x}_{\ell+1}^{+} ({  x}_{\ell}^{+})^{(n-1)} \, , \\ \non 
{[} ({  x}_{\ell}^{+})^{(n)}, {  x}_{1-\ell}^{-} {]} & = & 
({  x}_{\ell}^{+})^{(n-1)} {  h}_1 
+ {  x}_{1+\ell}^{+} ({  x}_{\ell}^{+})^{(n-2)} \, , \non \\ 
{[} {  h}_1, ({  x}_{1-\ell}^{-})^{(n)} {]} & = & 
(-2) \, {  x}_{2-\ell}^{-} 
({  x}_{1-\ell}^{-})^{(n-1)} \, , \non \\  
{[} {  x}_{\ell}^{+}, ({  x}_{1-\ell}^{-})^{(n)} {]} & = & 
({  x}_1^{-})^{(n-1)} {  h}_1 
- {  x}_{2-\ell}^{-} ({  x}_{1-\ell}^{-})^{(n-2)} \, .  
\label{four-rec}
\eea
Through the recursive relations we derive 
the following inductive formula for the product 
$({  x}_{\ell}^{+})^{(n-1)} 
({  x}_{1-\ell}^{-})^{(n)}$ with respect to $n$: 
\bea 
({  x}_{\ell}^{+})^{(n)} ({  x}_{1-\ell}^{-})^{(n+1)} 
& = & {  x}_{1-\ell}^{-} \,  
({  x}_{\ell}^{+})^{(n)} ({  x}_{1-\ell}^{-})^{(n)} 
+ {\frac 1 2} \, {[} {  h}_{1}, ({  x}_{\ell}^{+})^{(n-1)} 
({  x}_{1-\ell}^{-})^{(n)} {]} 
\non \\
& & \quad - ({  x}_{\ell}^{+})^{(n-1)} 
({  x}_{1-\ell}^{-})^{(n+1)} {  x}_{\ell}^{+} \, , \quad 
 {\rm for} \,\,  \ell \in {\bf Z} \, .  
\label{eq:ind} 
\eea
We now show three relations $({\rm A}_n)$, $({\rm B}_n)$ and $({\rm C}_n)$, 
inductively on $n$ as follows:  
We show $({\rm A}_1)$, $({\rm A}_2)$, $({\rm B}_1)$ and  $({\rm C}_1)$, 
directly. Relation $({\rm A}_n)$ is derived from $({\rm A}_{n-1})$ and 
$({\rm C}_{n-2})$. 
Here we make use of formula (\ref{eq:ind}). We derive  
relation $({\rm B}_n)$  from $({\rm A}_n)$,  
 $({\rm C}_{n-1})$ and $({\rm B}_{m})$ for $m \le n-1$, 
multiplying both hand sides of $({\rm A}_n)$ by ${  x}_{\ell}^{+}$ 
from the left.  We show 
$x_{\ell}^{+} (x_{\ell}^{+})^{(m)} (x_{1-\ell}^{-})^{(m)} 
\in U({\cal B}_{\ell}) {\cal B}_{\ell}^{+}$ for $m \le n-1$ 
by induction on $m$. Here we make use of $({\rm B}_{m})$ for $m \le n-1$ 
and $({\rm C}_{n-1})$.  Finally, 
 $({\rm C}_n)$ is derived from $({\rm B}_{n-1})$ and $({\rm C}_{n-1})$.  
Thus, the cycle of induction process, $({\rm A}_n)$, $({\rm B}_n)$ 
and $({\rm C}_n)$, is closed. 
\hfill \opensquare 

We remark that relation (\ref{eq:ind}) for $\ell=0$ 
is given by the classical limit of  formula 
$({\rm iv})_{r}$ in \S 3.5 of Ref. \cite{Chari-P1}. It is also 
reviewed in Appendix A of Ref. \cite{Criterion}.

\begin{lem2} 
Let $\ell$ be an integer. 
If a vector $\Psi$ in a finite-dimensional representation 
of $U({\cal B}_{\ell})$ satisfies 
\bea
x_{\ell+k}^{+} \, \Psi & = & 0 \, , \quad   
 h_{k} \, \Psi = d_k \Psi  \quad \mbox{for} 
\, \, \, k \in {\bm Z}_{\ge 0} \, , \non \\   
(x_{1-\ell}^{-})^{r} \, \Psi & \ne & 0 \, , 
\quad (x_{1-\ell}^{-})^{r+1} \, \Psi = 0 \, ,  
\eea
then we have the following: 
(i) For a given non-negative integer $n \le r$ and a given set of integers 
$k_j$ satisfying $1-\ell \le k_1 \le \cdots \le k_n$,  
we have with some $A_{k_1, \ldots, k_n} \in {\bm C}$     
\be 
(x_{1-\ell}^{-})^{r-n} x_{k_1}^{-} \cdots x_{k_n}^{-} \Psi  
= A_{k_1, \ldots, k_n} (x_{1-\ell}^{-})^{r} \Psi  \, . 
\label{eq:r-n}
\ee
(ii) The subspace of weight $-r$ of $U({\cal B}_{\ell}) \Psi$ 
is one-dimensional. Here we call eigenvalues of $h_0$ weights. 
(iii) Recall that eigenvalues $\lambda_n$ 
are defined by (\ref{diag-101}), i.e. (A.6) for $\ell=0$. Then, we have 
\be 
(x_{\ell}^{+})^{(n)}(x_{1-\ell}^{-})^{(n)} \Psi = \lambda_n \Psi \, , 
\quad 
\mbox{for} \, \, \, n = 1, 2, \ldots, r.  
\label{eq:lam-ell-n}
\ee
\label{lem:Psi} 
\end{lem2} 
{\bf Proof.} We show (i) by induction on $n$. The  case of $n=0$ is trivial. 
Let us assume the cases of $n-1$ and $n$. We have  
$$ 
x_{m}^{+} (x_{1-\ell}^{-})^{r+1-n} x_{k_1}^{-} \cdots x_{k_n}^{-}  \Psi 
 = A_{k_1, \ldots , k_n} x_{m}^{+}  \cdot 
 (x_{1-\ell}^{-})^{r+1} \Psi = 0 \non 
$$
Calculating $[x_{m}^{+}, \, 
(x_{1-\ell}^{-})^{(r+1-n)} x_{k_1}^{-} \cdots x_{k_n}^{-}]$ ,  
we derive (\ref{eq:r-n}) in the case of $n+1$.  
We now show (ii). 
Applying the Poincar{\' e}-Birkhoff-Witt theorem 
to $U({\cal B}_{\ell})$, we have that 
every vector $v$  in the subspace of weight $-r$ of $U \Psi$ 
is expressed as a linear combination of monomial vectors    
$x_{k_1}^{-} \cdots x_{k_r}^{-} \Psi$ 
for sets of integers $k_j$ satisfying $1- \ell \le k_1 \le  \cdots \le k_r$ 
 \cite{Criterion}. 
Thus,  we obtain (ii) from (i). 
We show (iii) by induction on $n$. 
We derive the $n=1$ case by $[x_{\ell}^{+}, x_{1-\ell}^{-}] = h_1$. 
Let us assume (\ref{eq:lam-ell-n}) for the cases of 
$1, 2, \dots, n-1$. 
We derive the case of $n$   
through $({\rm B}_n)$ of lemma \ref{lem:ABC}.  
\hfill \opensquare

Let ${\cal U}_{k}$ be 
the $sl_2$ subalgebra generated by ${  x}_{-k}^{+}$, 
${  x}_{k}^{-}$ and ${  h}_0$ for an integer $k$.

\par \noindent 
{\bf Proof of lemma \ref{lem:scheme} }
We show it in sequence as follows:  
(i) ${  h}_1 | \Phi \ra  =  {  d}_1 | \Phi \ra$ 
 where ${  d}_1 \in {\bm C}$; 
(ii) ${  x}_k^{+} |\Phi \ra  =  0$ ($k \in {\bm Z}_{>0})$;  
(iii) $ {  h}_{k} |\Phi \ra  =  {  d}_{k} |\Phi \ra$ 
($k \in {\bm Z}_{>0})$ where ${ d}_{k} \in {\bf C}$;    
\par  \noindent
(vi) ${  h}_{-1} | \Phi \ra  =  {  d}_{-1} | \Phi \ra$ 
where ${  d}_{-1} \in {\bm C}$;   
(v) $ {  x}_{-k}^{+} |\Phi \ra  =  0$  $(k \in {\bm Z}_{>0})$;  
(vi) ${  h}_{-k} |\Phi \ra  =  {  d}_{-k} |\Phi \ra$  
$( k \in {\bm Z}_{>0})$ where ${  d}_{-k} \in {\bm C}$.  
Let us first note that for $k = 0$ and 1, 
${\cal U}_k \, |\Phi \ra $  
corresponds to the $(r+1)$-dimensional irreducible 
representation of $U(sl_2)$. We therefore have 
$({  x}_{k}^{-})^{r} |\Phi \ra \ne  0$ and 
$({  x}_{k}^{-})^{r+1} |\Phi \ra = 0$ for $k=0, 1$.    
We derive (i) from (\ref{ann-10}) and (\ref{diag-101}), noting   
 ${  h}_1 = [{  x}_{0}^{+}, {  x}_{1}^{-}]$.  
 We show (ii) through induction on $k$: 
 \bea 
{  x}_{k+1}^{+} | \Phi \ra  & = & {\frac 1 2} 
\left({  h}_1 \, {  x}_k^{+} 
 - {  x}_k^{+} {  h}_1 \right) | \Phi \ra \non \\
 & = & {\frac 1 2} ({  h}_1 - {  d}_1) \, 
  {  x}_k^{+} | \Phi \ra \, , \quad \mbox{for} \, k \in {\bm Z}_{\ge 0} . 
 \eea
We derive (iii) by induction on $k$ 
using $({\rm B}_k)$ of lemma \ref{lem:ABC} for $\ell=0$:  
\be 
 h_k | \Phi \ra = \sum_{j=1}^{k-1} (-1)^{k-j+1} \lambda_{k-j} 
h_j | \Phi \ra + k (-1)^{k-1} \lambda_k | \Phi \ra 
\ee
In order to derive (iv), we first show that $\lambda_r \ne 0$ as follows: 
We note that vector $|\Phi \ra$ satisfies the necessary 
conditions of $\Psi$ in lemma \ref{lem:Psi}, since we have 
shown (ii) and (iii) in the above. 
From (iii) of lemma \ref{lem:Psi} for $\ell=1$, we have 
\be
({  x}_{1}^{+})^{(n)} ({  x}_{0}^{-})^{(n)} |\Phi \ra 
= {\lambda}_n^{+} |\Phi \ra 
\quad \mbox{for} \quad n = 1, 2, \ldots, r \, .  
\ee   
From (ii) of lemma \ref{lem:Psi}, we have  
$(x_1^{-})^r |\Phi \ra = A_1 (x_0^{-})^r |\Phi \ra$.  
Here $A_1 \ne 0$, since $(x_1^{-})^r |\Phi \ra \ne 0$ due to ${\cal U}_1$.  
We thus obtain that  ${\lambda}_r \ne 0$ \cite{Criterion}. 
We then consider $({\rm A}_{r+1})$ of lemma \ref{lem:ABC} 
for $\ell=1$: 
\bea  
({  x}_{1}^{+})^{(r)} ({  x}_{0}^{-})^{(r+1)}  
&  = &  \sum_{k=1}^{r+1} (-1)^{k-1} {  x}_{k-1}^{-} 
({  x}_{1}^{+})^{(r+1-k)} 
({  x}_{0}^{-})^{(r+1-k)} 
\, \mbox{\rm mod} \, U({\cal B}_1) {\cal B}_{1}^{+} \,. \non \\ 
\label{A(r+1)} 
\eea
Introducing ${\lambda}_{0}=1$, we have from (\ref{A(r+1)})  
\be 
\sum_{j=0}^{r} (-1)^{j} \, 
{\lambda}_{r-j} \, {  x}_{j}^{-} |\Phi \ra = 0 \, .   
\label{xk}
\ee
Applying ${  x}_{-1}^{+}$ to (\ref{xk}), we have 
\be 
{\lambda}_r
 \, {  h}_{-1} |\Phi \ra = 
\sum_{j=1}^{r} (-1)^{j-1} \, 
{\lambda}_{r-j} 
\, {  h}_{j-1} |\Phi \ra \, . 
\ee
Thus, we obtain ${  h}_{-1} |\Phi \ra = {  d}_{-1} | \Phi \ra$ where 
 ${  d}_{-1}$ is defined  by 
\be 
{ d}_{-1}  = {\frac 1 {{\lambda}_r} } \, 
\sum_{j=1}^{r} (-1)^{j-1} \, {\lambda}_{r-j} \, {  d}_{j-1} \, . 
\ee 
We show (v) inductively with respect to $k$:  
 \bea 
{  x}_{-(k+1)}^{+} | \Phi \ra  & = & {\frac 1 2} 
\left({  h}_{-1} \, {  x}_{-k}^{+} 
 - {  x}_{-k}^{+} {  h}_{-1} \right) | \Phi \ra \non \\
 & = & {\frac 1 2} ({  h}_{-1} - {  d}_{-1}) \, 
  {  x}_{-k}^{+} | \Phi \ra \quad \mbox{for} \, k \in {\bm Z}_{\ge 0}. 
 \eea
Multiplying (\ref{xk}) with ${  x}_{-k}^{+}$,  
we show (vi) through induction on $k$ by the following:  
\be 
  {h}_{-k} |\Phi \ra = {\frac 1 {{\lambda}_{r}}} 
\sum_{j=1}^{r} (-1)^{j-1} \, {\lambda}_{r-j} \, 
{  h}_{j-k} |\Phi \ra \, . 
\ee 
\hfill \opensquare 

%
%

 \setcounter{equation}{0} 
 \renewcommand{\theequation}{B.\arabic{equation}}


\section{Some combinatorial formulas}

\begin{lem2}[$q$-binomial theorem] 
For a positive integer $n$ we have 
\be 
\prod_{\ell = 0}^{n-1} (1 - q^{\pm 2 \ell} z ) = 
\sum_{j=0}^{n} (-1)^j q^{\pm (n-1)j} z^j \left[ 
\begin{array}{c} 
n \\
j  
\end{array}
\right]_q \, . 
\label{CFqb}
\ee
Here $q$ and $z$ are arbitrary. 
\end{lem2} 

\begin{lem2} 
Let $w_j$ be arbitrary parameters for $j = 1,2, \ldots, n$. 
 For an integer $n>0$  we have the following:   
\be 
\sum_{P\in {\cal S}(n)} 
\prod_{1 \le j < k \le n} f(w_{P j}-w_{P k}) =
\sum_{P\in {\cal S}(n)} 
\prod_{1 \le j < k \le n} f(w_{P k}-w_{P j}) = 
 [n]_q!  \, . 
\label{CF2}
\ee 
\end{lem2} 
\par \noindent {\bf Proof.} 
We express the sum in the left hand side of equation (\ref{CF2}) 
as $F(w_1, \cdots, w_n)$.   
Let us take an integer $j$ satisfying $1 \le j \le n$. 
As a function of variable $w_j$ the quantity $F(w_1, \cdots, w_n)$ 
is a meromorphic function with no poles, 
and it is bounded at infinity,  $w_j = \infty$. 
Therefore it is given by a constant with respect to the variable $w_j$.   
Similarly, we show that the quantity $F(w_1, \cdots, w_n)$ is a constant 
with respect to all variables $w_j$ for $j=1, 2, \ldots, n$. 
Let us evaluate the constant by substituting $w_j$ with $z_j$ of 
the complete $n$-string (\ref{N-string}). The summand of the sum 
in the left hand side of (\ref{CF2}) 
vanishes except for such a permutation $P$ that gives 
 $(P1, P2, \ldots, Pn)=(n, n-1, \ldots, 1)$.  
Thus, we have the equality (\ref{CF2}). 
\hfill \opensquare 

\begin{lem2}
 Let $m$ and $n$ be integers satisfying $m \ge n \ge 0$, and 
$w_j$ for $1 \le j \le m$ be arbitrary parameters. 
Then, we have   
\be 
\sum_{S_n \subseteq \Sigma_m}^{|S_n|=n} \prod_{j \in \Sigma_m \setminus S_n} 
\prod_{k \in S_n} f(w_{j}-w_{k}) =
\sum_{S_n \subseteq \Sigma_m}^{|S_n|=n} \prod_{j \in \Sigma_m \setminus S_n} 
\prod_{k \in S_n} f(w_{k}-w_{j}) =
 \left[ 
\begin{array}{c}
m \\
n 
\end{array} 
\right]_q  \, .  
\label{CF3} 
\ee
Here $\Sigma_m=\{ 1,2, \cdots, m \}$, and 
the sum is taken over all such subsets $S_n$ 
of  $\Sigma_m$ that have $n$ elements.  
Here $\prod_{j \in A} \prod_{k \in B} f(w_{j}-w_{k}) =1$ 
if $A$ or $B$ is empty.      
\end{lem2} 
\par \noindent {\bf Proof.}
We express the sum in the left hand side of equation (\ref{CF3}) 
as $G(w_1, \cdots, w_m)$.   
Let us take an integer $j$ satisfying $1 \le j \le m$. 
As a function of variable $w_j$ the quantity $G(w_1, \cdots, w_m)$ 
is a meromorphic function with no poles, 
and it is bounded at infinity,  $w_j = \infty$. 
Therefore it is a constant with respect to  $w_j$.   
Similarly we show that $G(w_1, \cdots, w_m)$ is a constant 
with respect to all variables $w_j$ for $j=1, 2, \ldots, m$. 
Let us evaluate the constant by substituting $w_j$ with $z_j$ of 
the complete $m$-string (\ref{N-string}). 
The product 
$\prod_{j \in \Sigma_m \setminus S_n} \prod_{k \in S_n} f(w_{j}-w_{k})$ 
vanishes except for the case of $S_n=\{1, 2, \ldots, n \}$.  
We thus have the equality (\ref{CF3}). 
\hfill \opensquare 

%
%

 \setcounter{equation}{0} 
 \renewcommand{\theequation}{C.\arabic{equation}}

\section{Proof of lemma  \ref{lem:MainCF} by induction on $n$}

The case of $n=1$ is trivial. 
Suppose that the case of  $n-1$ holds. 
We denote by ${\cal C}_n$ the cyclic group generated by 
a cyclic permutation $(12\cdots n)$,  
and by  ${\cal T}_{n-1}$ the symmetric group 
on the set $\{ 2, 3, \ldots, n \}$. 
Let us remark that any element $P$ of  ${\cal S}(n)$ is given by  
a product of an element $\sigma$ of ${\cal C}_n$ and an element 
$\tau$ of ${\cal T}_{n-1}$.    
%
%
Putting $P=\sigma \tau$ in the expression (\ref{CF0}), we have  
$\Delta_{S_n; \Sigma_M}^{\pm}$ as  
\bea 
& & \sum_{\sigma \in {\cal C}_n} \sum_{\tau 
\in {\cal T}_{n-1}} \prod_{\ell=1}^{n} 
\Bigg( \alpha_{\xi}^{\pm; \Sigma_M \setminus S_n}(w_{j_{\sigma \tau \ell}}) 
\prod_{k=\ell+1}^{n} q^{\pm 1} 
f(w_{j_{\sigma \tau \ell}}-w_{j_{\sigma \tau k}}) 
\non \\ 
& & \quad - {\bar \alpha}_{\xi}^{\pm; \Sigma_M \setminus S_n}
(w_{j_{\sigma \tau \ell}}) 
\prod_{k =\ell+1}^{n} q^{\mp 1} 
f(w_{j_{\sigma \tau k}}-w_{j_{\sigma \tau \ell}})  
\Bigg) \, . 
\eea
Here we note that $\sigma \tau 1= \sigma 1$. For the case of $\ell=1$ 
we have       
\be 
\prod_{k=2}^{n} f(w_{j_{\sigma \tau 1}}- w_{j_{\sigma \tau k}}) 
=\prod_{k=2}^{n} f(w_{j_{\sigma 1}}-w_{j_{\sigma \tau k}})  
= \prod_{k=2}^{n} f(w_{j_{\sigma 1}}-w_{j_{\sigma k}}) \, , 
\ee
and we have  $\Delta_{S_n; \Sigma_M}^{\pm}$ as follows   
\bea 
& & \sum_{\sigma \in {\cal C}_n} \Bigg\{
\Bigg( \alpha_{\xi}^{\pm; \Sigma_M \setminus S_n}(w_{j_{\sigma 1}}) 
\prod_{k=2}^{n} q^{\pm 1} 
f(w_{j_{\sigma 1}}-w_{j_{\sigma  k}})  
 - {\bar \alpha}_{\xi}^{\pm; \Sigma_M \setminus S_n}(w_{j_{\sigma 1}})
 \non \\ 
& \times  \prod_{k=2}^{n}& q^{\mp 1} 
f(w_{j_{\sigma k}}-w_{j_{\sigma 1}}) \Bigg) 
\sum_{\tau \in {\cal T}_{n-1}}  \prod_{\ell=2}^{n} 
\Bigg( \alpha_{\xi}^{\pm; \Sigma_M \setminus S_n}(w_{j_{\sigma \tau \ell}}) 
\prod_{k=\ell+1}^{n} q^{\pm 1} 
f(w_{j_{\sigma \tau \ell}} - w_{j_{\sigma \tau k}}) 
\non \\ 
& &  
- {\bar \alpha}_{\xi}^{\pm; \Sigma_M \setminus S_n}(w_{j_{\sigma \tau \ell}})
\prod_{k =\ell+1}^{n} q^{\mp 1} 
f(w_{j_{\sigma \tau k}}-w_{j_{\sigma \tau \ell}})  
\Bigg) \Bigg\} \, . \label{2L}
\eea
Here we note that ${\cal T}_{n-1}$ is equivalent 
to ${\cal S}(n-1)$.  
Let us define $j_{\sigma}$ by $ (j_{\sigma})_{k}=j_{\sigma k}$ 
($k= 1, \ldots, n$). Applying the induction assumption 
to the case where $S_{n-1}$ is given by  
$\{  j_{\sigma k} | k=2, \ldots, n \}$, 
the last line of (\ref{2L}) is given by  
\bea 
& & \sum_{\tau \in {\cal T}_{n-1}}  \prod_{\ell=2}^{n} 
\Bigg(\alpha_{\xi}^{\pm; \Sigma_M \setminus S_n}(w_{(j_{\sigma})_{\tau \ell}}) 
\prod_{k=\ell+1}^{n} q^{\pm 1} 
f(w_{(j_{\sigma})_{\tau \ell}}- w_{(j_{\sigma})_{\tau k}}) 
\non \\ 
& & \qquad - {\bar \alpha}_{\xi}^{\pm; \Sigma_M \setminus S_n}
(w_{(j_{\sigma})_{\tau \ell}})
\prod_{k =\ell+1}^{n} q^{\mp 1} 
f(w_{(j_{\sigma})_{\tau k}} - w_{(j_{\sigma})_{\tau \ell}}) \Bigg)
\non \\
& = & 
\sum_{\tau \in {\cal T}_{n-1}}  
\sum_{k=0}^{n-1} (-1)^k \left[ 
\begin{array}{c} 
n-1 \\
k 
\end{array}
 \right]_q q^{\pm (n-1)(n-2)/2} q^{\mp (n-2)k} \times \non \\ 
& & \times \, 
\prod_{2 \le \ell \le n-k} 
\alpha_{\xi}^{\pm; \Sigma_M \setminus S_n}
(w_{(j_{\sigma})_{\tau \ell}}) 
\prod_{n-k < \ell \le n} 
{\bar \alpha}_{\xi}^{\pm; \Sigma_M \setminus S_n}
(w_{(j_{\sigma})_{\tau \ell}})   
\non \\ 
& & \quad \times 
\prod_{2 \le \ell < m \le n} 
f(w_{(j_{\sigma})_{\tau \ell}}-w_{(j_{\sigma})_{\tau m}}) \, .  
\eea
Substituting it to (\ref{2L}), 
rewriting $\sigma \tau$ as $P \in {\cal S}(n)$, and
using  the recursive relations of the $q$-binomial coefficients,   
we obtain formula (\ref{CF1}) for the case of $n$.

%
%

 \setcounter{equation}{0} 
 \renewcommand{\theequation}{D.\arabic{equation}}
\section{Proof of lemma \ref{lem:main}}

We call a sequence of integers given by   
$n_k = n_1 + k-1$ for $k=1, \ldots, \ell$ with 
positive integers $n_1$ and $\ell$,  
a slope 1 increasing sequence of integers. 
Let us assume that $Z_{N_c} \setminus W$ is given by 
the union of sets of slope 1 increasing sequences of integers, 
$X_j$:  
$Z_{N_c} \setminus W = X_1 \cup X_2 \cup \cdots \cup X_m$.  
We have 
$Z_{N_c}= Y_0 \cup X_1 \cup Y_1 \cdots \cup X_m \cup Y_m$ 
where  $Y_j$ are slope 1 increasing sequences of integers,  
and $W= Y_0 \cup Y_1 \cup \cdots \cup Y_m$. 
Recall the notation ${\underline s}=s+R$.  
We now show that if $Y_1$ is given by    
$Y_1 = \{{\underline s}, {\underline s+1}, \ldots, {\underline t}\}$ 
with $s \le t$, we have 
\be
\Delta(\xi)^{\pm; \ldots, s-1, t+1, \ldots }_{j_1, \ldots, j_{\rho}} 
=0. 
\ee
Let us first consider the case of $s < t$. 
We express elements of $S_{N_c}$ as $i_1, i_2, \ldots, i_{N_c}$.  
Recall that each permutation $P$ gives 
sequence $(i_{P1}, i_{P2}, \ldots, i_{P{N_c}})$. 
From the fact  
\be
f(z_{s}-z_{s+1})= \cdots = f(z_{t-1}-z_{t}) =0 \, , 
\ee
it follows that product 
$\prod_{1 \le \ell < m \le {N_c}} f(w_{i_{P \ell}}-w_{i_{P m}})$ 
in formula (\ref{CF1}) vanishes unless integers ${\underline s}, 
{\underline s+1}, \ldots, {\underline t}$ appear 
in reverse order in sequence $(i_{P1}, i_{P2}, \ldots, i_{P{N_c}})$:  
if ${\underline a}=i_{P j(a)}$ and ${\underline a}+1=i_{P j(a+1)}$ then 
$ j(a+1) < j(a)$ for $s \le a  \le t$, i.e.  
${\underline s}$ comes later than ${\underline s+1}$ 
in $(i_{P1}, i_{P2}, \ldots, i_{P{N_c}})$, and so on. 
We next consider product $\prod_{1 \le \ell \le {N_c}-k} 
\alpha_{\xi}^{\pm; \Sigma_{R+{N_c}} \setminus S_{N_c}}(w_{i_{P \ell}})$.  
If subsequence $(i_{P1}, i_{P2}, \ldots, i_{P({N_c}-k)})$ 
contains integer ${\underline t}$, then the product vanishes. 
Similarly, if  subsequence 
$(i_{P({N_c}-k+1)}, \ldots, i_{P({N_c}-1)}, i_{P{N_c}})$ contains 
integer ${\underline s}$, then product 
$\prod_{{N_c}-k < \ell \le {N_c}} 
{\bar \alpha}_{\xi}^{\pm; \Sigma_{R+{N_c}} 
\setminus S_{N_c}}(w_{i_{P \ell}})$ vanishes.   
In order to make the products nonzero, 
integer ${\underline s}$ should be contained in subsequence 
$(i_{P1}, i_{P2}, \ldots, i_{P({N_c}-k)})$  and 
integer ${\underline t}$ 
in $(i_{P({N_c}-k+1)}, \ldots, i_{P({N_c}-1)}, i_{P{N_c}})$. 
However, it is not compatible with the constraint 
that integers ${\underline s}, {\underline s}+1, \ldots, {\underline t}$ 
should appear in reverse order 
in sequence $(i_{P1}, i_{P2}, \ldots, i_{P{N_c}})$.  
Thus, the summand of the sum 
$\Delta(\xi)^{\pm}_{S_{N_c}; \Sigma_{R+{N_c}}}$ in (\ref{CF1}) 
vanishes for any permutation $P$, 
and hence the sum $\Delta^{\pm}_{S_{N_c}; \Sigma_{R+{N_c}}}$ vanishes.   
In the case of $s=t$, we show that 
$\prod_{1 \le \ell \le {N_c}-k} 
\alpha_{\xi}^{\pm; \Sigma_{R+{N_c}} \setminus S_{N_c}}(w_{i_{P \ell}}) = 0 $
or $\prod_{{N_c}-k < \ell \le {N_c}} 
{\bar \alpha}_{\xi}^{\pm; \Sigma_{R+{N_c}} 
\setminus S_{N_c}}(w_{i_{P \ell}}) = 0$ for any $k$ and $P$.

For an illustration,   
we show $\Delta(\xi)_{j_1,j_2,j_3}^{\pm; 1,2,5}=0$ for ${N_c}=5$. 
Since $f(z_{3}-z_{4})=0$, 
product $\prod_{1 \le \ell < m \le 5} f(w_{i_{P \ell}}-w_{i_{P m}})$ 
vanishes  unless $\underline{4}$ comes earlier than $\underline{3}$ 
in $(i_{P1}, i_{P2}, i_{P3}, i_{P4}, i_{P5})$. 
When $Z_{N_c} \setminus W =\{\underline{1},\underline{2},\underline{5} \}$, 
 we have  
${\bar \alpha}_{\xi}^{\pm; \Sigma_{R+{N_c}} \setminus S_{N_c}}(z_{3}) = 0$ 
and $\alpha_{\xi}^{\pm; \Sigma_{R+{N_c}} \setminus S_{N_c}}(z_{4}) = 0.$
If  $\underline{4} \in \{i_{P1}, i_{P2}, \ldots, i_{P({N_c}-k)} \}$,   
product $\prod_{1 \le \ell \le {N_c}-k} 
\alpha_{\xi}^{\pm; \Sigma_{R+{N_c}} \setminus S_{N_c}}(w_{j_{P \ell}})$ 
vanishes.  
If $\underline{3} \in \{i_{P({N_c}-k+1)}, \ldots, i_{P{N_c}} \}$, 
then product 
$\prod_{{N_c}-k < \ell \le {N_c}} 
{\bar \alpha}_{\xi}^{\pm; \Sigma_{R+{N_c}} \setminus S_{N_c}}(w_{j_{P \ell}})$ 
vanishes. However, it is not compatible with 
the constraint that $\underline{4}$  comes earlier than 
$\underline{3}$ in sequence 
$(i_{P1}, i_{P2}, \ldots, 
i_{P5})$. Therefore, we have 
$\Delta(\xi)_{j_1,j_2,j_3}^{\pm; 1,2,5}=0$.

%
%

 \setcounter{equation}{0} 
 \renewcommand{\theequation}{E.\arabic{equation}}

\section{Derivation of equation (\ref{CF5})}

 Generalizing $G_{\ell}^{+; J_I,J_K}$ given  
by equation (\ref{defGC}), 
we define for $\ell \le r$
the following:  
\be 
C_{\ell}^{J_I,J_K}(m)=
\sum_{\ell_I = 0}^{\ell} 
q^{m (\ell_K - \ell_I) } 
\sum_{L_I \subset J_I}^{|L_I|= \ell_I} \exp(\sum_{j \in L_I} 2t_j)   
\sum_{L_K \subset J_K}^{|L_K|= \ell_K} \exp(\sum_{j \in L_K} 2t_j) \, ,      
\label{eq:dfC}
\ee
where $\ell_K=\ell-\ell_I$. 
When $m={\rho}-1$, the generalized coefficient gives the original one: 
$C_{\ell}^{J_I,J_K}({\rho}-1)= G_{\ell}^{+; J_I,J_K}$. 
In terms of $C_{\ell}^{J_I,J_K}(m)$ we introduce 
\be 
J({\rho},{\sigma}; \ell)_{m} = \sum_{J_I \cup J_K = J}^{|J_K|={\rho}-{\sigma}, |J_K|={\sigma}} 
\left( \prod_{j \in J_I} \prod_{k \in J_K} f(w_{k}-w_{j})  \right) 
C_{\ell}^{J_I,J_K}(m) \, ,  
\label{eq:dfJ}
\ee
Here, $w_j$ $(1 \le j \le r)$ are arbitrary parameters, and 
the sum is taken over all pairs of disjoint subsets $J_I$ and $J_K$ of $J$ 
with $J_I \cup J_K = J$.   
We note that $J({\rho},{\sigma}; \ell)_m$ vanishes when $\ell > \rho$ 
or $\sigma > \rho$ by definition. 
Let us denote by $\Sigma(J)_{\ell,m}^{\rho}$ 
the following sum over ${\sigma}$:  
\be 
\Sigma(J)^{\rho}_{\ell,m}= \sum_{{\sigma}=0}^{\rho} (-1)^{\sigma} q^{-({\rho}-1){\sigma}} x^{\sigma} J({\rho},{\sigma}; \ell)_m \, . 
\label{eq:SJ}
\ee
The main result of Appendix E is given as follows: 
\be 
\Sigma(J)_{\ell,m}^r
=  q^{- \ell m} \prod_{i=1}^{\ell} (1- x q^{2m - 2({\rho}-i)} ) 
\prod_{j=0}^{{\rho}-\ell-1} (1- x q^{-2j}) \, 
\sum_{L \subset J}^{|L|=\ell} \exp(\sum_{j \in L} 2t_j ) \, .  
\label{eq:mainC}
\ee
Let us derive (\ref{eq:mainC}). 
We first show that $J({\rho},{\sigma}; \ell)_{m}$ is expressed as follows:    
\be
J({\rho},{\sigma}; \ell)_m = \sum_{t=0}^{\ell} 
\left[ 
\begin{array}{c} 
{\rho}- \ell \\
{\sigma}-t
\end{array} 
\right]_q
\left[ 
\begin{array}{c} 
\ell \\
t
\end{array} 
\right]_q
q^{\ell({\sigma}-m) +(2m -{\rho})t} 
 \sum_{L \subset J}^{|L|=\ell} \exp(\sum_{j \in L} 2t_j ) \, .  
\label{Jrs}
\ee
Expression (\ref{Jrs}) is derived through induction on $\ell$. 
The case of $\ell=0$ is given by formula (\ref{CF3}). 
For the case of $\ell>0$, 
we note that we have from (\ref{eq:dfC})   
\bea 
\lim_{t_{j_{\rho}}\rightarrow \infty} 
J({\rho},{\sigma}; \ell)_m /\exp(2t_{j_{\rho}}) &= &
J({\rho}-1, {\sigma}; \ell-1)_m q^{{\sigma}-m} \non \\
& & \quad + J({\rho}-1, {\sigma}-1; \ell-1)_m q^{-{\rho}+{\sigma}+m} \, ,  
\label{recJ}  
\eea
and also that $J({\rho},{\sigma}; \ell)_m$ is symmetric with respect to 
$\exp{2t_{j_1}}$, $\exp{2t_{j_2}}$, $\ldots$, $\exp{2t_{j_{\rho}}}$,  
by  definition.  
Making use of (\ref{recJ}) and the symmetric property,  
we have (\ref{Jrs}). 
We now evaluate  coefficient $Z({\rho},{\sigma}; \ell)_m$ defined 
in the following:       
\be 
J({\rho},{\sigma}; \ell)_m = Z({\rho},{\sigma}; \ell)_m  \, 
\sum_{L \subset J}^{|L|=\ell} \exp(\sum_{j \in L} 2t_j ) \, .  
\ee
 From the recursion relation (\ref{recJ}) we have  
\be 
Z({\rho},{\sigma}; \ell)_m  = 
Z({\rho}-1, {\sigma}; \ell-1)_m \, q^{{\sigma}-m} 
+ Z({\rho}-1, {\sigma}-1; \ell-1)_m \, q^{-{\rho}+{\sigma}+m} \, .  
\label{recZ}  
\ee
Let us define  $\Sigma(Z)^{\rho}_{\ell,m}$ by 
\be 
\Sigma(Z)^{\rho}_{\ell,m}
= \sum_{{\sigma}=0}^{\rho} (-1)^{\sigma} 
q^{-({\rho}-1){\sigma}} x^{\sigma} Z({\rho},{\sigma}; \ell)_m \, . 
\label{eq:SJ2}
\ee 
We reformulate the sum over ${\sigma}$  as follows: 
 \be 
%
\Sigma(J)^{\rho}_{\ell,m}
= 
%
\Sigma(Z)^{\rho}_{\ell,m}
\sum_{L \subset J}^{|L|=\ell} \exp(\sum_{j \in L} 2t_j )  \, . 
\ee
Here we recall that $\ell \le r$. 
Through induction on $\ell \ge 0$ and ${\rho}-\ell \ge 0$  using the recursion 
relation (\ref{recZ}), we have  
\be 
\Sigma(Z)^{\rho}_{\ell,m}
= q^{- \ell m} \prod_{i=1}^{\ell} (1- x q^{2m - 2({\rho}-i)} ) 
\prod_{j=0}^{{\rho}-\ell-1} (1- x q^{-2j}) \, . 
\ee

\vskip 24pt 
\par \noindent 
\, {\bf References} 
\vskip 12pt




\begin{thebibliography}{[99]}

\bibitem{Baxter73} R. Baxter, 
Eight-Vertex Model in Lattice Statistics and One-Dimensional Anisotropic 
Heisenberg Chain. I. Some Fundamental Eigenvectors, 
Ann. Phys. {\bf 76} (1973) 1-24. 
%


\bibitem{Baxter} R.~J. Baxter, 
Completeness of the Bethe ansatz for the six and eight vertex models, 
J. Stat. Phys. {\bf 108} (2002) 1--48.  

\bibitem{Andrei} D. Braak and N. Andrei, 
 On the Spectrum of the XXZ-Chain at Roots of Unity, 
J. Stat. Phys. {\bf 105} (2001) 677--709. 

\bibitem{Chari} V. Chari, 
Integrable representations of affine Lie-algebras, 
Invent. math. {\bf 85} (1986) 317--335.  

\bibitem{Chari-P0} V. Chari and A. Pressley, 
New Unitary Representations of Loop Groups, 
Math. Ann. {\bf 275} (1986) 87--104.  


\bibitem{Chari-P1} V. Chari and A. Pressley, 
Quantum Affine Algebras, 
Commun. Math. Phys. {\bf 142} (1991) 261--283.  

\bibitem{Chari-P2} V. Chari and A. Pressley, 
Quantum Affine Algebras at Roots of Unity, 
Represent. Theory {\bf 1} (1997) 280--328. 


\bibitem{Chari-P3} V. Chari and A. Pressley, 
Weyl modules for classical and quantum affine algebras, 
Represent. Theory {\bf 5} (2001) 191--223. 


\bibitem{Missing} T. Deguchi,   
Construction of some missing eigenvectors of the XYZ spin chain 
at the discrete coupling constants 
and the exponentially large spectral degeneracy 
of the transfer matrix, 
J. Phys. A {\bf 35} (2002) 879--895.  


\bibitem{CSOS} T. Deguchi,  
 The 8V CSOS model and the $sl_2$ loop algebra symmetry 
 of the six-vertex model at roots of unity, 
Int. J. Mod. Phys. B {\bf 16} (2002) 1899--1905. 

\bibitem{twisted} T. Deguchi, 
The $sl_2$ loop algebra symmetry of the twisted transfer 
matrix of the six-vertex model at roots of unity, 
J. Phys. A {\bf 37} (2004) 347--358. 

\bibitem{Poincare} T. Deguchi, 
The Six-Vertex Model at Roots of Unity 
and some Highest Weight Representations of the 
$sl_2$ Loop Algebra, 
Ann. Henri Poincar{\'e} {\bf 7}(2006) 1531--1540 (cond-mat/0603112).   

\bibitem{Criterion} T. Deguchi, 
Irreducibility criterion for a highest weight representation of 
the $sl_2$ loop algebra and the dimensions of reducible representations, 
 J. Stat. Mech (2007) P05007 (math-ph/0610002). 

\bibitem{DFM} T. Deguchi, K. Fabricius and B.~M. McCoy, 
The $sl_2$ Loop Algebra Symmetry of the Six-Vertex Model at 
Roots of Unity, 
J. Stat. Phys. {\bf 102} (2001) 701--736.   

\bibitem{Drinfeld} V.G. Drinfeld, A new realization of Yangians and 
quantized affine algebras, Soviet Math. Doklady {\bf 36} (1988) 212--216.  

\bibitem{FM1} K. Fabricius and B.~M. McCoy, 
Bethe's Equation Is Incomplete for the XXZ Model at Roots of Unity,    
J. Stat. Phys. {\bf 103} (2001) 647--678.   

\bibitem{FM2} K. Fabricius and B.~M. McCoy, 
Completing Bethe's Equations at Roots of Unity,  
J. Stat. Phys. {\bf 104} (2001) 573--587.  

\bibitem{Odyssey} K. Fabricius and B.~M. McCoy, 
Evaluation Parameters and Bethe Roots for the Six-Vertex Model 
at Roots of Unity, 
in Progress in Mathematical Physics Vol. 23 
 ({\it MathPhys Odyssey 2001}), edited by 
  M. Kashiwara and T. Miwa, (Birkh{\"a}user, Boston, 2002) 119--144. 


\bibitem{FM-8vertex} K. Fabricius and B.~M. McCoy, 
New Developments in the Eight-Vertex Model, 
J. Stat. Phys. {\bf 111} (2003) 323--337; 
Functional Equations and Fusion Matrices for the Eight-Vertex Model, 
Publ. of RIMS, Kyoto Univ. {\bf 40} (2004) 905--932. 

\bibitem{FM-8vertex2} K. Fabricius and B.~M. McCoy, 
New Developments in the Eight-Vertex Model II. Chains of odd length, 
J. Stat. Phys. {\bf 120} (2005) 37--70.  


\bibitem{Elliptic-Current} K. Fabricius and B.~M. McCoy, 
An elliptic current operator for the eight-vertex model, 
J. Phys. A: Math. Gen. {\bf 39} (2006) 14869--14886.   

\bibitem{Jimbo} M. Jimbo, 
A $q$-analogue of $U(gl(N+1))$, Hecke algebra 
and the Yang-Baxter equation, 
 Lett. Math. Phys. {\bf 11} (1986) 247--252. 


\bibitem{Jimbo-review} M. Jimbo, 
Topics from Representations of $U_q(g)$--An Introductory 
Guide to Physicists, in {\it Nankai Lectures on Mathematical 
Physics} (World Scientific, Singapore, 1992) pp. 1-61.   


\bibitem{Korepin} 
V.E. Korepin, N.M. Bogoliubov and A.G. Izergin, 
{\it Quantum Inverse Scattering Method 
and Correlation Functions} (Cambridge University Press,  Cambridge, 1993) 


\bibitem{Korff-twisted} C. Korff, 
The twisted XXZ chain at roots of unity revisited, 
J. Phys. A {\bf 37} (2004) 1681--1689. 

\bibitem{Korff-Q} C. Korff, 
Auxiliary matrices for the six-vertex model at roots of 1 
and a geometric interpretation of its symmetries, 
J. Phys. A {\bf 36}  (2003) 5229--5266.  


\bibitem{Korff-Q05} C. Korff, 
Auxiliary matrices on both sides of the equator, 
J. Phys. A {\bf 38} (2005) 47--67. 

\bibitem{Korff-McCoy} C. Korff and B.~M. McCoy, 
Loop symmetry of integrable vertex models at roots of unity, 
Nucl. Phys. {\bf B} (2001) {\bf 618} [FS] 551--569. 


\bibitem{Modular} G. Lusztig, Modular representations and 
quantum groups, Contemp. Math. {\bf 82} (1989) 59--77. 

\bibitem{Lusztig} G. Lusztig, {\it Introduction to Quantum Groups} 
(Birkh{\"a}user, Boston, 1993). 
 

\bibitem{8VABA} L. Takhtajan and L. Faddeev, 
The Quantum Method of the Inverse Problem and the 
Heisenberg XYZ Model, 
 Russian Math. Surveys {\bf 34}(5) (1979) 11--68. 

\bibitem{TF} L. Takhtajan and L. Faddeev, 
Spectrum and Scattering of Excitations 
in the One-Dimensional Isotropic Heisenberg Model, 
J. Sov. Math. {\bf 24} (1984) 241--267. 


\bibitem{Tarasov-cyclic} V.~O. Tarasov, 
 Cyclic Monodromy matrices for the $R$-matrix of the six-vertex model 
and the chiral Potts model with fixed spin boundary conditions,  
Int. J. Mod. Phys. A {\bf 7} Suppl. {\bf 1B} (1992) 963--975. 
%

\bibitem{Tarasov} V. Tarasov, On the Bethe vectors 
for the XXZ model at roots of unity, 
J. Math. Sciences {\bf 125} (2005) 242--248 
(math.QA/0306032).  



\end{thebibliography}
\end{document}